\documentclass[reqno,12pt]{article}
\pdfoutput=1
\usepackage{ae}
\usepackage[T1]{fontenc}
\usepackage[ansinew]{inputenc}
\usepackage{mathrsfs}
\usepackage{amsmath}
\usepackage{amssymb}
\usepackage{graphicx}
\usepackage{color}
\definecolor{darkblue}{cmyk}{0.9,0.9,0,0}
\definecolor{darkgreen}{rgb}{0,0.55,0}
\usepackage[colorlinks=true,linkcolor=darkblue,citecolor=darkblue,urlcolor=darkblue]{hyperref}
\usepackage{epsfig}
\usepackage{cite}

\usepackage{amsmath,amsfonts,amssymb,amsbsy}
\usepackage{graphicx,color}
\usepackage{graphics}
\usepackage{epsfig}
\usepackage{verbatim}
\usepackage{stmaryrd}
\usepackage{epsf}

\usepackage{amsmath, amssymb,graphicx}
\usepackage{epsfig}
\usepackage{verbatim}
\usepackage{amsfonts}
\usepackage{ulem}
\usepackage{wrapfig}

\usepackage{tikz}
\usetikzlibrary{arrows,shapes}
\usetikzlibrary{patterns,fadings}

\newcommand{\btp}{\begin{tikzpicture}[baseline=0pt,scale=0.9,line width=0.7pt]}
\newcommand{\btpp}{\begin{tikzpicture}[baseline=-5pt,scale=0.25,line width=0.7pt]}
\newcommand{\etp}{\end{tikzpicture}}

\def\bc{\begin{center}}
\def\ec{\end{center}}

\usepackage{epsf}

\newcommand{\be}{\begin{equation}}
\newcommand{\ee}{\end{equation}}
\newcommand{\ba}{\begin{eqnarray}}
\newcommand{\ea}{\end{eqnarray}}
\newcommand{\nn}{{\nonumber}}

\newcommand{\beaa}{\begin{eqnarray}}
\newcommand{\eeaa}{\end{eqnarray}}
\newcommand{\etap}{\tilde\eta}

\newcommand{\alf}{{\textstyle{\frac{1}{2}}}}

\newcommand{\Refl}{{\mathcal R}}

\newcommand{\sign}{{\rm sign}}

\newcommand{\eps}{\epsilon}

\newcommand{\calpha}{{\check\alpha}}
\newcommand{\cbeta}{{\check\beta}}

\definecolor{darkgreen}{rgb}{0.1,0.7,0.1}

\newcommand{\Blue}[1]{{\color{blue}#1\color{black}}}
\newcommand{\Red}[1]{{\color{red}#1\color{black}}}
\newcommand{\la}[1]{\label{#1}}

\DeclareFontFamily{OT1}{pzc}{}
\DeclareFontShape{OT1}{pzc}{m}{it}{<-> s * [1.10] pzcmi7t}{}
\DeclareMathAlphabet{\mathpzc}{OT1}{pzc}{m}{it}

\def\({\left(}
\def\){\right)}
\def\[{\left[}
\def\]{\right]}

\def\<{\langle}
\def\>{\rangle}

\def\bY{{\bf Y}}
\def\bT{{\bf T}}
\def\cR{{\cal R}}
\def\cB{{\cal B}}

\def\cO{{\cal O}}
\def\cY{{\cal Y}}
\def\cX{{\cal X}}

\def\baY{{\overline Y}}

\def\mC{{\mathbb C}}
\def\ms{{\mathpzc s}}
\def\s*{\ *_{\!\!\!\!\!\!\!\!\!\,_{\,_\text{\scriptsize{sym}}}}}
\def\hs*{\ \hat{*}_{\!\!\!\!\!\!\!\!\!\,_{\,_\text{\scriptsize{sym}}}}}
\def\d{\partial}

\def\nref#1{(\ref{#1})}

\def\cusp{ {\rm cusp}}

\begin{document}

\thispagestyle{empty}

\renewcommand{\thefootnote}{\fnsymbol{footnote}}
\setcounter{footnote}{0}
\setcounter{figure}{0}
\begin{center}
$$$$
{\Large\textbf{\mathversion{bold}
The quark anti-quark potential
\\
and the cusp anomalous dimension
\\
from a TBA equation}\par}

\vspace{1.0cm}

\textrm{Diego Correa$^a$,   Juan Maldacena$^b$ and  Amit Sever$^{b,c}$ }
\\ \vspace{1.2cm}
\footnotesize{

\textit{$^{a}$  Instituto de F\'{\i}sica La Plata, Universidad Nacional de La Plata, \\ C.C. 67, 1900 La Plata, Argentina} \\
\texttt{} \\
\vspace{3mm}
\textit{$^{b}$ School of Natural Sciences,\\Institute for Advanced Study, Princeton, NJ 08540, USA.} \\
\texttt{} \\
\vspace{3mm}
\textit{$^c$
Perimeter Institute for Theoretical Physics\\ Waterloo,
Ontario N2J 2W9, Canada} \\
\texttt{}
\vspace{3mm}
}

\par\vspace{1.5cm}

\textbf{Abstract}\vspace{2mm}
\end{center}

\noindent

We derive a set of integral equations of the TBA type for the generalized cusp
anomalous dimension, or the quark antiquark potential on the three sphere,
 as a function of the angles. We do this by considering a family
of local operators on a Wilson loop with charge $L$. In the large $L$ limit the problem
can be solved in terms of a certain boundary reflection matrix. We determine this reflection
matrix by using the symmetries and the boundary crossing equation. The cusp is   introduced
through a relative rotation   between the two boundaries. Then the TBA trick of exchanging space and
time leads to an exact equation for all values of $L$. The $L=0$ case corresponds to the
cusped Wilson loop with no operators inserted. We then derive a slightly simplified
integral equation which describes the small angle limit. We solve this equation up to three
loops in perturbation theory and match the results that were obtained with more direct approaches.

\vspace*{\fill}

\setcounter{page}{1}
\renewcommand{\thefootnote}{\arabic{footnote}}
\setcounter{footnote}{0}

\newpage
\tableofcontents

\section{Introduction}

\begin{figure}[h]
\centering
\def\svgwidth{12cm}
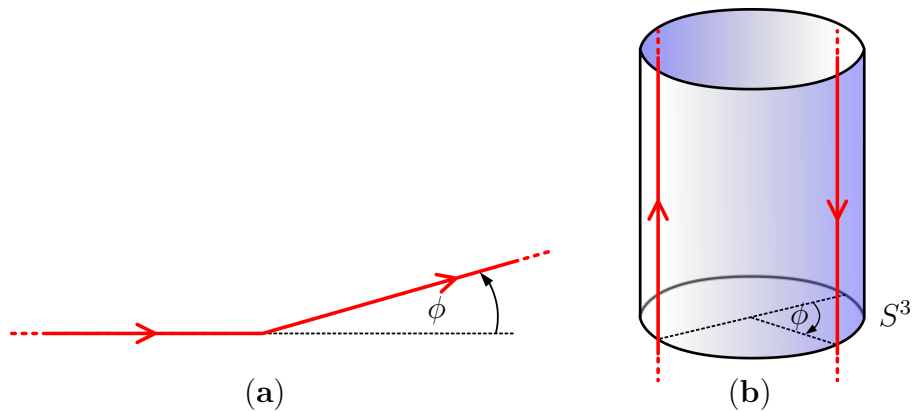
\caption{({\bf a}) A Wilson line with a cusp angle $\phi$. ({\bf b}) Under the plane to cylinder map
 the two half lines in (a) are
  mapped to a quark anti-quark pair sitting at two points on $S^3$ at a relative angle of $\pi -\phi$.  The
  quark anti-quark lines are extended along the time direction. }
\label{CuspDiagram}
\end{figure}

In this article we derive an equation for the cusp anomalous dimension for all angles
and for all values of the 't Hooft coupling $\lambda =g_{YM}^2 N$ in the planar limit
of ${\cal N} =4$ super Yang Mills. We obtain a system of non-linear integral equations of the form
of a  Thermodynamic Bethe Anstaz (TBA) system. The value of the cusp anomalous dimension can
be obtained from a solution of the TBA system. This is also equal to the quark/anti-quark potential
on the three sphere, see figure \ref{CuspDiagram}.

The cusp anomalous dimension is associated with the logarithmic divergence arising from
  a Wilson loop with a cusped contour \cite{PolyakovCusp}
\be \la{cuspdef}
 \langle W \rangle \sim e^{ - \Gamma_{\cusp}(\phi,\lambda) \log { L_{\rm IR} \over \epsilon_{\rm UV}} }\,,
\ee
where $L_{\rm IR}$ and $\epsilon_{\rm UV}$ are IR and UV cutoffs respectively.

The locally supersymmetric Wilson loop in ${\cal N}=4$ super Yang Mills also includes a coupling
to the scalar fields specified by a direction in the internal space $\vec n$ (with $\vec n^2 =1$)
\be  \la{wildef}
 W \sim {\rm Tr}\left[ P  e^{ i \oint A\cdot dx + \oint |dx| \vec n\cdot \vec \Phi } \right]\,.
\ee
Instead of considering the same vector $\vec n$ on the two lines that make the cusp, we can
take two vectors $\vec n$ and $\vec n'$. This introduces a second angle $\cos \theta = \vec n\cdot\vec n'$. Thus we have the generalized cusp anomalous dimension
 $\Gamma_{\cusp}(\phi, \theta , \lambda)$
 \cite{Drukker:1999zq}.
 $\Gamma_{\cusp}(\phi, \theta)$
can be computed in terms of a solution of the TBA system of equations presented in this article. We  can
also consider the continuation
$\phi = i \varphi$, where  $\varphi$ is a boost angle in Lorentzian signature.
Before describing the computation, let us make some general remarks.

\subsection{Remarks on the cusp anomalous dimension }

$\Gamma_{\cusp}$ is related to a variety of physical observables:

\begin{itemize}

\item
 It characterizes the IR divergences that arise when we scatter massive colored particles. Here
 $\varphi$ is the boost angle between two external massive particle lines. For each consecutive pair of lines in the color ordered diagram we get a factor of the form \nref{cuspdef}, where
 $L_{IR}$ is the IR cutoff and  $\epsilon^2_{\rm UV}$ is the given by the square of the
 sum of the momenta of the two consecutive particles.   More explicitly, the angle is given by
 $ \cosh \varphi   =- { p_1 . p_2 \over \sqrt{ p_1^2 p_2^2 } } $. This relation is general for
 any conformal gauge theory. See \cite{Korchemsky:1991zp,Becher:2009kw} and references therein.
  In ${\cal N}=4$ super Yang Mills the massive particles can be
 obtained by setting some Higgs vevs to be non-zero $\vec \Phi$. Then the angle $\theta$ is
 the angle between the Higgs vevs associated to consecutive massive particles \cite{LadderPaper}.

 \item
 The IR divergences of massless particles are characterized by $\Gamma^{\infty}_{\cusp}$ which
is the coefficient of the large $\varphi$ behavior  of the cusp anomalous dimension, $\Gamma_{\cusp} \propto \varphi \Gamma_{\cusp}^\infty $.   $\Gamma^{\infty}_{\cusp}$ was computed in the seminal paper
 \cite{BES}.  Note that $\Gamma^{\infty}_{\cusp}$ is also sometimes called the ``cusp
 anomalous dimension'' though it is a particular limit of the general, angle dependent ``cusp
 anomalous  dimension'' defined in \nref{cuspdef} .

 \item
 By the plane to cylinder map this quantity is identical with the energy of a static quark and
 anti-quark sitting on a spatial three sphere at an angle $  \pi - \phi$.
 \be
 \Gamma_{\cusp}(\phi,\theta) = V( \phi , \theta )\,.
 \ee
 See figure \ref{CuspDiagram} . This potential depends on the angle $\phi$ as well as on the internal
 orientations of the quark and anti-quark, which define the second angle $\theta$.

 \item
 In particular, in the small $\delta = \pi -\phi $ limit we get the same answer as the quark-anti-quark
 potential in flat space\footnote{This limit does not  commute with the perturbative expansion in $\lambda$. So  \nref{qqpot} is correct if $ \delta  \ll \lambda $. If we expand first in $\lambda$ and then
 take the $\delta  \to 0$ limit we get a different answer due to IR divergences that arise
 in the naive perturbative expansion. These also arise in QCD,  The origin of these logs are discussed in \cite{Pineda,LadderPaper}. }
  \be \label{qqpot}
  \Gamma_{\cusp}( \phi , \lambda ) \sim  { v( \theta, \lambda ) \over \delta }\,,~~~~~~{\rm when }~~~~~\delta = \pi - \phi \to 0  \,,
  \ee
  where $v(\lambda)$ is the coefficient of the quark-anti-quark potential, $V = { v(\theta,\lambda) \over r }$, for a quark and an anti-quark at distance $r$ in flat space and
  couplings to the Higgs fields which are rotated by a relative  angle $\theta$.

  \item
  In the small $\phi$ limit the cusp anomalous dimension goes as $\phi^2$ and one can
   define a Bremsstrahlung function $B$ by
   \be
    \Gamma_{\cusp} \sim - (\phi^2 - \theta^2 ) B(\lambda)  ~~~~~~~~~ \phi , \theta \ll 1\,.
    \ee
    This function $B$ can be computed exactly using localization, see \cite{Correa:2012at} and \cite{Fiol:2012sg}.
    Here we will derive a set of integral equations that also determines $B$. In this way
    we can link the localization and integrability exact solutions. This function
    $B$ is also related to a variety of observables, see \cite{Correa:2012at,Fiol:2012sg}
     for further discussion.

  \end{itemize}

Another motivation to study the cusp anomalous dimension is the study of amplitudes.
Amplitudes are also functions of the angles between particles. Here we get a very simple function
of one angle which has a structure very similar to amplitudes, since it is related to
amplitudes of massive particles. Thus, obtaining exact results for this quantity is useful
to learn about the general structure of the amplitude problem.

\subsection{Method }

The method to obtain the equation is a bit indirect and we need several preliminary results
that are interesting in their own right. Just for orientation we will outline the main
idea and method for its derivation.

\begin{figure}[h]
\centering
\def\svgwidth{10cm}
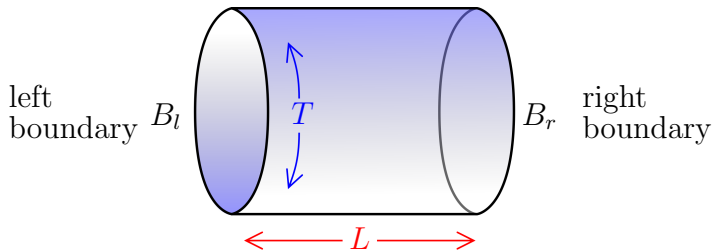
\caption{The BTBA trick. The same partition function can be viewed in two ways  \nref{openclosed}.
In the open string channel it is a trace over all states in the open string Hilbert space. In this case
Euclidean time runs along the $T$ arrow. Alternatively we can view it as the propagation of a closed
string along the $L$ arrow. The closed string has length $T$ and propagates over a Euclidean time $L$. The
two boundary conditions, now lead to two boundary states that create the closed strings that propagate
along the closed string channel.  }
\label{TBAfigure}
\end{figure}

The method consists of the following steps
\begin{itemize}
\item
We first consider  the problem of computing the  spectrum of local operators
 on a Wilson line. We consider the particular case of operators with a large charge,
i.e. operators containing a large number, $L$, of the complex scalar field $Z$ insertions.  These
insertions create a BMN vacuum \cite{BMN}.

\item
In the large $L$ limit the problem can be solved using an asymptotic Bethe Ansatz that involves the propagation of certain ``magnons''. These equations describe magnons moving on a long strip of length $L$ with two boundaries associated to the Wilson loop on each of the ``sides'' of the operator, see
\cite{DrukkerKawamoto}. The propagation of the magnons in the bulk is the usual one \cite{Review}.
The new feature is the existence of a boundary. The magnons are reflected at the boundary and one
needs the boundary reflection matrix. This is fixed in two steps.

\item
We determine the matrix structure of the reflection matrix from group theory, as in
\cite{BeisertDynamic,HMopen,CY,CRY}. This reflection matrix is such that it obeys the
boundary Yang Baxter equation \cite{Ghoshal:1993tm}. This is evidence that the boundary
condition preserves integrability.

\item
We derive  a crossing equation for the reflection phase and we find  a solution.

\item
Doing a time/space flip, so that now we have the mirror theory between two
boundary states separated by a mirror ``time'' $L$. See figure \ref{TBAfigure}.
 We can apply a symmetry generator that
rotates one boundary relative to the other, so that we introduce the two angles.

\item
We compute this overlap using TBA equations for any $L$,
 focusing on the ground state
energy, which is extracted by taking
 the large $T$ limit of the computation in figure \ref{TBAfigure}. These boundary
 TBA equations
 can be derived following a method similar to the relativistic case  \cite{LeClair:1995uf}.

\item
We set  $L=0$ we get the cusp anomalous dimension.

\end{itemize}

Let us discuss these steps in a bit more detail. First we should note that the exact integrability methods, as currently understood, work best to compute energies of states. Thus, we should phrase the computation of the cusp anomalous dimension as the computation of an energy. This is very simple.  Under the usual plane to cylinder map, the cusp on the plane maps into two static quark and anti-quark
lines on $S^3 \times R$. The quark and anti-quark lines are extended along the time direction, and
they are separated by an angle $\pi -\phi$ on the $S^3$, see figure \ref{CuspDiagram}. The case $\phi=0$, which is the
straight line in the plane, is mapped to a quark-anti-quark pair at opposite points on the sphere.
If $\theta=0$, this is a BPS configuration and the cusp anomalous dimension vanishes exactly for all $\lambda$. In fact,
for $\theta = \pm \phi$ we continue to have a BPS configuration \cite{Zarembo:2002an} and the cusp anomalous dimension
continues to vanish. In general, the cusp anomalous dimension is the energy of this quark-anti-quark configuration, as a function of
the two angles, $\phi$ and $\theta$.

The configuration with $\theta = \phi =0$ preserves 16 supercharges which, together with the bosonic symmetries,
give rise to a $OSp(4^*|4)$ symmetry group. This is important to determine the boundary reflection matrix.

Let us begin by considering an apparently unrelated problem which is the problem of computing the
anomalous dimension of operators inserted along a Wilson loop. First we consider a straight
Wilson loop and we insert an operator at the point $t=0$.
For example, we can consider an insertion of a complex scalar field $Z$ on the contour
\be
\la{simplein}
 P e^{i \int_{-\infty}^0 ( A + i \Phi_4)} Z(0) e^{i \int_0^\infty( A_t + i \Phi_4) }  = B_l Z(0) B_r\,.
 \ee
 These are operators that live on the loop and should not be confused with closed string
 operators. We denote these operators as $ B_l Z B_r $, where $B_{l,r}$ stands for the usual
 path ordered exponentials of the gauge field. The operator considered above is BPS if $Z$ is constructed out of
  scalars that do not appear in $B_{l,r}$  \nref{simplein}.
   To be definite, we consider $Z = \Phi_5 + i \Phi_6$.
  We can similarly consider operators of the form $B_l Z^L B_r$ which continue to be BPS. The straight Wilson loop
  is invariant under dilatations, so we can characterize the operators by their dimension
  under dilatations. These operators have dimension $\Delta = L$.

 Determining the scaling dimension of operators of this type, but with more general insertions,  is easier in the large $L$ limit. Then, we can solve this problem by considering impurities propagating along a long chain of $Z$'s. The impurities are the same as the ones that were used to solve the closed string problem in a similar regime \cite{BeisertDynamic,BeisertStaudacher}. The new aspect is that the impurities can be reflected from the boundaries at the end of the chain. This picture was discussed at the 1-loop order in the weak coupling limit in \cite{DrukkerKawamoto}. To proceed, we need to determine the boundary reflection matrix to all orders in the coupling. The matrix structure can be determined by group theory, as in \cite{BeisertDynamic}. The phase factor is more subtle. We write a crossing equation for it and we solve it following the strategy outlined in \cite{Volin,VieiraVolin}. At this stage we have completely solved the problem for operators with large $L$. Up to corrections of order $e^{- ({\rm const}) L }$, we can find the energy of any open string state by solving the appropriate Asymptotic Bethe equations.

  After we have found the boundary reflection matrix we can then consider the possibility of rotating  the half Wilson line that is associated with it. This rotation will simply act on the indices of the reflection
  matrix via a global transformation. Now we can consider states of the form $B_l Z^L B_r(\theta,\phi)$, where we have rotated one of the sides of the Wilson line. This operator is no longer BPS but its energy is very small   when $L$ is very large, i.e. it has zero energy up to $e^{- ({\rm const}) L }$ corrections. These are called Luscher (or wrapping) corrections.

  Before writing down the Thermodynamic Bethe Ansatz that describes the most general finite $L$ state,
  we will make some checks on the phase that has been obtained. As a non-trivial
  check one can get the first corrections to the ground state energy for large $L$.
  Namely, we are interested in the anomalous dimension of the operator of the form
  $B_l Z^L B_r(\theta,\phi)$. This correction is given by a Luscher-type formula.
  This formula can be most simply understood by considering the problem in the mirror picture.
  Namely, we exchange space and time in the open string picture.
  In other words, we have the equivalence
 \be
 \la{openclosed}
   Z_{B_l,B_r}^{\rm open} =  {\rm Tr}_{\rm open} [ e^{ - T  H^{\rm open}_{B_l ,B_r}} ] = \langle B_l | e^{ - L H_{\rm closed} }|B_r \rangle\,,
  \ee
  where $H^{\rm open}_{B_l ,B_r}$ is the open chain Hamiltonian on a strip of length $L$ and $H_{\rm closed}$ is
  the closed chain Hamiltonian of the mirror theory on a circle of size $T$. So now we have a closed string exchanged between two boundary states. The analytic continuation of the boundary reflection matrix gives us the probability of emitting a pair of particles from the boundary state. It turns out that this continued reflection matrix has a pole at zero mirror momentum which implies that we can create
  single particles \cite{Ghoshal:1993tm}. The coefficient of the pole in the reflection
  matrix at zero mirror momentum determines the prefactor of the Luscher correction \cite{Bajnok:2004tq}.
  We compute this at strong coupling and we find agreement with a direct string theory computation.
  Furthermore, the {\it leading order } correction at weak coupling, going like $g^2$, also comes from this Luscher type term. In this way we match the leading corrections at weak and strong coupling. This constitutes a test of the boundary reflection matrix. In particular  the very existence of the pole at zero mirror momentum is due to the phase factor of the matrix,
  which we derived by solving the crossing equation.

  Finally, one can write down a Thermodynamic Bethe Ansatz equation that describes  the finite
  $L$ situation. This follows the standard route for getting the energies of states of an integrable field theory with a boundary. The derivation of these equations is very similar to the derivation of the equations for closed string states. The new element is that instead of a thermodynamic partition function we have the overlap between two boundary states, as in
  \nref{openclosed}. The derivation of TBA equations for integrable systems with a boundary
  was considered in \cite{LeClair:1995uf}.
  The boundary states are given in terms of the analytic continuation of the boundary reflection matrix.
  The TBA system of equations arises from evaluating this exact overlap between the two boundary states in an approximate way by
  giving the densities. Most of the TBA equations come from the entropy terms, which are the
  same in our case. Thus the boundary TBA equations are very similar in structure to the
  closed ones. We obtain
 \be
   \log Y_A = \log (\kappa^l_A \kappa^r_A) - 2 L E_{m,A} + K_{AB} * \log ( 1 + Y_B )\,.
   \ee
   The cusp anomalous dimension, or quark/anti-quark potential
    is given schematically   by
   \be
   {\cal E} = - { 1 \over 2 \pi } \sum_A \int\limits_0^\infty  d q_A \log(1 + Y_A )\,.
   \ee
   Here $E_{m,A}$ and $q_A$ are the energies and momenta of the excitations in the mirror
   theory. The equations will be given below in their full detail,  \nref{1stTBA}-\nref{lastTBA}.
    The information about the boundary
  is contained in $\kappa_A$ which comes from the reflection phase of the theory and depends
  on the boundary state.

{\bf Note:} We were informed that similar ideas were pursued in \cite{DrukkerTBA}.

\section{ Spectrum of operators on a Wilson line }\la{sec2}

Let us first discuss the symmetries preserved by a straight Wilson line. Let us
start with the bosonic symmetries. It preserves an $SL(2) \times SU(2) \times SO(5)$ symmetry
group. The $SO(5)$ is the subset of $SO(6)$ that leaves $\Phi^4$ invariant, where
$\Phi^4$ is the scalar that couples to the Wilson line.
The $SU(2)$ factor corresponds to  the spatial rotations around the loop. The $SL(2)$ factor
 contains time translations,
dilatations and special conformal transformations along the time direction.
In addition, we preserve half of the supercharges. The full supergroup is $OSp(4^*|4)$. The
star means it is the real form  of $SO(4)$ such that $SO(4^*) \sim SL(2) \times SU(2)$.

Now we can consider the insertion of an operator of the form $Z^L$ on the Wilson loop,
we can denote this as $B_l Z^L B_r$. Here we choose $Z$ to be $Z = \Phi^5 + i \Phi^6$.

The operator $Z$ inserted at the origin
preserves an $SU(2|2)^2$ subgroup of the full symmetry group of the
theory.
The Wilson loop, together with the $Z$ insertions at the origin preserve an
$SU(2|2)_D $ subgroup of all the symmetry groups we mentioned. This is a diagonal
combination of the two $SU(2|2)$ factors preserved by $Z$.
This common preserved symmetry is very useful for analyzing this problem.
These operators are BPS, and they have protected anomalous dimension,
${\cal E}  \equiv \Delta - J_{56} =0$.

Note that on $S^3$ we have a flux tube that goes between the quark and the anti-quark. These operators
inserted on the Wilson loop are mapped to
  to various excitations of the flux tube.

 \subsection{The boundary reflection matrix }

 Recall that the bulk excitations are in a fundamental representation of each of the two
 $\widetilde{su}(2|2)$ factors of the $\widetilde{su}(2|2)^2$ symmetry of the $Z$-vacuum.
  The tilde means that we are considering the momentum dependent
 central extensions discussed in \cite{BeisertDynamic,Beisertnonlin}.
 In other words, we can think of them as particles with  two indices
 $\Psi_{A,\dot B}$,
  where $A$ labels the fundamental of the first $\widetilde{su}(2|2)$ and $\dot B$ labels the {fundamental} of the second $\widetilde{su}(2|2)$ factor of  the $\widetilde{su}(2|2)^2$ symmetry of the infinite chain. This central extension determines
  the dispersion relation for the excitations
  \ba \la{constraint}
   { i \over g } & = & x^+ + { 1 \over x^+} -  x^- - { 1 \over x^-}~,
    \\
   e^{ i p } &= & { x^+ \over x^- } ~,~~~~~~~~~~
   \epsilon =  i g  \({ 1 \over x^+ }- { 1 \over x^-} -x^++x^-\)
 = \sqrt{ 1 +  16   g^2  \sin^2{\tfrac{p}{2}} }\,, \label{enfromx}
\ea
Throughout this paper we define $g$ as\footnote{Note that $g \not = g_{YM}$.}
\be \la{defofg}
g   \equiv   {\sqrt{  \lambda} \over 4 \pi } = { \sqrt{ g^2_{YM} N } \over  4 \pi }\,.
   \ee
 The scattering matrix between two particles has the form
 $S_{A \dot A , B \dot B}^{C \dot C D \dot D} = S_0^2  \hat S_{A B}^{ CD} \hat S_{\dot A \dot B}^{\dot C \dot D } $\cite{BeisertDynamic}.
  Namely, it is the product of a phase factor $S_0^2$ and two identical matrices, one for each $\widetilde{su}(2|2)$ factor. These matrices are fixed (up to an overall factor) by the $\widetilde{su}(2|2)$ symmetry of the theory \cite{BeisertDynamic,Beisertnonlin}. These matrices depend on the
  two momenta, $p_1$ and $p_2$, of the scattered variables. The phase factor $S_0(p_1,p_2)$ was guessed in \cite{BHL,BES}, and
 a nice derivation was given in \cite{Volin,VieiraVolin}.

\begin{figure}[h]
\centering
\def\svgwidth{12cm}
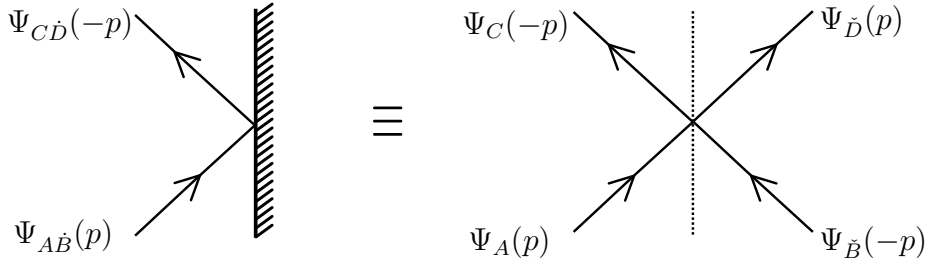
\caption{Unfolding of $R(p)$ into $S(p,-p)$. There is a non-trivial map between dotted and checked indices. See appendix \ref{Rmatrix}  for details.}
\label{unfolding}
\end{figure}

 In our problem we need to fix a reflection matrix of the form $R_{A \dot B}^{C \dot D }(p) $.
 Let us consider first the reflection from the right boundary, see figure \ref{unfolding}.
 This matrix depends on only one momentum $p$, the momentum of the incident magnon.
 The boundary is invariant under an $\widetilde{su}(2|2)_{\rm D} $ symmetry group, which is diagonally
 embedded in the $\widetilde{su}(2 |2)^2$ symmetry group of the bulk of the spin chain (see appendix \ref{Rmatrix}).
 A similar problem was studied in \cite{CRY} and the matrix part of the reflection is the same.
 Thus the symmetries constraining the reflection matrix are exactly the same as those
 constraining the bulk scattering matrix for each of the $\widetilde{su}(2|2)$ factors.
 From this argument we expect that the matrix structure should be completely fixed. In fact,
  the matrix structure should be essentially the same as what we encounter
 in the matrix $\hat S_{AB}^{CD}(p,-p)$, or  $
 R_{A \dot A}^{C \dot C }(p)   \propto   \hat S_{A \dot A}^{C \dot C }(p,-p) $.
 One is tempted to say that the scattering phase factor would be $S_0(p,-p)$. However, this is not fixed by the
 symmetries, and will not be true as we discuss below. In the presence of a boundary, we can do a kind of ``unfolding''
 of the spin chain. Here each bulk magnon is viewed as a pair of magnons of $\widetilde{su}(2|2)_{\rm D}$, one with momentum $p$ to the left
 of the boundary and one with momentum $-p$ to the right of the boundary. See figure \ref{unfolding} .

 This completely solves the problem of fixing the matrix structure of the reflection matrix.
 The full reflection matrix, in complete detail, is given in appendix \ref{Rmatrix}. One can
 also check that it obeys the boundary Yang Baxter equation. But this is clear from the
 ``unfolded'' picture in terms of a single chain. We should emphasize that we have assumed
 that there are no boundary degrees of freedom.
  We do not see any evidence of  any boundary degrees
  of freedom at either weak or strong coupling, so this is a reasonable assumption.

  Before we determine the phase, let us make a side remark. There is a variety
  of problems that give rise to a spin chain with boundaries and preserve the same symmetries,
  $OSp(4^*|4)$. We can consider an open string ending on a D5 brane that wraps $AdS_4 \times S^2$, or
  $AdS_2 \times S^4$. In fact,  there is a whole family of BPS branes of this kind that
  arises by adding flux for the $U(1)$ gauge field on the brane worldvolume on the $S^2$ or $AdS_2$.
  In fact, in the limit of large electric flux on the $AdS_2 \times S^4$ brane we get a boundary condition
  like the Wilson loop one. In fact the $AdS_2 \times S^4$ branes can be interpreted as Wilson loops in
  the $k$-fold  antisymmetric representation of $U(N)$ \cite{Yamaguchi:2006tq}.
  In all these cases one can choose the BMN vacuum (or choose the field $Z$) in such a way
  that we preserve the $\widetilde{su}(2|2)_{\rm D}$ of the spin chain.
  Therefore, we would get the same matrix structure for the reflection matrix, again assuming that there
  are no boundary degrees of freedom. However, they would differ in the choice of a phase factor.
  Below we get a phase factor which has all the right properties to correspond to the one of the Wilson loop.
  It would be interesting to fix the phase factor also for these other cases, but we leave this to the future.

  In order to fix the phase factor we write a crossing equation. We derive this by writing
  the identity state of \cite{BeisertDynamic}, scattering it through the boundary and
  demanding that the full phase is  equal to one. Denoting the phase factor as $R_0$,
  defined more precisely in appendix \ref{Rmatrix}, we obtain the crossing equation
  \be
  \la{bdycross}
  R_0(p) R_0(\bar p) =  \sigma(p, - \bar p)^2\,,
  \ee
  where the bar indicates the action of  the crossing transformation. Here $\sigma(p_1,p_2)$ is the
  bulk dressing phase, discussed in \cite{BES,VieiraVolin}. We are going to $\bar p$ along the
    the same contour in
  momentum space that we choose in the formulation of the bulk crossing equation.

  In addition,  we also should impose the unitarity condition
  \be
   R_0(p) R_0(-p) =1\,.
   \ee
  We now write the ansatz
  \be \la{fullphase}
  R_0(p) = { 1 \over \sigma_B(p) \sigma(p,-p)} \left( { 1 + { 1 \over (x^-)^2 } } \over 1 + { 1 \over (x^+)^2 } \right)\,.
 \ee
 Here $\sigma$ is the bulk dressing phase. This would be our naive choice for a phase
 factor. The explicit factors of $x^\pm$ have been chosen only to simplify the final formula.
 We have an unknown   factor $\sigma_B(p)$.
 Now \nref{bdycross} becomes
 \be\la{crossingequation}
 \sigma_B(p) \sigma_B(\bar p) = { x^- + { 1 \over x^-}  \over x^+ + { 1 \over x^+} }\,.
 \ee

 We can now solve this equation using the method proposed in \cite{Volin,VieiraVolin}.
 We give the details in appendix \ref{crossingsolution}. We obtain
 \ba \la{phasefa}
  \sigma_B & = &  e^{ i \chi(x^+) - i \chi(x^-) }\,,
\\ \la{chiintegral}
 i \chi(x) & = & i \Phi(x) =  \oint\limits_{|z|=1} { d z \over 2 \pi i } { 1 \over x - z } \log \left\{ \sinh[ 2 \pi g ( z + { 1 \over z} ) ] \over 2 \pi g ( z + { 1 \over z } ) \right\}\,,~~~~~~|x|>1\,.
 \ea
 This expression is valid   when $|x|>1$. The value for $\chi$ in other regions
is given by analytic continuation. We have also introduced the function $\Phi(x)$ which is
given by the integral for all values of $x$. When $|x|<1$ these two functions differ by
\be \la{phasefacon}
 i \chi(x) = i \Phi(x) +  \log \left\{ \sinh[ 2 \pi g ( x + { 1 \over x} ) ] \over 2 \pi g ( x + { 1 \over x } ) \right\}\,, ~~~~~~~~|x|< 1
 \ee
The ambiguities in the choice of branch cuts for the logarithm cancel out when we compute
$\sigma_B$ in \nref{phasefa}.  Note that $\chi(x) = \chi(-x)$.

So far, we have found {\it a particular}
solution of the boundary crossing equation. Still, the true dressing phase
might require the inclusion of further CDD factors.
 In order to make a conjecture for the exact
boundary dressing phase, we need to compare against some explicit computations.

Before doing so, let us observe that, given $\sigma_B(p)$, we can define an infinite family of solutions by taking
\be
\sigma_B^{(s)}(p) = \left(\frac{x^-+\frac{1}{x^-}}{x^++\frac{1}{x^+}}\right)^s\left[\sigma_B(p)\right]^{1-2s}\,.
\label{othersols}
\ee
By computing the dressing phase in the physical regime we will be able to show that  $s=0$ is
the solution we want.

The proposal for the phase factor for the reflection matrix, given in \nref{fullphase}, \nref{phasefa} is one of the important results of this paper.
We will perform various checks on its validity.

\subsection{Checks of the boundary reflection phase in the physical region }

\begin{figure}[h]
\centering
\def\svgwidth{12cm}
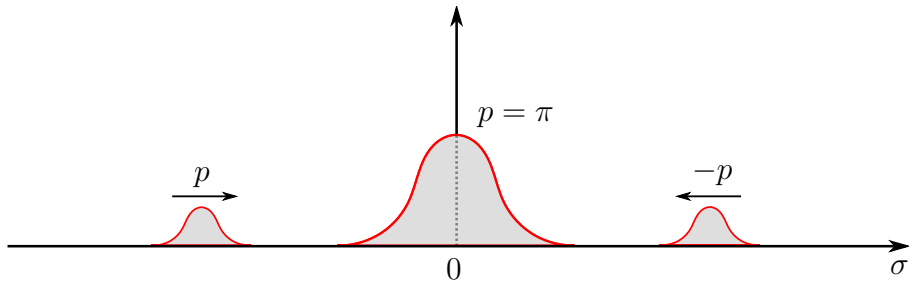
\caption{Computation of the reflection phase at strong coupling. We have a soliton at
the boundary, which is at rest at $\sigma=0$. There is also an image soliton coming from
the right. Then the soliton with momentum $p$ scatters through the soliton at rest and the
one with momentum $-p$, leading to a certain time delay. From the time delay we can
compute the derivative of the reflection phase with respect to the energy.  }
\label{SolitonScattering}
\end{figure}

Let us describe   how
 to compute the boundary dressing phase at strong coupling.
We have to consider the open string solution that corresponds to a 1/2 BPS Wilson line carrying a large
$J_{56}$ charge given in \cite{DrukkerKawamoto}. This solution describes the transition from the boundary
Wilson line to an infinite  BMN vacuum.
It is convenient to understand this solution in the conformal gauge, when we set the
stress tensor on the $S^5$ equal to one, and the stress tensor of the $AdS_5$ to minus one.
In these variables, the problem only involves an $AdS_2 \times S^2$ subspace and we can
perform the Pohlmeyer reduction in each factor. The $S^2$ part gives rise to a sine gordon
theory and the solution is just half of a  soliton at rest. More precisely, the center of
the sine gordon soliton sits at the boundary. In the $AdS_2$ part we have a sinh gordon
theory, and the solution is a sinh-gordon ``soliton''. This is a singular solution which is
the direct analytic continuation of the sine gordon soliton. The singularity reflects the
fact that the string goes to the $AdS$ boundary. If we compute the energy, there is a divergent
part and the finite part is zero. The setup is explained in more detail in \cite{DrukkerKawamoto}. The fact that the finite part of the energy  is zero is consistent with
the absence of a boundary impurity transforming non-trivially under $\widetilde{su}(2|2)$.
A bulk magnon is a sine gordon soliton, and leaves the $AdS$ part of the solution unperturbed.
In the presence of a boundary, we need to put also the ``image'' of this soliton and the
configuration looks as in figure \ref{SolitonScattering}.
 The reflection involves the scattering of the soliton
with the image soliton as well as the scattering with the soliton at rest. These
soliton scattering phases were computed in \cite{HM,HMopen}. So the strong coupling limit of the right
boundary scattering phase $R_0(p) = e^{i\delta_{\rm R}(p)}$ is given by
\be
\delta_{\rm R}(p) = -8g\cos\tfrac{p}{2}\log\cos\tfrac{p}{2} -4g\cos\tfrac{p}{2}\log\left(\frac{1-\sin\tfrac{p}{2}}{1+\sin\tfrac{p}{2}}\right)\,.
\label{deltaR}
\ee
The first term in (\ref{deltaR}) is exactly what one gets from the strong coupling limit of the factor $1/\sigma(p,-p)$ \cite{HM}.
 We will see that the second term
 corresponds to $\sigma_B(p)^{-1}$. At strong coupling can expand \nref{chiintegral} as
\be \la{chistr}
i \chi(x ) \sim  4 g \left[ - 1 + ( x + { 1 \over x } ) { 1 \over 2 i } \log {( x + i ) \over (x-i)}
\right] + {\cal O}(1)\,,
\ee
which leads, for physical excitations, to 
\be \la{sigstrong}
{1\over i}\log \sigma_B(p) =  4 g \cos \tfrac{p}{2} \log \left( {1 - \sin { p \over 2} \over 1 + \sin { p \over 2} } \right)\,.
\ee
This indicates that we must pick the case $s=0$ from the family of solutions (\ref{othersols}).

Finally, let us discuss the behavior at weak coupling. The bulk dressing phase $\sigma$
has its first contribution at order $g^6$, leading  to four loop corrections to anomalous dimensions.
 On the other hand, the boundary dressing phase, $\sigma_B$,
receives its first contribution at $g^4$, so that it will start modifying anomalous
dimensions of operators inserted on the Wilson loop (dual to open string states) at three loops.

\subsection{ Reflection matrix for a Wilson line at  general angles }
\label{Rwithangles}

We will need the boundary reflection matrix for a Wilson line sitting at general angles,
$\phi$ and $\theta$. In particular, we want the left and right boundaries
of the open chain to be rotated by   relative angles.
We can obtain the boundary state of the Wilson line at
a different position on the $S^3$, or the $S^5$, by applying a symmetry transformation on $B_r$.
This should be a
 symmetry that is broken by $B_r$. So for example, we can apply an $SU(2)_L$ rotation on
the $S^3$ which is in one of the $SU(2)$ factors in the $SO(4)$ group  of rotations of the 3-sphere. If we
apply  an $SU(2)_L$ rotation with an angle $2 \phi$, we will get that the quark is rotated by
an angle $\phi$ on the $S^3$, away from the south pole. See figure \ref{CuspDiagram}(b).  Note that the
$SU(2)_L$ we are considering is a symmetry of the $Z$ vacuum.
We have a similar feature on the $S^5$. We can also apply a rotation in an $SU(2)_{L'}$ factor inside
$SO(4)\subset SO(6)$ (this $SO(4)$ leaves the $Z$ vacuum invariant).
 The reflection matrix is very easy to obtain.  We pick these two $SU(2)$ generators so that they sit in
 the   bosonic part of one of the $\widetilde{ su}(2|2) $ factors of the $\widetilde{ su}(2|2)^2 $ symmetry of the bulk. Then they will simply introduce some
phases of the form $e^{ i \phi}$ or $e^{ i \theta}$ when a state is reflected from the boundary and
its $SU(2)$ quantum number changes. The $SU(2)_L$ or $SU(2)_{L'}$ quantum numbers can change because they are not symmetries of the boundary state. More explicitly, the reflection matrix from a boundary state  at  angles $\phi,\theta$ is
given by
\be
R^{B \dot B }_{A \dot A } (\theta, \phi) = ( m^{-1}) ^{B}_{D} m_{A}^{C} R^{D \dot B }_{C \dot A }(0,0)
 ~,~~~~~~~{\rm with}~~~~~ \la{mone}
m = { \rm diag}(e^{i \theta } , e^{ - i \theta} , e^{ i \phi} , e^{ - i \phi } )\,.
\ee
Note that the matrix $m$ acts only on the undotted indices since we did a rotation inside only one of the
$ \widetilde{ su}(2|2)$ factors.

\subsection{ Luscher computations and checks in the mirror region }
\la{Luschersub}

In this subsection we start considering the problem with two boundaries. In other
words the operator $B_l Z^L B_r(\phi, \theta)$. Here $\phi, ~\theta$ are the relative orientations of
the two boundaries. On the plane, this corresponds to a cusp, plus an operator of the form $Z^L$ at
the tip.
In the limit $L \gg 1$ we get the naive superposition of the two boundaries and the
energy of the state is zero (${\cal E} = \Delta - L =0$), regardless of the orientation of the two boundaries.
The leading correction is of the form $e^{  - ({\rm constant}) L }$.
These corrections come from the exchange of particles
 along the ``mirror'' channel. The boundary sources
particles, which then travel to the other boundary. These  corrections sometimes go under
the name of Luscher corrections. Of course the familiar Yukawa potential is a simple example where
the leading correction comes from the exchange of a single massive particle.

In order to derive the precise correction formula it is convenient to describe in more detail the
mirror theory. In the bulk of the worldsheet the
 mirror theory  was discussed in various papers, see \cite{ArutyunovFrolov}
for example. This theory is defined by exchanging the space and time directions of the spin chain we have
been considering so far. Thus, instead of
\nref{enfromx} we define $  q = i \epsilon $ and $E_m = i p $, and use the same formulas
as in \nref{enfromx}. Here $q$ is the mirror momentum and $E_m$ is the mirror energy.
In order for these to be real we will need to pick a solution of  \nref{constraint} with
$|x^+|>1$ and $|x^-|<1$. From the expression for $q$, we can write
\ba
z^{[\pm a]} &=& \frac1{4g}\left(\sqrt{1+\frac{16g^2}{a^2+q^2}}\pm1\right)\left(q+ia\right)\,,
\label{zpm}
\\
  E_m &= &2 { \rm{ arcsinh }} { \sqrt{ q^2 + a^2} \over 4 g }\,. \la{mirrdis}
 \ea
 Here $z^\pm$ just denote the values of $x^\pm$ in the mirror region. We have also written
 the dispersion relation in the mirror region, for an arbitrary
 bound state. The elementary mirror magnon has $a=1$.

When we have a boundary, this time/space flip turns the boundary into a boundary state, see figure
\ref{TBAfigure}.
Then a suitable analytic continuation of the boundary reflection matrix characterizes the
boundary state. The boundary state creates a supersposition of many particles. The total
mirror momentum should be zero since it is translational invariant. So, schematically the
state has the form
\be \la{BdyState}
| B \rangle = |0\rangle + \int\limits_0^\infty { d q \over 2 \pi }  K^{A \dot A , B\dot B } (q) a^{ \dagger}_{-q \, A \dot A}  a^\dagger_{q \, B \dot B } | 0 \rangle + \cdots
\ee
with
\be \la{kfromr}
K^{A \dot A , B\dot B }(q) = \left[ R^{-1}(z^+,z^-)\right]^{A \dot A }_{  D\dot D } {\cal C}^{D \dot D ,B\dot B}\,,
   \ee
where we put the mirror values \nref{zpm}.
Here $R$ is the right reflection matrix, with $z^\pm$ continued
to the mirror region \nref{zpm}. This amounts to an analytic
continuation of the reflection matrix. Here $a^\dagger_{q \, A \dot A}$ is the creation operator of
a magnon with momentum $q$. ${\cal C}$ is a charge conjugation matrix.
In the case of a relativistic model with a single particle \nref{kfromr} reads $K(\theta) = { 1/ R(\theta - i { \pi \over 2 } ) } $, \cite{Ghoshal:1993tm}. The formula \nref{kfromr} can be obtained by performing a
$\pi/2$ rotation of the boundary condition.
 Due to the independence of reflection events from a boundary, we
can exponentiate \nref{kfromr} to get the full boundary state \cite{Ghoshal:1993tm,LeClair:1995uf}.
Similarly, we can form a future boundary state.
 This is a
boundary state that annihilates the particles. It is given by
\be \la{BdyStatetwo}
 \langle B | = \langle 0 |  +  \langle 0 | \int\limits_0^\infty {  d q \over 2 \pi }
\bar  K_{A \dot A , B\dot B } (q) a ^{A \dot A} _q a_{-q}^{B \dot B }   + \cdots
\ee
with
\be \la{barkdef}
\bar K_{A \dot A , B\dot B }(q) = \left[ R^{-1}\left(-{ 1\over z^-},-{ 1 \over z^+}\right)\right]^{D\dot D }_ { B\dot B }
{\cal C}_{D \dot D A \dot A}\,.
\ee
In the relativist case \nref{barkdef} would be $\bar K(\theta) = { 1 \over R(- i { \pi \over 2} -
\theta ) } $.

When $L$ is very large the leading $L$-dependent contribution comes from the exchange of
this pair of particles and we can write the corresponding contribution as
\be \la{luschn}
\delta {\cal E}= - \int\limits_0^\infty {dq \over 2 \pi }   e^{ - 2 L E_m(q) }  t(q) ~,~~~~~~ t(q) = {\rm Tr}[  K(q) \bar K(  q ) ]\,.
\ee
This formula is correct whenever the integral is finite.

In our case, the phase factor $\sigma_B$ has a pole at $q=0$. In the physical region $\sigma_B(p)$ was
perfectly finite. This pole in the mirror region is
 crucial for obtaining the correct answer. But first we need
  to generalize \nref{luschn} to the situation when we have a pole at
$q=0$.  The physical interpretation of this pole at $q=0$ is that the boundary state is
sourcing single  particles states in the mirror theory \cite{Ghoshal:1993tm}.
For a similar case in the AdS/CFT context see \cite{CYLuscher}.
Obviously such source has to contain only zero momentum particles.

A  careful analysis leads to the formula  \cite{Bajnok:2004tq}
\be \label{luscherfull}
 {\cal E}  \sim
  -  \int\limits_0^\infty { d q \over 2 \pi } \log \left\{
  1 + e^{ - 2 L E_m(q) } {\rm Tr}[ K(q) \bar K(  q ) ] \right\}
  \sim  - { 1 \over 2 }
 e^{ -  L E_m(0)}  \sqrt{  q^2 {\rm Tr}[ K(q) \bar K(  q ) ] |_{q=0} }\,.
\ee
In the last equality we extracted the leading term in the integral, which comes only
from the coefficient of the pole. Notice that the $L$
dependence is precisely what we expect from the exchange of a single particle.
We  should  sum over all the particles that can be exchanged. The mirror theory
contains bound states indexed by an integer $a=1,2, \cdots$, and we should sum over them.

\begin{figure}[h]
\centering
\def\svgwidth{14cm}
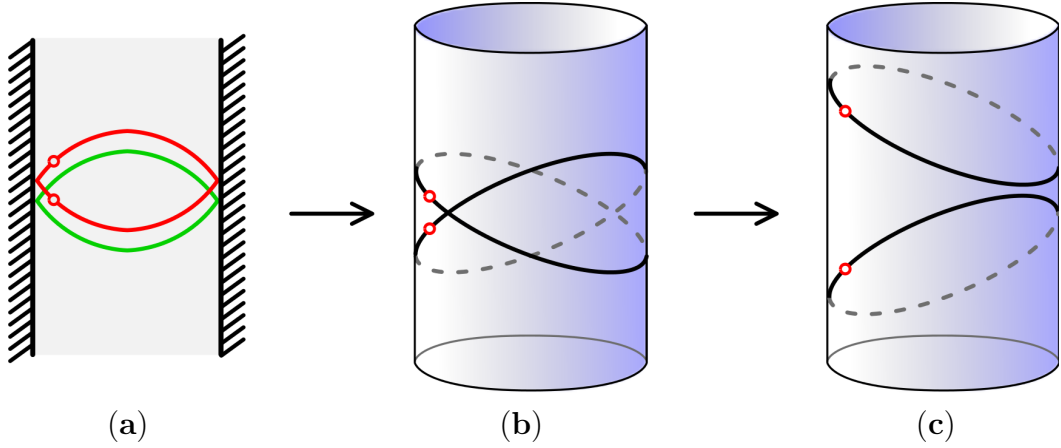
\caption{({\bf a}) We have a strip with pairs of particles being exchanged. The two colors represent
the two types of indices. In ({\bf b}) we unfolded this into a cylinder computation. The $K$ matrices
became $S$ matrices for a single $\widetilde{su}(2|2)$. ({\bf c}) Using crossing we have moved the
lines. The red circles indicates the action of the matrix $m$. }
\label{Unfolding}
\end{figure}

In appendix \ref{ComputationOft} we show that we can evaluate $t(q)$ for a fundamental mirror particle and
we obtain
\be \la{texpr}
t(q) = \sigma_B(z^+,z^- )
 \sigma_B\left(- { 1\over z^-} ,- { 1 \over z^+}  \right) \left( { z^- \over z^+ } \right)^2
 \left( {\rm Tr}[ (-1)^F] \right)^2\,,
 \ee
 where the trace is over the four states of a single $\widetilde { su} (2|2)$ magnon.
Let us now give a simple explanation for this formula, for more details see appendix \ref{ComputationOft}. We can write  the reflection
matrices that appear in $K$ and $\bar K$ \nref{kfromr} \nref{barkdef}
in terms of bulk S-matrices for the
unfolded theory, namely in terms of bulk  $S$ matrices for a single $\widetilde{su}(2|2)$ factor.
The matrix in $K$ is essentially ${\cal S}(-p,p)$ and the one in $\bar K$ is
${\cal S}(  \bar p , -\bar p)$. When we multiply these matrices we can use the bulk
crossing equation to get the identity. Here we should use the full bulk matrix, including
the bulk $\sigma$ factor. This is the reason that the bulk $\sigma$ factor disappears
from the final formula \nref{texpr}, but the boundary one remains. The factor of $z^-/z^+$ arises
from the  factor in parenthesis in
 \nref{fullphase}. Finally, the $(-1)^F$ is related to the
fact that  we have fermions. Here $F$ is the fermion number.
 When we perform the TBA trick, we get periodic fermions in
Euclidean time if we started with periodic fermions in the spatial directions. Of course
a periodic fermion in Euclidean time is the same as the trace with a $(-1)^F$ inserted.
The operations that lead to \nref{texpr} can be understood graphically as in figure \ref{Unfolding}.

Of course, for a fundamental magnon ${\rm Tr}[ (-1)^F] =0$. This is good,  since it is saying
that the correction vanishes in the BPS situation.
If we rotate one boundary relative to the other then we need to perform the replacement
\be \la{texprmone}
 {\rm Tr}[ (-1)^F] \longrightarrow {\rm Tr}[ (-1)^F m ] = - 2 ( \cos \phi - \cos \theta )\,.
\ee
where $m$ is given in \nref{mone}. Again,  we see that it vanishes in the BPS case
 $\phi = \pm \theta$.

To write down the full Luscher formula we need to compute $t(q)$ also for the bound states
of the mirror theory. One can first use the standard fusion procedure to get the bound
state reflection matrix. Then one can use the same argument as above to eliminate the
bulk $S$ matrices, as in figure \ref{Unfolding}.
 The final formula is

\ba \la{texprall}
t_a(q) &=&  \sigma_B(z^{[+a]},z^{[-a]} )
 \sigma_B\left(- { 1\over z^{[-a]}} ,- { 1 \over z^{[+a]}}  \right) \left( { z^{[-a]} \over z^{[+a]} } \right)^2 \left( {\rm Tr}[ (-1)^F m_a ] \right)^2\,,~~
 \\ \la{tracealla}
 &~&~~~~~~~~   {\rm Tr}[ (-1)^F m_a ]  = (-1)^a 2 ( \cos \phi - \cos \theta){ \sin a \phi \over \sin \phi }\,,`
 \ea
 where now the trace is over all the states of a magnon boundstate in a single copy of $\widetilde{su}(2|2)$, see equation \nref{mfora} in appendix \ref{ComputationOft}.
 As anticipated, an important property of $\sigma_B$ is that it has a  pole at $q=0$.
 More precisely the combination of $\sigma_B$ in \nref{texprall} becomes
 \ba \la{chifact}
    e^{ i \chi(z^{[+a]}) - i \chi(z^{[-a]}) + i \chi (1/z^{[-a]}) -i \chi(1/z^{[+a]})}
\!\!&=&\!\!
\frac{2\pi g(z^{[-a]}+\tfrac{1}{z^{[-a]}})}{\sinh[2\pi g(z^{[-a]}+\tfrac{1}{z^{[-a]}})]}
\frac{2\pi g(z^{[+a]}+\tfrac{1}{z^{[+a]}})}{\sinh[2\pi g(z^{[+a]}+\tfrac{1}{z^{[+a]}})]}\nn\\
&& \times e^{i(\Phi(z^{[+a]})-\Phi(z^{[-a]})+\Phi(1/z^{[-a]})-\Phi(1/z^{[+a]} ) ) }\,.
\ea
Here we used that $z$ is  in the mirror kinematics and we used \nref{phasefacon} to evaluate
$\chi(x)$ when $|x|<1$. We have also used that $\chi(-x) = \chi(x)$.
Each of the sinh factors leads to a pole at $q=0$. Namely, using
  \nref{zpm}
we get
\be
2 \pi g (z^{[\pm a] } + { 1 \over z^{ [\pm a ]}  }) =\pm  i \pi a + \pi q \sqrt{ 1 + { 16 g^2 \over a^2 } } + {\cal O}(q^3)\,.
\ee
for small $q$.  We then can write the pole part of  \nref{chifact} as
\ba
   e^{ i \chi(z^{[+a]}) - i \chi(z^{[-a]}) + i \chi (1/z^{[-a]}) -i \chi(1/z^{[+a]})}
&\sim  &  { 1 \over q^2}  { a^4 \over ( a^2 + 16 g^2 ) }  F(a,g) ^2  + \cO(1)\,,
\\
\la{ffun}
{\rm with}~~~~~~~F(a,g)^2 &\equiv & e^{i(\Phi(z^{[+a]})-\Phi(z^{[-a]})+\Phi(1/z^{[-a]})-\Phi(1/z^{[+a]}))} |_{q=0}\,,
\ea
where the last factor is evaluated at $q=0$.

Then we find  the coefficient of the double pole of $t$ as
\be \la{smallqt}
\lim_{q\to 0 }  q^2\, t_a (q) = 4 { (\cos\phi-\cos\theta )^2 \over
\sin^2 \phi }  \sin^2(a\,\phi) { a^4 \over (a^2 + 16 g^2 ) } \left( -a + \sqrt{ a^2 + 16 g^2} \over
 a + \sqrt{a^2 + 16 g^2} \right)^2 F(a,g) ^2\,.
\ee
 The factor in parenthesis is $(z^{[-a]}/z^{[a]} )^2 $.
Finally, inserting this into the expression for the energy \nref{luscherfull}, we find
\be
\Delta {\mathcal E} \sim -  { (\cos\phi-\cos\theta ) \over
\sin\phi } \sum_{a=1}^\infty  (-1)^a \left( -1 + \sqrt{ 1 + 16 g^2/a^2} \over
 1 + \sqrt{1 + 16 g^2/a^2} \right)^{ 1 + L }\!\!\!\!\! \sin(a\,\phi) { a  \over \sqrt{ 1 + 16 g^2/a^2 } } F(a,g)\,.
\label{luschergen}
\ee
The factor in parenthesis is just  $e^{ - E_m(a) (L+1) }$, representing the
exchange of a bound state in the mirror channel.
The sign  $(-1)^a$ is a bit subtle and has to do with the correct sign we should pick for
the square root in \nref{luscherfull}. The correct sign is easier to understand for
an angle of the form $\phi = \pi - \delta$, for small $\delta$. In this case we have a
quark antiquark configuration and it is clear that we should get a negative contribution
to the energy. In fact, we can think of the overlap of the two boundary states as computing a
kind of norm or inner product. We see that in terms of $\delta$ the expression has the
expected sign. In other words, for small $\delta$ we get the positive sign of the square root
in \nref{luscherfull}.
 Of course, once we get the expression for small $\delta$ we can write it in
terms of $\phi$, or even analytically continue $\phi = - i \varphi$.

 \subsubsection{Leading Luscher correction at weak coupling}

The expression \nref{luschergen} gives the leading Luscher correction at all values of the coupling
for large $L$. Let us now examine it at weak coupling. Then the factor in parenthesis in \nref{luschergen}
is of order $g^2$. So, at leading order, we get a term of the form $g^{2 + 2 L }$. This has the
interesting implication that this leading ``wrapping'' correction appears at $L+1$ loops. In particular for
$L=0$, the one loop contribution comes from such a term!. In fact, expanding \nref{luschergen} to leading order
in $g^2$, and setting $L=0$, we can set $F =1$ to this order and obtain
\ba
\Gamma_{\cusp} &=&-4 g^2  { (\cos\phi-\cos\theta ) \over
\sin \phi }  \sum_{a=1}^\infty (-1)^a   { \sin a \phi \over a }
\\
&=& 2 g^2  { (\cos\phi-\cos\theta ) \over
\sin \phi }   \phi\,,
\ea
which coincides exactly with the leading 1-loop contribution to $\Gamma_{\cusp} (\phi,\theta)$
computed in \cite{Drukker:1999zq}.

We can also do the computation of the leading order term for any $L$, we get
\be
{\cal E} =  -    g^ { 2 + 2 L}  { (\cos\phi-\cos\theta ) \over
\sin \phi } { (-1)^L ( 4 \pi )^{ 1 + 2 L } \over (1 + 2 L)! }
    B_{1 + 2 L} \left( { \pi - \phi \over 2 \pi } \right)  + \cO( g^{ 4 + 2 L } )
\ee
 where $B_n(x)$ is the Bernoulli polynomial, which is a polynomial of degree $2L+1$. In \cite{LadderPaper} a
 particular class of diagrams was identified which produced the same expression.

\subsubsection{Leading Luscher correction at strong coupling}

We can also compute the leading large $L$ correction at strong coupling. We simply evaluate
 the large $g$ limit of \nref{luschergen}. First we note that
  \be
  \left( { z^{[-a]} \over z^{[+a]}  } \right)^{L+1} \sim e^{ - { a \over 2 g } L } = e^{ - L  E_m(q=0)  }\,.
  \ee
  This implies that to leading order in $e^{ - L}$ we only need to consider the case $a=1$.
  The expansion of the function $F$ is done in appendix \ref{funfexp} eqn.  \nref{ffunres}.
  Putting everything together we find that the leading strong coupling
  correction goes as
  \be \la{luschstrong}
{ \cal E } =    ( \cos \phi - \cos \theta ) { 16 g \over e^2 } e^{ - { L \over 2 g } }\,.
  \ee
This agrees precisely with the result computed directly from  classical string theory
in appendix \ref{luschapp}, see \nref{finres}. This constitutes a nontrivial check of the
reflection phase.  Notice, the funny factor of $e^{-2}$ which is correctly matched.

\section{The open Asymptotic  Bethe Ansatz equations}

We will now write down the asymptotic Bethe ansatz (ABA) equations that
 describe the spectrum of operators with large $L$ inserted on the Wilson loop.
These give rise to a spin chain with two boundaries, which are separated by a large distance $L$.
Moreover, the ABA equations are used to derive the BTBA system by embedding them into the closed equations,
as we do in appendix \ref{embedingderivation}.

In order to obtain the ABA equations we have to diagonalize the way the bulk and boundary scattering matrices act.
This can be done by formulating a nested Bethe ansatz, which defines impurities at different levels
of nesting. Here we just sketch the computation, which is a straightforward generalization of the case
with periodic boundary conditions studied in \cite{BeisertDynamic}.

Consider an asymptotic state with $N^{\rm I}$ bulk magnons, or level I excitations, on the half-line
with a right boundary. We will introduce a second boundary and relative angles later, when writing down
the Bethe equations. In particular, we can consider a state whose  level I impurities all  carry the
same $SU(2|2)_{\rm D}$ index\footnote{The choice of  index is arbitrary.}.
Say, for example, in the unfolded notation,
\be
|\Psi_3(p_1)\cdots \Psi_3(p_{N^{\rm I}})\Psi_{\check 3}(-p_{N^{\rm I}})\cdots \Psi_{\check 3}(-p_1)\rangle \equiv |0\rangle^{\rm II}\,,
\la{levelIIvac}
\ee
which is regarded as the level II vacuum state. Of course, we could also consider states
where $N^{\rm II}$ out of the $N^{\rm I}$ level I impurities have different indices.
Those should be understood as $N^{\rm II}$ impurities in the level II vacuum state. In total,
such states will contain $N^{\rm I}$ level I impurities and $N^{\rm II}$ level II impurities.
In general, we have $|\Psi_{a_1}(y_1)\cdots \Psi_{a_{N^{\rm II}}}(y_{N^{\rm II}})\rangle^{\rm II}$ for $a_k=1,2$, where $y_k$ are auxiliary parameters associated with the level II impurities.

Similarly, a third level of nesting can be defined. If all the level II excitations carry the same index,
for instance $|\Psi_1(y_1)\cdots \Psi_1(y_{N^{\rm II}})\rangle^{\rm II}\equiv|0\rangle^{\rm III}$,
we can define a level III vacuum state.  Then, magnons $\Psi_2$ will be treated as level III impurities
propagating in $|0\rangle^{\rm III}$. For the kind of $SU(2|2)$ spin chain we are considering, this level III
is the final level of nesting\footnote{$\Psi_4$ are not considered as elementary but as double excitations.}.

Then, to formulate a coordinate Bethe ansatz, bulk and boundary scattering factors among excitations of different levels
have to be introduced to write the nested wavefunctions. Those can be determined by imposing certain compatibility conditions.
Namely, that the action of the bulk and boundary scattering matrices on wavefunctions with higher level impurities
just pulls out the same factor as when acting on the level II vacuum state. Naturally, the bulk scattering factors are exactly the same as the ones obtained in the periodic case \cite{BeisertDynamic},
\ba
S^{\rm I,I}(x_1^\pm,x_2^\pm) \!\!& = &\!\! - S_0(p_1,p_2)\,,
\\
S^{\rm I,II}(x^\pm,y) \!\!& = &\!\! 1/S^{\rm II,I}(y,x^\pm) = -\frac{y-x^{-}}{y-x^{+}}\,,
\\
S^{\rm III,II}(w,y)\!\!& = &\!\! \frac{w-y-{1\over y}+\frac{i}{2g}}{w-y-{1\over y}-\frac{i}{2g}} = \frac{w-v+\frac{i}{2g}}{w-v-\frac{i}{2g}}\,,
\\
S^{\rm III,III}(w_1,w_2) \!\!& = &\!\! \frac{w_1-w_2-\frac{i}{g}}{w_1-w_2+\frac{i}{g}}\,,
\ea
where
\be
S_0(p_1,p_2)^2 =  \frac{(x_1^+-x_2^-)(1-\frac{1}{x_1^-x_2^+})}{(x_1^--x_2^+)(1-\frac{1}{x_1^+x_2^-})} {1\over\sigma(p_1,p_2)^2}
\ee
is the bulk dressing factor and $v = y +{1\over y}$. All other bulk scattering factors are trivial.

\begin{figure}[h]
\centering
\def\svgwidth{14cm}
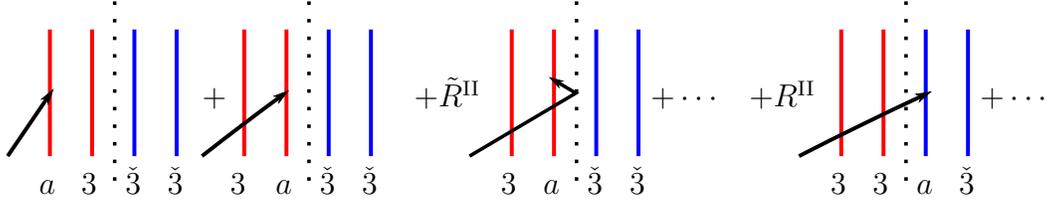
\caption{Propagation of a single level II impurity across the defect.}
\label{levelIIdefect}
\end{figure}

The reflection factors can be derived in the same way. The level II vacuum \nref{levelIIvac} containing
$N^{\rm I}$ magnons becomes a lattice with $2N^{\rm I}$ sites. Consider a single level II
impurity propagating in this vacuum  from left, i.e. propagating along the left (undotted) indices
of the bulk magnons. Undotted and dotted indices can only mix by the reflection of the rightmost
bulk magnon. That could make us think there exists a defect in the middle of the level II vacuum lattice which
separates  the 3  and $\check 3$ indices of the rightmost level I impurity. In principle, the level
II impurity could be reflected and transmitted across such defect, see figure \ref{levelIIdefect}.

However, and because the boundary scattering matrix $R(p)\propto S(p,-p)$, the compatibility condition we obtain
from the reflection of the rightmost level I impurity is analogous to the ones we obtain from the scattering of
two level I impurities. In this way, the compatibility conditions imply that level II impurities are purely transmitted.
In other words $\tilde R^{\rm II}=0$ and $R^{\rm II}=1$. Analogously, the reflection of level III impurities
is determined. In summary, we have
\be
R^{\rm I}(x^\pm) = R_0(p)\,, \quad  R^{\rm II}(y) = 1\,, \quad  R^{\rm III}(w) =  1\,,
\ee
where $R_0(p)$ is the boundary phase  factor \nref{fullphase}.

Let us now put the system in a finite strip by introducing another boundary. We will then
have certain quantization conditions on the rapidities for all kind of excitations,
namely the Bethe ansatz equations. We will introduce the left boundary with relative angles
with respect to the right one, by using the rotation discussed in section \ref{Rwithangles}.
To understand how this rotation affects the factors $R^{\rm I}$, $R^{\rm II}$ and $R^{\rm III}$
it is enough to consider the action of $m$, defined in \nref{mone}, on the following key
components of the reflection matrix
\ba
R_{3\check 3}^{3\check 3} \mapsto R_{3\check 3}^{3\check 3}\,,
&\Rightarrow &  R^{\rm I} \mapsto R^{\rm I}\,,\nn\\
R_{1\check 3}^{3\check 1} \mapsto e^{i\theta-i\phi} R_{1\check 3}^{3\check 1}\,,
&\Rightarrow & R^{\rm II} \mapsto e^{i\theta-i\phi} R^{\rm II}\,,\\
R_{2\check 1}^{1\check 2} \mapsto e^{-2i\theta}R_{2\check 1}^{1\check 2}\,,
&\Rightarrow & R^{\rm III} \mapsto e^{-2i\theta} R^{\rm III}\,,\nn
\ea

\begin{figure}[h]
\centering
\def\svgwidth{17cm}
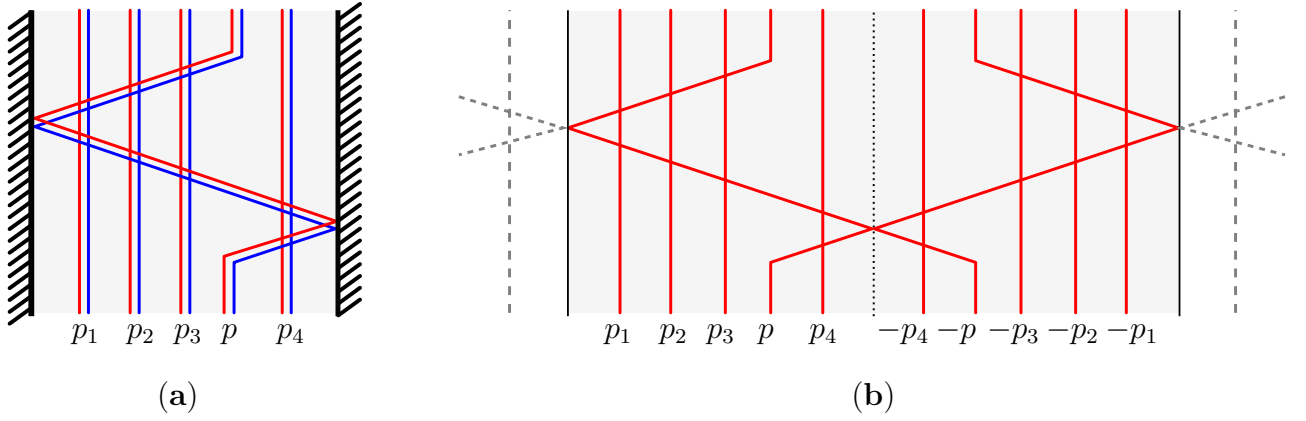
\caption{ Bethe equation for the open chain.  ({\bf a}) The original picture with boundaries. The particle goes to one boundary, then the other, and finally back to the origina position. ({\bf b}) The unfolded picture.  We
have a closed circle. The leftmost solid line is identified the rightmost one.
The motion that leads to the Bethe equations involves moving the magnon with momentum $p$ around
the closed circle and at the same time we also move its partner which has momentum $-p$ around
the circle in the opposite direction.  }
\label{openbethe}
\end{figure}

Let us finally write down the nested Bethe ansatz equations. They are obtained
by picking an impurity of any level of nesting and moving it through all the
other impurities twice and reflecting it from both boundaries as it is
shown in the left picture of figure \ref{openbethe}.

If we go to the unfolded picture what we have is periodic chain of length $2L$, where for every level I excitation of momentum $p_k$ there exists a mirrored one of momentum $-p_k$, figure \ref{openbethe}. Such duplication does not occur for higher levels of nesting, for which the excitations do not necessarily come in pairs. When moving around the level I excitations to derive the Bethe equations,
 we have to recall that their duplication is an artifact of the unfolding. Every pair represents a single magnon in the original picture. When we move the original magnon, it looks like moving the pair simultaneously in the unfolded picture. Then, for level I impurities we  pick up the factors that correspond to simultaneously moving around the pair with momentum $p_k$ and $-p_k$ in opposite directions.
For level II impurities, we have to collect the factors corresponding to going through all the level I pairs and all the level III impurities (scattering between level II particles is trivial). Finally, for level III impurities we get the factors of going through all the level II impurities and all the other level III impurities (scattering between level III and level I particles is trivial).
The resulting set of open Bethe ansatz equations is the following
\begin{align}
1 &= \left(\frac{x_{k}^{+}}{x_{k}^{-}}\right)^{2L}
\left( { 1 + { 1 \over (x^-)^2 } } \over 1 + { 1 \over (x^+)^2 } \right)^2 { 1\over \sigma_B(p_k)^2 \sigma(p_k,-p_k)^2}
\prod_{l=1}^{N^{\rm II}}\frac{y_{l}-x_{k}^{-}}{y_{l}-x_{k}^{+}}\frac{y_{l}+x_{k}^{-}}{y_{l}+x_{k}^{+}}, \label{BI}
\\
& \quad \prod_{l\neq k}^{N^{{\rm I}}}
\frac{(x_k^+ - x_l^-)(1-\frac{1}{x_k^-x_l^+})}{(x_k^--x_l^+)(1-\frac{1}{x_k^+ x_l^-})}
\frac{(x_l^+ + x_k^+)(1+\frac{1}{x_l^-x_k^-})}{(x_l^-+x_k^-)(1+\frac{1}{x_l^+ x_k^+})}
{1\over \sigma(p_k,p_l)^2 \sigma(p_l,-p_k)^2}\nn
\\
1 & = e^{i\theta-i\phi} \prod_{l=1}^{N^{{\rm I}}}\frac{y_{k}-x_{l}^{+}}{y_{k}-x_{l}^{-}}\frac{y_{k}+x_{l}^{-}}{y_{k}+x_{l}^{+}}
    \prod_{l=1}^{N^{\rm III}}\frac{w_{l}-v_{k}-\frac{i}{g}}{w_{l}-v_{k}+\frac{i}{g}}
    \label{BII}\\
1 &= e^{-2i\theta}  \prod_{l=1}^{N^{\rm II}}\frac{w_{k}-v_{l}+\frac{i}{g}}{w_{k}-v_{l}-\frac{i}{g}}
    \prod_{l\neq k}^{N^{\rm III}}\frac{w_{k}-w_{l}-\frac{2i}{g}}{w_{k}-w_{l}+\frac{2i}{g}}
     \label{BIII}.
\end{align}
Eq. \nref{BI} can be re-written, including $l=k$ in the second product, as
\begin{align}
1 &= -\left(\frac{x_{k}^{+}}{x_{k}^{-}}\right)^{2L}{x^-+{1\over x^-}\over x^++{1\over x^+}}
{ 1\over \sigma_B(p_k)^2 }
\prod_{l=1}^{N^{\rm II}}\frac{y_{l}-x_{k}^{-}}{y_{l}-x_{k}^{+}}\frac{y_{l}+x_{k}^{-}}{y_{l}+x_{k}^{+}}, \label{BI2}
\\
& \quad \prod_{l=1}^{N^{{\rm I}}}
\frac{(x_k^+ - x_l^-)(1-\frac{1}{x_k^-x_l^+})}{(x_k^--x_l^+)(1-\frac{1}{x_k^+ x_l^-})}
\frac{(x_l^+ + x_k^+)(1+\frac{1}{x_l^-x_k^-})}{(x_l^-+x_k^-)(1+\frac{1}{x_l^+ x_k^+})}
{1\over \sigma(p_k,p_l)^2 \sigma(p_l,-p_k)^2}\nn
\end{align}

As usual, the energy is given by
\be
{\cal E} = \sum_{k=1}^{N^I } \epsilon( p_k)
\ee

\section{The boundary TBA equations}
\la{BdyTBA}

The Bethe equations (\ref{BI})-(\ref{BIII}) presented in the previous section are the correct description of the spectrum for large chains, $L\gg 1$. As $L$ becomes small,
 wrapping effects come into play and the Bethe equations are no longer valid. Moreover, in this paper,  we are mainly interested in  $L=0$.
  A description of the spectrum that is valid for any $L$
   is the Boundary Thermodynamic Bethe Ansatz (BTBA) equations. These are a set of integral equations that govern the dynamics in the mirror channel. That is, the dynamics of excitations after exchanging the two dimensional space and time directions \cite{Zamolodchikov:1989cf,Ghoshal:1993tm}, see figure \ref{TBAfigure}.
The TBA equations can be derived from the knowledge of the spectrum of states and bound
states in the mirror channel.  This  spectrum was derived in \cite{Arutyunov:2009zu}.
The derivation of the TBA equations then follows the standard route given in
\cite{Yang:1968rm,Zamolodchikov:1989cf,Bombardelli:2009ns,Arutyunov:2009ur,Gromov:2009bc}. In the case that we have a boundary we can
follow essentially the same route. We use the boundary state defined in
section \ref{sec2}, and the untangling of boundary reflection matrices described
in figure \ref{Unfolding}. Then we
get a TBA which looks very similar to what we would obtain for a closed chain
 of twice the length $L$, except for the fact that for each particle
 of momentum $q$ we get one of momentum $-q$,
since the boundary state creates such a pair of particles. The consequence of this is
that the $Y$ functions obey a reflection property
\be \la{FLIP}
Y_{a,s} (u) = Y_{a,-s}(-u)
\ee
  The set of $Y_{a,s}$ functions is the same as the one we have for
  the closed string problem \cite{Bombardelli:2009ns,Arutyunov:2009ur,Gromov:2009bc}.
 However, due to \nref{FLIP} we can restrict our attention to the ones with $s\geq 0$.
  The boundary data appears as chemical potentials which depend
 on the angles, $\theta,~\phi$,
 as well as a $u$ dependent chemical potential given by the boundary
 dressing phase $\sigma_B$. The precise form of  the equations is derived in appendix
 \nref{TBAderivation}.
\begin{figure}[h]
\centering
\def\svgwidth{16cm}
 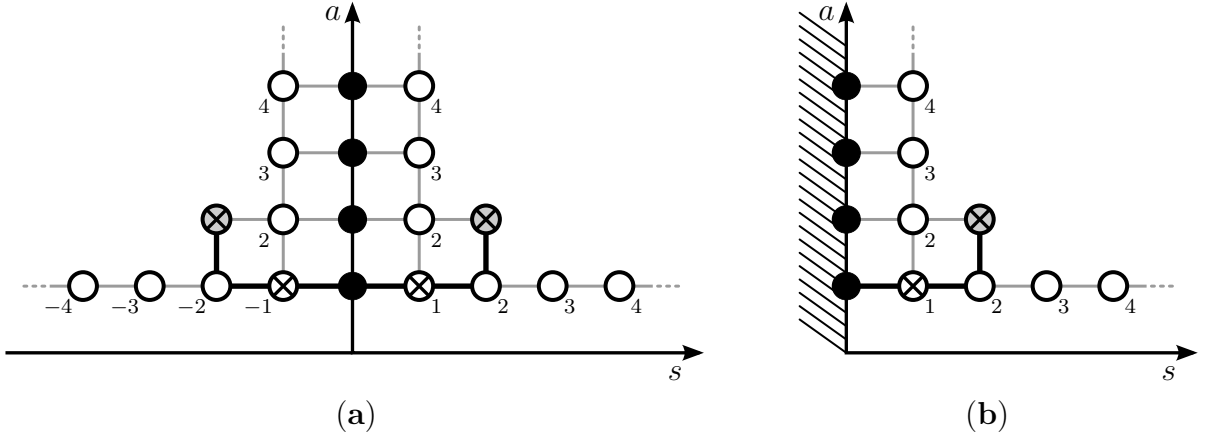
\caption{
(a) Set of $Y_{a,s}$ functions for the closed string problem. Here we have the same set but
the additional condition \nref{FLIP} implies that we can restrict to the set in (b).  }\label{Ysystem}
\end{figure}

Let us summarize the final equations
\ba
\label{1stTBA}
\log{Y_{1, 1}\over\bY_{1,1}}\!\!&=&\!\! K_{m-1}*\log{1+{\baY_{1, m}}\over1+\overline\bY_{1, m}}{1+\bY_{m,1}\over1+Y_{m,1}}+\cR^{(01)}_{1\,a}* \log(1+Y_{a,0}) \\
\label{2ndTBA}
\log{\baY_{2, 2}\over\overline\bY_{2,2}}\!\!&=&\!\!\ \ K_{m-1}*\log{1+{\baY_{1, m}}\over1+\overline\bY_{1, m}}{1+\bY_{m,1}\over1+Y_{m,1}}+\cB^{(01)}_{1\,a}* \log(1+Y_{a,0})
\\
\label{3rdTBA}
\log{\baY_{1, s}\over\overline\bY_{1, s}}\!\!&=&\!\!- K_{s-1,t-1}*\log{1+\baY_{1, t}\over1+\overline\bY_{1, t}}-K_{s-1}\hat *\log{1+Y_{1,1}\over1+\baY_{2,2}}
\\
\log{Y_{a,1}\over\bY_{a,1}}\!\!&=&\!\!- K_{a-1,b-1}*\log{1+Y_{b,1}\over1+\bY_{b,1}}-K_{a-1}\hat*\log{1+Y_{1,1}\over 1+\baY_{2,2}}\nn
\\
&&\qquad\qquad\qquad+\[\cR^{(01)}_{ab}+\cB^{(01)}_{a-2,b}\]*\log(1+Y_{b,0})
\label{4thTBA}
\\
\label{lastTBA}
\log{Y_{a,0}\over\bY_{a,0}}\!\!&=&\!\!\[2{\cal S}_{a\,b}-{\cal R}_{a\,b}^{(11)}+{\cal B}_{a\,b}^{(11)}\]*\log(1+Y_{b,0})+2\[{\cal R}_{a\,b}^{(1\,0)}+{\cal B}_{a,b-2}^{(1\,0)}\]\s*\log{1+Y_{b,1}\over1+\bY_{b,1}}\nn
\\
&&+2{\cal R}_{a\,1}^{(1\,0)}\hs*\log{1+Y_{1, 1}\over1+\bY_{1, 1}} -2{\cal B}_{a\,1}^{(1\,0)}\hs*\log{1+\baY_{2,2}\over1+\overline\bY_{2,2}}
\ea
where we used the conventions of \cite{GKV,GKKV} for the kernels and integration contours\footnote{
The convolutions of terms depending on $Y_{1,1}$ or $\baY_{2,2}$ are over a finite range $|u| \leq 2 g$. We use $\hat*$ as a reminder of that.}.
We have also defined the barred $Y$'s as   $\baY_{a,s}^{(\text{here})}=1/Y_{a,s}^{(\text{there})}$, (see appendix \ref{kernels} for a summary). Here, the momentum carrying $Y_{a,0}$ functions
 are defined as symmetric functions $Y_{a,0}(-u)=Y_{a,0}(u)$ and $\s*f(v)=[*f(v)+*f(-v)]/2$ is a symmetric convolution\footnote{For the ground state,  we expect all functions to be symmetric, $Y_{a,s} (u) = Y_{a,s}(-u)$.  But for excited states \nref{FLIP} only requires
  the $Y_{a,0}$ functions  to be symmetric.
   The equation for excited states could in principle be obtained by analytic continuation from these equations \cite{Dorey:1996re,Bazhanov:1996aq,GKV,GKKV}. }.
    There are implicit sums over one of the
 indices of the kernels\footnote{ The indices of $Y_{1,m}$ or $Y_{m,1}$ run over $m\geq 2 $. For $Y_{b,0}$ they run over $b\geq 1 $. The same as in \cite{GKKV}. }.
  The bold face $\bY$'s represent the asymptotic large $L$ solution. This is the solution
  we obtain  when  the convolutions with the momentum carrying $Y_{a,0}$'s are dropped. These asymptotic solutions are the only place
   where the angles and the boundary dressing phase enter. They are given by
\ba\la{asymptotic1}
\bY_{1,1}\!\!&=&\!\!-\frac{\cos\theta}{\cos\phi}\ ,\qquad\quad\overline\bY_{1,s}=\frac{\sin^2\theta}{\sin[(s+1)\theta]\sin[(s-1)\theta]}
\\
\overline\bY_{2,2}\!\!&=&\!\!-\frac{\cos\theta}{\cos\phi}\ ,\qquad\quad\bY_{a,1}=\frac{\sin^2\phi}{\sin[(a+1)\phi]\sin[(a-1)\phi]}\\
\la{asymptotic2}
\bY_{a,0}&=&4
{ e^{i\chi(z^{[+a]}) + i \chi(1/z^{[-a]})} \over e^{ i \chi(z^{[-a]}) + i \chi(1/z^{[+a]})} } \left(\frac{z^{[-a]}}{z^{[+a]}}\right)^{2L+2}\!\!\!\!\!\!\!\! (\cos\phi-\cos\theta)^2\,\frac{\sin^2 a\, \phi}{\sin^2\phi}\,.\la{asymptotic3}
\ea
where $\chi$ is the function defining  the boundary dressing phase (\ref{phasefa}).
Notice that the length $L$ appears only  in \nref{asymptotic2}.
Here $z^{[\pm a]}$ are the solutions of
\be \la{uExpression}
u = g \left( z^{[+a]} + { 1 \over z^{[+a]}} \right)  - i { a \over 2  } =  g \left( z^{[-a]} + { 1 \over z^{[-a]}}
\right)  + i { a \over 2  } = { q \over 2} \sqrt{ 1 + { 16 g^2 \over a^2 + q^2 } }
\ee
in the mirror region with $ |z^{[+ a]}|>1$ and $|z^{[- a]}|<1$.

 Once we solve this  system of equations, we can compute the ground state energy
 as
\be
{\cal E}  =
-\sum_{a=1}^{\infty}
\int\limits_0^{\infty}\frac{dq}{2\pi} \log(1+Y_{a,0})\,,
\label{exactE}
\ee
where $q$ is the mirror
momentum of each magnon bound state
\be
q = g \left[ z^{[+ a]} -  z^{[- a]}- { 1 \over  z^{[+ a]} }  + { 1 \over
  z^{[- a]} } \right]
\ee

\subsection{Recovering the Luscher result }
\la{reclu}

As a simple check of these equations let us rederive the results of section
\ref{Luschersub}. In the large $L$ limit we see that
the factor $ \left(\frac{z^{[-a]}}{z^{[+a]}}\right)^{2L+2} =  e^{ - E_m 2 (L+1) }$ is very small.
This implies that the $\bY_{a,0}$ in \nref{asymptotic2} are very small. So we expect
 that the $Y_{a,0}$ are also small and that we  can
 set  them to zero in all the convolution terms of the TBA equations. In this limit,
 the energy is given by inserting the asymptotic form $\bY_{a,0}$, \nref{asymptotic2}, in
 the expression for the energy \nref{exactE}. One would be tempted to expand
 the logarithm in \nref{exactE}, since  $\bY_{a,0}$ is very small. However,
 $\bY_{a,0}$ has a double  pole a $u=0$, or $q=0$,  coming from the boundary
 dressing phase. In other words, it
 behaves as
 \be \la{smallq}
 \bY_{a,0} \sim { G_a^2  \over q^2 }  + \cO(1)
 \ee
 for small $q$.
 We can then write the integrals  in \nref{exactE} as
 \ba
 \int\limits_0^{\infty}\frac{dq}{2\pi} \log(1+\bY_{a,0}) & = &
  \int\limits_0^{\infty}\frac{dq}{2\pi} \log\left(1+{ G_a^2 \over q^2 } \right) +
   \int\limits_0^{\infty}\frac{dq}{2\pi} \log{ (1+\bY_{a,0}) \over
   \left(1+{
G_a^2 \over q^2 } \right)}
   \ea
   In the second term we can certainly expand to first order in $\bY_{a,0}$ and $G_a^2$,
   which produces a result which is of order $e^{ - 2 E_m ( L +1)}$. The first term, however, gives $G_a/2 \sim e^{ - E_m (L+1)}$, which is bigger. So  we get
   \be \la{enget}
   {\cal E} \sim  - { 1 \over 2} \sum_{a=1}^\infty { G_a }
   \ee
   But this is precisely the same as what we got in section \ref{sec2}. Namely,
   \nref{luscherfull} is the same as \nref{enget}  after we realize that $G_a$ defined
   in \nref{smallq} is essentially
     the same as \nref{smallqt}, using \nref{asymptotic2}. This is not too surprising
      since \cite{Bajnok:2004tq} derived \nref{luscherfull} by appealing to
       TBA equations. In summary,
      \nref{enget} agrees precisely with \nref{luschergen}.

 In the next  section we will perform a weak coupling check of the equations.
 We will derive  a simplified
set of equations that describe the small angle limit $\theta, ~ \phi \sim 0 $ and
we will expand and solve the resulting equations up to order $g^6$.

\section{The near BPS limit}
\label{smallphi}

When $\phi=\theta$ the Wilson loop is BPS and the energy vanishes.
As we deform the angles away from this supersymmetric configuration, the energy behaves as
\be \la{DevBPS}
\Gamma_{\cusp}(\phi,\theta)  = - ( \phi^2 -\theta^2 ) {  1  \over 1 - { \phi^2 \over \pi^2 } } B( \tilde \lambda ) + {\cal O}((\phi^2  -\theta^2)^2) ~,~~~~~~~~~ \tilde \lambda = \lambda ( 1 - { \phi^2 \over \pi^2 } )\,.
\ee
The function $B$, also known as the ``Bremsstrahlung function", is related to a variety of physical quantities
\cite{Correa:2012at,Fiol:2012sg}. It was computed exactly in \cite{Correa:2012at,Fiol:2012sg}
  using localization. In the planar limit we get
\be
B={ 1 \over 4 \pi^2 } { \sqrt{\tilde\lambda } I_2( \sqrt{\tilde\lambda} ) \over I_1( \sqrt{\tilde\lambda} ) }  + {\cal O}( 1/N^2 ) \la{bplanar}
\ee
On the one hand, this  allows us to test the BTBA equation to high loop orders by penetrating deep into almost all parts of the equation. On the other hand,  the simplicity of \nref{bplanar}
suggests that,    in the near BPS limit, the BTBA equations can be drastically simplified.
The equations we will find in this limit are not that simple.
We hope that understanding how to simplify them  will teach us  how to simplify  TBA equation in general.

In this section we will study the BTBA equations in this limit. We will show that the BTBA equations can be reduced to a simplified set of equations. We will then solve them to 3-loop order.
 Here we restrict the discussion to $\theta =0$\footnote{The general near BPS case, with $\theta \not =0$,
  has a similar degree of complexity. In fact, we have explicitly expanded the equations up to second order in $\lambda$ and verified the corresponding expansion in \nref{bplanar}.  But we will not give the details here.}, so that
 $\tilde\lambda=\lambda$ and $\Gamma_{\cusp}(\phi,\theta)  = -\phi^2 B(\lambda ) + {\cal O}(\phi^4)$.
  We also set $L=0$ to extract  the cusp anomalous dimension. It is important to note that now $\phi$ is the smallest parameter. In particular, it is smaller than $\lambda$.

In this small angle limit, the momentum carrying Y-functions are of order $Y_{a,0}={\cal O}(\phi^4)$ and therefore very small. This limit  reminds us of the large $L$ asymptotic limit where the momentum carrying $Y_{a,0}$'s are exponentially suppressed. However, as opposed to the asymptotic limit, in the small angle limit,
 we cannot drop the convolutions with the momentum carrying $Y_{a,0}$'s. Instead, we remain with a simplified set of non linear equations. The reason  is that the large value of $\log Y_{a,0}$ is not due to
the sources in the BTBA equations. Instead, it is due to the fact that
  the fermionic $Y$-functions ($Y_{1,1}$ and $Y_{2,2}$)  approach  $-1$ and lead to a big contribution through the    $\log(1+Y_{1,1})$ and $\log(1+Y_{2,2})$ terms in the convolutions.

\subsection{The simplified equations at small angles}

As the momentum carrying Y-functions are small, they only contribute to $B(\lambda)$ through their double pole. We define $\mathbb C_a$ as the coefficient of the double pole at $u=0$,
\be
\lim_{\substack{q\to0\\ \phi\to0}} Y_{a,0}
=\[- {\phi^2\over 2 u}{\mathbb C}_{a}\]^2
\la{defC}
\ee
The energy, which  is dominated by the value of
$Y_{a,0}$  at the double pole, reduces to
\be
{\cal E}=\frac{\phi^2}{2}\sum_{a=1}^\infty{\mC_a\over\sqrt{1+{16g^2/ a^2}}}\,,
\label{EfromC}
\ee
where square root factor comes from the $q\to0$ limit of $(q/2u)$, see \nref{uExpression}.

In this small $\phi$-limit, the other $Y$-functions can be expanded as
\ba\la{expansion}
Y_{1,1}&=&-1-\phi^2\,\Psi+{\cal O}(\phi^4)\,,\qquad\quad
Y_{m,1}=\cY_m\[1+\phi^2(\Omega_m-\cX_m)/2\] + {\cal O}(\phi^4)\,,\\
\overline{Y}_{2,2}&=&-1-\phi^2\,\Phi+{\cal O}(\phi^4)\,,\qquad\quad
\overline{Y}_{1,m}=\cY_m\[1+\phi^2(\Omega_m+\cX_m)/2\] +{\cal O}(\phi^4)\,.\nn
\ea
where we assumed that to leading order $Y_{1,1}=Y_{2,2}=-1$ and $Y_{m,1}=Y_{1,m}$. It is not difficult to see that this assumption is consistent with the BTBA equations. Moreover, we find that the functions $\Omega_m$ drop out of the equations.

We find that the BTBA equations (\ref{1stTBA})-(\ref{asymptotic3}) reduce to
\ba
\Psi\!\!\!&=&\!\!\!{1\over2}+K_{m-1}*\[\cX_m{\cY_m\over 1+\cY_m}+{1\over3}\]-\pi\,\mC_{a}\,\cR^{(01)}_{1\,a}(u,0)
\la{wlanyc1}
\\
\la{wlanyc2}\\
\Phi\!\!\!&=&\!\!\!{1\over2}+K_{m-1}*\[\cX_m{\cY_m\over 1+\cY_m}+{1\over3}\]-\pi\,\mC_{a}\,\cB^{(01)}_{1\,a}(u,0)
\\
\log \cY_m\!\!\!&=&\!\!\!-K_{m-1,n-1}*\log\(1+\cY_n\)-K_{m-1}\hat*\log{\Psi\over\Phi}
\la{wlanyc3}\\
\cX_m\!\!\!&=&\!\!\!-{m^2\over3}-K_{m-1,n-1}*\[\cX_n{\cY_n\over 1+\cY_n}+{1\over3}\]+\pi\,\mC_{n}\[\cR^{(01)}_{mn}+\cB^{(01)}_{m-2,n}\](u,0)
\la{wlanyc4}
\\
  \Delta_\text{conv}  \!\!\!&=&\!\!\! \left. \left\{ {\cal R}_{a\,1}^{(1\,0)}\hat*\log({\Psi\over 1/2})
  -{\cal B}_{a\,1}^{(1\,0)}\hat*\log({\Phi\over 1/2})+\[{\cal R}_{a\,b}^{(1\,0)}
  +{\cal B}_{a,b-2}^{(1\,0)}\]*\log\({1+\cY_{b}\over1+{1\over b^2-1}}\) \right\}\right|_{u=0}  \la{deltaconv}
\\
\mC_a \!\!\!&=&\!\!\! (-1)^a  a^2 F(a,g){z_0^{[-a]}\over z_0^{[+a]}} e^{  \Delta_\text{conv} }  \la{defc}
\ea
where $z_0^{[\pm a ]}$ denote the values of $z^{[\pm a ]}$ at $q=0$ \nref{zpm}. In \nref{deltaconv} we
are evaluating the non-convoluted variable of the kernels at $u=0$. The hat on $\hat *$ is a convolution
over the range $|u| \leq 2 g $. $F(a,g)$ is given in \nref{ffun}.
These equations are derived by implementing the expansion of Y-functions \nref{expansion} in the TBA system of equations \nref{1stTBA}-\nref{lastTBA}.
Let us make a couple of comments.  First, the factors of $1/2$, $1/3$, $m^2/3$

 stand for the subtraction of the asymptotic solutions. These read
\be
\underline\Psi=\underline\Phi= {1\over2}\,, \qquad
\underline{\cY_{m}}={1\over m^2-1}\,, \qquad
\underline{\cX_m} = -{m^2\over3 }\ee

Second, note that in the BPS vacuum where $\phi=0$, the TBA equations are not well defined and need a regularization. A regulator commonly used is a twist for the fermions \cite{Frolov:2009in}.
Here, the angle $\phi$  can be viewed as a physical  regulator.
As opposed to other regulators,
the leading order solution $\cY_m$ is a non trivial function of the coupling.

\subsection{Weak coupling expansion of the small $\phi$ TBA}

To test the BTBA equations, we have solved the small angle simplified equations, \nref{wlanyc1}-\nref{defc},
up to three loops. In this section
 we will present the results.  The  derivation is given in appendix \ref{smallphisolution}.

The small $\phi$ TBA equations, \nref{wlanyc1}-\nref{defc},   are certainly simpler than the general TBA equations \nref{1stTBA}-\nref{lastTBA}, but they continue to be non-linear.
 However, if we make a weak coupling expansion we obtain a linear system of integral equations order by order.

To solve these linear equations we find it useful to first simplify the TBA equations as in \cite{Arutyunov:2009ux,Arutyunov:2009ax}.
To simplify   \nref{wlanyc1} and
\nref{wlanyc2}, we take a convolution of the equations with $\ms*\ms^{-1}$ where
\be
\ms(u) = \frac{1}{2\cosh(\pi u)}
\ee
The other equations can also be simplified as shown in the appendix \ref{smallphisolution}.
Then   \nref{wlanyc1}-\nref{defc} become
\ba
\Phi-\Psi\!\!\! &=&\!\!\!\pi\; \mC_a \hat K_{y,a}(u,0)\,,
 \label{ferdiftba}\\
\Phi + \Psi\!\!\! &=&\!\!\!
- 2 {\mathpzc s}*{\cX_{2}\over1+\cY_{2}}
+ 2 \pi {\mathpzc s}*\cR^{(01)}_{2\,n}(u,0)\mC_n
-\pi\,\mC_{a}\,K_a(u,0)\,,
\label{feraditba}
\\
\log\cY_m\!\!\! &=&\!\!\!{\mathpzc s}* I_{m,n}\log{\cY_n \over1+\cY_n }+\delta_{m,2}\, {\mathpzc s}\hat*\log{\Phi\over\Psi}\,,\label{yosimtba}\\
\cX_m\!\!\! &=&\!\!\!{\mathpzc s}*I_{m,n}{\cX_n \over 1+\cY_n }+\pi {\mathpzc s}\ \mC_m
+\delta_{m,2}\,{\mathpzc s}\hat*(\Phi-\Psi)\,,
\label{xosimtba}
\\
  \Delta_\text{conv}  \!\!\!&=&\!\!\! \left. \left\{ {\cal R}_{a\,1}^{(1\,0)}\hat*\log({\Psi\over1/2})
  -{\cal B}_{a\,1}^{(1\,0)}\hat*\log({\Phi\over 1/2})+\[{\cal R}_{a\,b}^{(1\,0)}+{\cal B}_{a,b-2}^{(1\,0)}\]*\log\({1+\cY_{b}\over1+{1\over b^2-1}}\)\right\}\right|_{u=0}  \la{deltaconvnew}
\\
\mC_a \!\!\!&=&\!\!\! (-1)^a  a^2 F(a,g){z_0^{[-a]}\over z_0^{[+a]}} e^{  \Delta_\text{conv} }  \la{defcnew} \ea
where $I_{m,n}=\delta_{m+1,n}+\delta_{m-1,n}$ and $\hat K_{y,a}$ is defined in appendix \ref{smallphisolution}.

Now expanding the functions $\Psi$, $\Phi$, ${\cal Y}_n$ and ${\cal X}_n$ in powers of $g^2$,
we can obtain them order by order by solving a linear system of equations. Up to three loops
(see appendix \ref{smallphisolution} for details) we find that
\be
\mC_a = 4(-1)^a g^2 + 8(-1)^a \[\pi^2-{4\over a^2}\]g^4 +16(-1)^a \[{\pi^4\over 3}-{4\pi^2\over a^2}+{20\over a^4}\]g^6  +{\cal O}(g^8)\,,
\ee
Finally,  the relation \nref{EfromC}, we obtain the expression for energy up to 3-loop order\footnote{
We encounter the sum  $\sum_{a=1}^{\infty} (-1)^a = - { 1 \over 2}$. This can be understood by regularizing it
 as $ \lim_{ \phi \to 0 } \left[ \sum_{a=1}^{\infty} (-1)^a { \sin a \phi \over a \phi } \right] = - { 1 \over 2 }$. }
\be
{\cal E}=-\phi^2 \[g^2 -g^4\frac{2 \pi ^2 }{3}+g^6\frac{2 \pi ^4 }{3}+ O(g^8)\]=-\phi^2 \[\frac{\lambda }{16 \pi ^2}-\frac{\lambda ^2}{384 \pi ^2}+\frac{\lambda^3}{6144 \pi ^2}
+{\cal O}(\lambda^4)\]\,,
\ee
In perfect agreement with the expansion of (\ref{bplanar}).

\section{Conclusions and discussion}

In this paper we have considered the problem of computing the
quark anti-quark potential on the 3-sphere in ${\cal N}=4 $ super Yang
Mills in the planar approximation. Since the planar theory
is integrable \cite{Review}, we expected to be able to derive an
exact expression.
Indeed, we found a system of boundary TBA equations \nref{1stTBA}-\nref{lastTBA}
 which determines the potential
as a function of three parameters:  the planar coupling $\lambda$,
the geometric angle $\phi$, which sets the angular separation on the 3-sphere and
an internal angle $\theta$ which is the relative orientation of the coupling to
the scalar field for the quark and the anti-quark.

This  quark and anti-quark  configuration gives rise to  an integrable
system with a boundary. This is most clearly seen in the string theory picture
where we have a string going between the two lines on the boundary. One might be surprised
that we have a boundary since the string is infinitely long. However, note that the local
 geometry of the string near the boundary is $AdS_2$,  which indeed has a boundary.
 The energy is then the
ground state energy , or Casimir energy,  on the strip and it is given in terms of the solution of the
 TBA equations \nref{exactE}.
This is  the energy of the flux tube connecting the quark and anti-quark. These TBA equations
should also enable one to compute the energies of excitations of the flux tube. These correspond to
operators that are inserted on the  Wilson loop.

The quark anti-quark potential on $S^3$ is the same as the cusp anomalous dimension as a function of
the angles, $\Gamma_{cusp}(\phi,\theta,\lambda)$.

The derivation of the  boundary TBA equations is similar to the one in other
integrable models with boundary \cite{LeClair:1995uf}.
A crucial step is the determination of
the boundary reflection matrix. The matrix part is fixed by the symmetries and
the dressing phase was found by solving the boundary
crossing equation  and the final answer is in  \nref{phasefa},
 \nref{chiintegral}.
Since there is always a certain amount of guesswork in determining the dressing
phase, we have checked it at strong coupling and we have seen that it
gives the right value both   in the physical and mirror regions.
A crucial feature  of the dressing phase is that it contains a pole at zero
mirror momentum. This is crucial for the proposed phase to work at weak coupling.
Note that the boundary dressing phase is responsible for the
 leading order contribution in the mirror picture, while it
 only starts contributing at three loops for
 anomalous dimensions in the physical picture.
   The pole simply means that the boundary
is sourcing single particle states.

The BTBA equations were written in \nref{1stTBA}-\nref{lastTBA}. They look very similar to the bulk
TBA equations \cite{Gromov:2009bc,Arutyunov:2009ux,Bombardelli:2009ns},
except that the boundary conditions for large $u$ are different. They now
depend on the angles. In addition, for the momentum carrying nodes, the $Y_{a,0}$,
there is an extra source term involving the boundary dressing phase.

We have obtained a simplified set of equations, \nref{ferdiftba}-\nref{defcnew},
 which describes the small
angle region, $\phi , ~\theta \ll 1 $. In this region,  the simplest way to solve the
problem is through supersymmetric localization, as explained in \cite{Correa:2012at}.
 The planar answer is
\be \la{brem}
 \Gamma_{cusp}(\phi, \theta=0, \lambda)  = - \phi^2 B  + \cO(\phi^4) ~,~~~~~~~~~~~~~~~
  B =    { 1 \over 4 \pi^2 } { \sqrt{\lambda } I_2( \sqrt{\lambda} ) \over I_1( \sqrt{\lambda} ) }
 \ee
So, we know the answer by independent means. Thus, these simplified BTBA equations
should reproduce \nref{brem}. Indeed,  directly expanding these
simplified equations up to third order in the coupling  we reproduced the expansion of \nref{brem}.
However, these ``simplified'' equations are vastly
more complex than the simple Bessel functions in \nref{brem}!.
Thus, there should be a way to simplify these equations much further and directly get the
simple answer \nref{brem}. Hopefully, the methods used to simplify the equation will also
be useful in order to simplify the full BTBA equations for general angles.
 Note that in \cite{Gromov:2011cx} the TBA system for closed strings was reduced to a  set of
 equations involving a finite number of functions. It is very
  likely that the same method  works in
 our case.

Notice that the simplified small angle  equations
 connect the integrability and the localization exact solutions.
In particular, computing the function $B$ by both methods would  enable us to see whether
 the coupling constant $\lambda$ that
appears in both approaches is the same or not. Of course, we expect them to be the same for ${\cal N}=4$
super Yang Mills.
However, if one could generalize the discussion in this paper to Wilson loops in ABJM theory  \cite{Aharony:2008ug}, then
this small angle region could enable us to compute the undetermined function $h(\lambda)$ that appears
in the integrability approach to the ABJM theory \cite{Gromov:2008qe}.

In principle, one might wonder whether the Wilson loop leads to an integrable boundary condition.
We have found that the reflection matrix obeys the boundary Yang Baxter equation. The TBA equations were
derived assuming integrability. So all the checks we performed on them are further evidence that the
Wilson loop boundary condition is indeed integrable.

There are further checks of the equations that one should be able to do.
In particular,
one would like to reproduce the BES   equation \cite{BES} for $\varphi \to \infty$.

It would also be nice to take the small $\delta = \pi -\phi$ limit. In this limit the answer should
go like $1/\delta $ and probably  one can obtain again a simplified equation for the coefficient. This
determines the quark anti-quark potential in the flat space limit.

One should also be able to take the strong coupling limit of the equations and reproduce the
result derived from classical strings in $AdS_5 \times S^5$  in \cite{Forini:2010ek,arXiv:1105.5144}.
It is likely that the ideas in \cite{Gromov:2009tq,Gromov:2009at}
 would enable this. 

Though solving the TBA equation analytically looks difficult, it should be possible
to solve the equations numerically. The problem should be very similar to the one solved in
 \cite{Gromov:2009zb}.

It would also be nice to study the problem of determining the open string spectrum on
the $AdS_4 \times S^2$ or $AdS_2 \times S^4$ D-branes which also preserve the same amount of
symmetry. The only difference with the current paper should be a different choice for the
boundary dressing phase. For this reason, the TBA equations would be the same,  except
for the choice of the boundary dressing phase.

The study of perturbative amplitudes at weak coupling has found remarkably
simple underlying structures. It would be interesting to study these structures in the
context of the cusp anomalous dimension, where we have a function of a single angle $\phi$.
In particular, it would be nice to see how to connect those structures with the TBA approach
described here. This would most probably lead to both a simplification of this TBA approach as well
as some hints on the exact structure underlying the amplitude problem.

Throughout this paper we have considered the locally BPS Wilson loop which contains the
coupling to the scalar, as in \nref{wildef}. Of course, one can also consider the
Wilson loop which does not couple to the scalars, $W = tr P e^{ i \oint A } $. It would be interesting
to see whether this leads to an integrable boundary condition. At strong coupling this loop leads to
a Neumann boundary condition on the $S^5$ \cite{Alday:2007he},  which is classically integrable.\footnote{See \cite{Dekel:2011ja} for a systematic study of classically integrable boundary conditions.}

{\bf Note:} We were informed that similar ideas were pursued in \cite{DrukkerTBA}.

{\bf Acknowledgements }

We would like to thank N. Arkani-Hamed, B. Baso,
S. Caron-Huot, N. Drukker, D. Gaiotto, N. Gromov, I. Klebanov,   P. Vieira and A. Zamolodchikov for discussions.

A. S. would like to thank Nordita for warm hospitality. This work was supported in part by   U.S.~Department of Energy grant \#DE-FG02-90ER40542.
Research at the Perimeter Institute is supported in part by the Government of Canada through NSERC and by the Province of Ontario through MRI.  The research of  A.S.
 has been supported in part by the Province of Ontario through ERA grant ER 06-02-293.
D.C would like to thanks IAS for hospitality. The research of D.C has been supported in part by a CONICET-Fulbright fellowship and grant PICT 2010-0724.

\appendix

\section{Reflection Matrix}
\label{Rmatrix}
With the conventions we are using,  the (canonical) diagonal symmetry generators are
\begin{align}
&{L_{\rm D}}^{\!\!\check +}_{~\check +} =
{L}^{+}_{~+} - \tilde{{L}}^{\dot +}_{~\dot+}\,,\qquad
 {R_{\rm D}}^{\!\!\check a}_{~\check b} = {R}^{a}_{~b} + \tilde{{R}}^{\dot a}_{~\dot b}\,,\hspace{-1cm}
& {Q_{\rm D}}^{\!\!\check \pm}_{\check a} = {Q}^\pm_{~a} \mp  i \tilde{Q}^{\dot\mp}_{~\dot a} \,,\nn
\\
&{L_{\rm D}}^{\!\!\check \pm}_{~\check \mp} =
{L}^{\pm}_{~\mp} - \tilde{{L}}^{\dot\mp}_{~\dot\pm}\,,
& {S_{\rm D}}^{\!\!\check a}_{\check \pm} = {S}^a_{~\pm} \pm i  \tilde{S}^{\dot a}_{~\dot\mp} \,.
\label{Ls3wl}
\end{align}
The generators (\ref{Ls3wl}) give rise to the diagonal $\widetilde{su}(2|2)_{\rm D}$ residual symmetry. Now we should determine how a bulk magnon transforms under the diagonal $\widetilde{su}(2|2)_{\rm D}$. A bulk magnon transforms in a representation $(\boxslash_{(a,b,c,d)},\tilde{\boxslash}_{(a,b,c,d)})$ of the bulk symmetry $\widetilde{su}(2|2)_L\times{\widetilde{su}(2|2)}_R$.
The quantum numbers
\begin{equation}
a=\sqrt{g}\eta,\quad
b=\sqrt{g}\frac{i\zeta}{\eta}\left(\frac{x^{+}}{x^{-}}-1\right),\quad
c=-\sqrt{g}\frac{\eta}{\zeta x^{+}},\quad
d=-\sqrt{g}\frac{x^{+}}{i\eta}\left(\frac{x^{-}}{x^{+}}-1\right)\,,
\label{abcd}
\end{equation}
characterize the action of the fermionic generators. For the left fundamental $(\phi^1,\phi^2,\psi^+,\psi^-)$
\begin{align}
Q_{\;a}^{\alpha}|\phi^b\rangle & =  a\,\delta_{a}^{b}|\psi^\alpha\rangle ,
&& S_{\;\alpha}^{a}|\phi^{b}\rangle = c\,\epsilon_{\alpha\beta}\epsilon^{ab}|\psi^{\beta}\rangle ,\nonumber \\
Q_{\;a}^{\alpha}|\psi^{\beta}\rangle & =
 b\, \epsilon^{\alpha\beta}\epsilon_{ab}|{\phi}^{b}\rangle ,
&& S_{\;{\alpha}}^a|\psi^{\beta}\rangle = d\, \delta_{\alpha}^{\beta}|\phi^{a}\rangle ,
\end{align}
and similarly for the right generators $\tilde Q_{\;\dot a}^{\dot \alpha}$ and $\tilde S_{\;{\dot\alpha}}^{\dot a}$ and  on a right fundamental $(\tilde\phi^{\dot 1},\tilde\phi^{\dot 2},\tilde\psi^{\dot +},\tilde\psi^{\dot -})$. $\zeta$ is a phase and unitarity requires $|\eta|^2= i\left(x^{-}-x^{+}\right)$.

The left part of a bulk magnon, $\boxslash_{(a,b,c,d)}$ is also $\boxslash_{(a,b,c,d)}^{{\rm D}}$, with identical quantum numbers
However, the right $\tilde\boxslash_{(a, b, c, d)}$ needs to be re-arranged to transform canonically under the action of the diagonal symmetry generators. That can be achieved defining,
\be
(\tilde \phi^{\check 1},\tilde \phi^{\check 2},\tilde \psi^{\check +},\tilde \psi^{\check -})
:=
(\tilde \phi^{\dot 1},\tilde \phi^{\dot2}, - i\tilde \psi^{\dot -},+i \tilde \psi^{\dot +})\,,
\label{change2}
\ee
which turns out to be a  $\boxslash^{\rm D}_{(a,- b,- c,  d)}$. Due to the signs in $-b$ and $-c$  we should interpret the right part as a magnon with quasi-momentum $-p$ and phase $\zeta e^{ip}$ under
 ${su}(2|2)_{{\rm D}}$.
Therefore, with the change of basis (\ref{change2}), the original left and right parts of the bulk magnon transforms in the following tensor representation of ${su}(2|2)_{{\rm D}}$
\begin{equation}
\boxslash_{(a,b,c,d)}\otimes \boxslash_{(a,-b,-c,d)} = \mathcal{V}\left(p,\zeta\right) \otimes \mathcal{V}\left(-p,\zeta e^{ip}\right)\,.
\label{diagrep}
\end{equation}

Diagonal $\widetilde{su}(2|2)_{\rm D}$ is preserved during the reflection, which fixes the boundary scattering matrix
up to a phase factor. We should take into account that the multiplet labels change with the reflection according to,
\be
\mathcal{V}\left(p,\zeta\right) \otimes \mathcal{V}\left(-p,\zeta e^{ip}\right)
\rightarrow
\mathcal{V}\left(-p,\zeta\right) \otimes \mathcal{V}\left(p,\zeta e^{-ip}\right)\,.
\label{change}
\ee
The reflection matrix $\mathcal{R}_{\rm R}(p,-p)$ intertwines the same representations as a bulk $S$-matrix $\mathcal{S}(p,-p)$, and therefore the two must be equal up to a phase \cite{CRY}. The resulting reflection matrix is given by
\ba
\Refl_{\rm R} \;|\phi_p^{\check a} \times \tilde\phi_{-p}^{\check b} \rangle \!&=& \!
A_{\rm R}(p) |\phi_{-p}^{\{\check a} \times \tilde \phi_{p}^{\check b\}} \rangle +B_{\rm R}(p) |\phi_{-p}^{ [\check a} \times \tilde\phi_{p}^{\check b ]}\rangle
+ \alf C_{\rm R}(p) \eps^{\check a \check b} \eps_{\calpha\cbeta} |\psi_{-p}^\calpha \times\tilde \psi_{p}^\cbeta \rangle\,, \nn
\\
\Refl_{\rm R} \;|\psi_p^\calpha \times \tilde\psi_{-p}^\cbeta \rangle \!&=& \!
D_{\rm R} (p)|\psi_{-p}^{\{\calpha}\times \tilde\psi_{p}^{\cbeta\}}\rangle
+ E_{\rm R}(p) |\psi_{-p}^{ [\calpha} \times \tilde\psi_{p}^{\cbeta ]}\rangle
+ \alf F_{\rm R}(p) \eps_{\check a\check b} \eps^{\calpha\cbeta}|\phi_{-p}^{\check a} \times\tilde\phi_{p}^{\check b}\rangle\,,
\nn\\
\Refl_{\rm R} \;|\phi_p^{\check a} \times \tilde\psi_{-p}^\cbeta\rangle \!&=& \!
G_{\rm R}(p) |\psi_{-p}^\cbeta\times\tilde \phi_{p}^{\check a}\rangle + H_{\rm R}(p) |\phi_{-p}^{\check a} \times\tilde \psi_{p}^\cbeta\rangle\,,
\nn\\
\Refl_{\rm R} \;|\psi_p^\calpha \times\tilde \phi_{-p}^{\check b}\rangle \!&=& \!
K_{\rm R}(p) |\psi_{-p}^\calpha \times \tilde\phi_{p}^{\check b}\rangle +  L_{\rm R}(p) |\phi_{-p}^{\check b} \times\tilde \psi_{p}^\calpha\rangle \,.
\label{Rche}
\ea
where
\begin{align}
& A_{\rm R} = R_0(p) \frac{x^-}{x^+}{\frac{\eta_1\eta_2}{\etap_1\etap_2}}\,,
&& D_{\rm R} = - R_0(p)\,,\nn
 \\
& B_{\rm R} = -R_0(p) \frac{x^-(x^-+(x^+)^3)}{(x^+)^2(1+x^-x^+)}{\frac{\eta_1\eta_2}{\etap_1\etap_2}}\,,
&& E_{\rm R} = R_0(p)  \frac{x^++(x^-)^3}{x^-(1+x^-x^+)}\,,\nn
\\
& C_{\rm R} =  -R_0(p) \frac{i{\eta_1\eta_2}(x^-+x^+)}{\zeta x^+(1+x^-x^+)}\,,
&& F_{\rm R} = -R_0(p) \frac{i\zeta(x^-+x^+)(x^- -x^+)^2}{{\etap_1\etap_2} x^+(1+x^-x^+)}\,,\nn
\\
& G_{\rm R} =  R_0(p)  \frac{x^- + x^+}{2x^+} {\frac{\eta_1}{\etap_1}}\,,
&& H_{\rm R} = R_0(p)  \frac{x^- - x^+}{2x^+} {\frac{\eta_1}{\etap_2}}\,,\nn
\\
& K_{\rm R} =  R_0(p)  \frac{x^- - x^+}{2x^+} {\frac{\eta_2}{\etap_1}}\,,
&& L_{\rm R}  = R_0(p)  \frac{x^- + x^+}{2x^+} {\frac{\eta_2}{\etap_2}}\,.
\label{Rchecoe}
\end{align}
$\tilde\eta_i$ are the values of $\eta_i$ after the reflection. The choice for $\eta$ in the so-called
string theory basis is
\be
\eta(p,\zeta) = \zeta^{\frac12} e^{\frac{i p}{4}}\sqrt{ix^--ix^+}\,,
\label{etastring}
\ee

\section{Solution to the boundary crossing equation}
\label{crossingsolution}

In this appendix we solve the crossing equation for the boundary dressing phase
\be\la{crun}
\sigma_B(p) \sigma_B(\bar p) = \frac{x^-+\frac{1}{x^-}}{x^++\frac{1}{x^+}}\,,
\qquad \sigma_B(p) \sigma_B(-p) = 1\,.
\ee
We follow a procedure similar to the one described in \cite{Volin,VieiraVolin}.
First we introduce the Zhukovski variable $u$,
\be
x(u)+\frac1{x(u)} = \frac{u}{g}\,,\qquad x^{\pm} := x(u\pm \tfrac{i}{2})\,,
\label{Zhuko}
\ee
so that the crossing equation can be written as
\be\la{crosu}
\sigma_B(u)\sigma_B^\gamma(u) = \frac{u-\tfrac{i}{2}}{u+\tfrac{i}{2}}\,.
\ee
The index $\gamma$ means the analytical continuation along a closed contour $\gamma$ that crosses
both $x^+(u)$ and $x^-(u)$ cuts. We are also going to assume that
\be \la{ratioG}
\sigma_B(x^+(u),x^-(u)) = \frac{G(x^+)}{G(x^{-})}\,.
\ee
In terms of the shift operator $D:=e^{\tfrac{i}{2}\partial_u}$, the crossing equation reads
\be
\left[G(x(u))G(1/x(u))\right]^{D-D^{-1}} = u^{D^{-1}-D}\,.
\ee
Therefore, our crossing equation is of the form
\be
G(x(u))G(1/x(u)) = u^{F(D)}\,.
\label{crossf}
\ee
Naively one would say that $F(D)=-1$, but that would associate the cuts of $x(u)$ to
$G(x(u))G(1/x(u))$. Instead, we will use
\be
F_k(D)= \frac{D^k}{1-D^k} + \frac{D^{-k}}{1-D^{-k}} = \sum_{n=1}^\infty D^{kn} + \sum_{n=1}^\infty D^{-kn}\,,
\label{fk}
\ee
for some integer $k$. Different values of $k$ would lead to different expressions
for $u^{F_k(D)}$. A posteriori we will analyze what values of $k$ are consistent with the crossing condition.
With this $F_k(D)$, we get
\be
u^{F_k(D)}
=\exp\left(\sum_{n=1}^\infty\log(u^2+\tfrac{k^2n^2}{4})\right)\,.
\label{ufdiv}
\ee
As in \cite{VieiraVolin}, we regulate the divergent sum in (\ref{ufdiv}), by taking the derivative of the exponential's argument, performing the sum and integrating back. As a result, up to an irrelevant integration constant, we obtain
\be
u^{F_k(D)} =  \frac{\sinh(\tfrac{2\pi}{k} u)}{\tfrac{2\pi}{k} u}
= \frac{1}{\Gamma(1+\frac{2 u i}{k})\Gamma(1-\frac{2 u i}{k})}\,.
\label{ufk}
\ee
We should now check consistency with original crossing equation, so we compute:
\be
\left( \frac{\sinh(\tfrac{2\pi}{k} u)}{\tfrac{2\pi}{k} u}\right)^{D-D^{-1}}
= \frac{\sinh(\tfrac{2\pi}{k} (u+\tfrac{i}{2}))}{\sinh(\tfrac{2\pi}{k} (u-\tfrac{i}{2}))}\frac{u-\tfrac{i}{2}}{u+\tfrac{i}{2}}\,.
\ee
Consistency with the crossing condition (\ref{crosu}) requires
\be
\frac{\sinh(\tfrac{2\pi}{k} (u+\tfrac{i}{2}))}{\sinh(\tfrac{2\pi}{k} (u-\tfrac{i}{2}))}= 1\,,
\ee
and we then choose the value $k=1$.

We still need to solve for the function $G(x)$. Let us define $G(x):=e^{i\chi(x)}$. The crossing condition (\ref{crossf}) imposes
\be
\chi(x(u+i0))+\chi(x(u-i0)) = \frac{1}{i}\log\left(\frac{\sinh(2\pi u)}{2\pi u}\right)\,.
\label{boundRH}
\ee
The kernel introduced in the bulk case can also be used to solve our problem. Indeed, the kernel
\be
K\star f := \int\limits_{-2g+i0}^{2g+i0} \frac{dw}{2\pi i}\frac{x(u)-\frac{1}{x(u)}}{x(w)-\frac{1}{x(w)}}
\frac{1}{w-u} f(w)\,,
\ee
satisfies
\be
(K\star f)(u+i0) + (K\star f)(u-i0) =f(u)\,,
\ee
if $|u|<2g$. Thus, equation (\ref{boundRH}) is solved by
\be
\chi(x(u)) = -i K\star\log\left(\frac{\sinh(2\pi u)}{2\pi u}\right)\,.
\ee
Up to a term that cancels when we compute the difference $\chi(x^+)-\chi(x^-)$, we can write
$\chi(x)$ as a contour integral,
\be\la{lohsigma}
\chi(x) = \Phi(x) =-i \oint\limits_{|z|=1} \frac{dz}{2\pi i} \frac{1}{x-z} \log\left(\frac{\sinh[2\pi g(z+\tfrac{1}{z})]}{2\pi g(z+\tfrac{1}{z})}\right)\,.
\ee
This `DHM' representation \cite{DHM} of the solution for the crossing equation is valid for $|x|>1$. As $x$ moves towards the interior
of the unit circle, the function defined by the contour integral is discontinuous, as it picks up a residue at $x$.
Since we want the analytic continuation to the interior of the disk to be continuous, for $|x|<1$, we have instead
\be
\chi(x) = \Phi(x) - i \log\left(\frac{\sinh[2\pi g(x+\tfrac{1}{x})]}{2\pi g(x+\tfrac{1}{x})}\right)\,.
\label{chi<}
\ee
This analogous to the analysis of \cite{AF0904.4575} for the bulk dressing phase.
As $x$ moves inside the unit disk, some of the branch cuts of the logarithmic term
in (\ref{chi<}) can be crossed. There are infinitely many of those branch cuts, corresponding to the zeros of $\sinh(2\pi g(x+\tfrac{1}{x}))$. However, crossing such branch cuts could only produce $2\pi$ terms in $\chi$ which are, in any case, irrelevant to the boundary dressing factor $\sigma_B(x^+,x^-) = e^{i\chi(x^+)-i\chi(x^-)}$.

~

Being careful about the crossing contour one can check that the final $\sigma_B$ obtained after
this procedure indeed solve the crossing equation \nref{crun}. The unitarity condition is a consequence
of the fact that $\chi(-x) = \chi(x)$ plus the the fact that
 $\sigma_B$ is a  ratio of a function of $x^+$ and the same function of $x^-$  \nref{ratioG}.

We can expand the contour integral $\Phi(x)$ in negative powers of $x$ for large $|x|$
values,
\be
\la{expansion}
\Phi(x) = - \sum_{r=1}^\infty \frac{c_r(g)}{x^r}\,.
\ee
The coefficients $c_r(g)$ can be expanded either in the weak or in the strong coupling limit.
In the weak coupling limit we obtain
\be
c_r(g) = \sum_{n=1}^\infty\frac{i (-4)^n g^{2 n} \zeta (2 n) \Gamma
   \left(-\frac{r}{2}-n\right)}{\Gamma (1-2 n) \Gamma
   \left(-\frac{r}{2}+n+1\right)}\,.  \qquad
   \label{expaPhi}
\ee
Notice that for odd $r$ the coefficient $c_r(g)$ is vanishing, while for even $r$ is order $g^r$.

In the strong coupling limit we find
\be
c_r(g) = 2 g i \int\limits_{0}^{2\pi} dt e^{itr} |\cos t| +{\cal O}(1) = \frac{4 g i^{r+1} (1+(-1)^r)}{1-r^2} +{\cal O}(1)\,.
\ee

To evaluate the contour integral for $|x|<1$ we can use the identity
\be
\la{idePhi}
\Phi(x) + \Phi(1/x) = \Phi(0)\,,\qquad {\rm for\ |x|\neq 1}\,,
\ee
and the expansion
\be
\Phi(0) =  -i \sum_{n=1}^\infty g^{2 n} (-16)^n {(2n-1)!!\over n (2n)!!}\zeta (2 n)\,.
\label{expaPhi0}
\ee

\section{Luscher correction at strong coupling}
\la{luschapp}

In this appendix we consider an open string operator of the form $B_l Z^L B_r(\theta,\phi)$
and compute the leading correction to the energy for large $L$, this correction goes as
$e^{ - ({\rm constant}) L }$.
We will compute the correction for $\phi =0$, $\theta \not =0$ at leading order in the
strong coupling expansion.
In this case we have a  string that moves on
$AdS_2 \times S^3 $. It is convenient to decouple the $AdS_2$ and $S^3$ problems by
choosing a worldsheet gauge where $- T_{\pm \pm }^{AdS} = T^{S}_{\pm \pm } = 1$.
 We fix the solution on the $AdS_2$ part. This $AdS_2$
   solution is completely characterized by the  extent of the
   spatial worldsheet coordinate $\sigma$, which we take to run between $[ - s/2,s/2]$. The other
   worldsheet coordinate is $\tau$.
   In particular the spacetime energy $\Delta$ of the solution is fixed, once  $s$ is fixed.
   As we vary the parameters we will see that $L$ will change, $\theta$ will change, and so will $\Delta - L$.

   So we now  concentrate on the solution on the $S^3$, which we parametrize as
   \be
   x_1 + i x_2 = e^{i \gamma \tau } \sqrt{ 1 - \rho^2(\sigma) } ~,~~~~~~~~~~~~~ x_3 + i x_4 =
   \rho(\sigma ) e^{ i \varphi(\sigma) }
 \ee
Inserting this in the Euler Lagrange equations for the string and imposing the Virasoro
 constraints, $T_{\pm \pm } =1$ one finds two integrals of motion, $\ell$ and $\gamma$.
 They are given by
 \be
\ell =   \rho^2  \varphi'
\ee
and
\be  \la{firstin}
 {  \rho^2 (\rho')^2 \over 1 - \rho^2  } =  - \ell^2 - ( \gamma^2 -1) \rho^2 + \gamma^2 \rho^4
 \ee
The boundary conditions are $\rho'(0) =0$,  $\rho(s/2) =1$.
Let us define $\rho_0$ to be the value of $\rho$ at $\sigma =0$ where the derivative vanishes.
It is a   root of
 \be \la{rhozero}
  0 =  - \ell^2 - ( \gamma^2 -1) \rho_0^2 + \gamma^2 \rho_0^4
  \ee
  By using \nref{firstin} we can write the following expressions
\ba
{ s \over 2 } & =& \int_{\rho_0}^1 d\rho { \rho \over \sqrt{1 - \rho^2 } \sqrt{D} }
\\
\la{thetaint}
{ \theta \over 2 } & = & \int_{\rho_0}^1 d\rho { \ell \over \rho   \sqrt{1 - \rho^2 } \sqrt{D} }
\\
{ L \over 2  } &=& 2g
\int_0^{s/2}  d\sigma  \gamma  |x_1 + i x_2|^2 =   2g   \gamma  \int_{\rho_0}^1 d\rho {  \rho
 \sqrt{1 - \rho^2 } \over \sqrt{D} }
 \\
 D &=&  - \ell^2 - ( \gamma^2 -1) \rho^2 + \gamma^2 \rho^4 = ( \rho^2 -\rho_0^2 ) [ \gamma^2(\rho^2 + \rho_0^2 ) - (\gamma^2 -1) ]
\ea
From the first two equations we should find $\rho_0$ and $\gamma$ as a function of $s$ and
$\theta$, and then we can find the expression for $L$ and for the energy.
We want to find a solution where $L$ is very large.

This happens when $\rho_0 \to 0$ and $\gamma \to 1$ and $\ell \to 0$.
More precisely, we need to scale them as
\be \la{scaledvar}
\gamma = 1 + \epsilon/2 ~,~~~~~~~~~~ \ell = \epsilon   { \hat \ell \over 2 }
 ~,~~~~~~ \rho = \sqrt{\epsilon} v
\ee
where $v$ is a new rescaled variable and $\hat \ell$ is fixed as $\epsilon \to 0$.
Now, to leading  order in $\epsilon$ we find that \nref{rhozero} becomes
\be
\la{vzero}
0=- { \hat \ell^2 \over 4 } - v^2_0 + v^4_0 ~,~~~~~{\rm or} ~~~~~~~~~ v_0^2 = { 1 + \sqrt{ 1 + \hat \ell^2 } \over 2 }
\ee
The integral  for  $\theta$, \nref{thetaint},
 becomes negligibly small away from $\rho \sim \rho_0$ since there
a factor of $\ell$ multiplying. So it receives all its contribution from the small
$\rho$ region, namely the finite $v$ region, see \nref{scaledvar}. We can write
\ba \la{thetall}
{ \theta \over 2 } & = & { \hat \ell \over 2 } \int_{v_0}^{\infty} { 1 \over v \sqrt{\tilde D  } } ~,~~~~~~ \Longrightarrow ~~~~~~~~~~ \hat \ell = \tan \theta
\\
\tilde D &=&  ( v^2 -v_0^2) ( v^2 + v_0^2 -1) \la{defdtilde}
\ea
We can similarly compute the integral for   $s$,  
\ba
{ s \over 2 } &=& \int_{\rho_0}^1 d\rho { \rho \over \sqrt{ 1 - \rho^2 } }\left( { 1 \over \sqrt{ D}}  - { 1 \over \rho^2 }\right)  + \int_{\rho_0}^1 d\rho {1 \over \rho \sqrt{ 1 - \rho^2 } }
\\
&=& \int_{v_0}^{\infty}dv\ v \left( { 1\over  \sqrt{ \tilde D } }- {1 \over v^2 } \right) + \log 2 - \log \rho_0
\\
{ s \over 2 } &=& \log 4 - { 1 \over 2} \log ( { \epsilon \over \cos \theta }  )  ~,~~~~~~ \Longrightarrow
{ \epsilon \over \cos \theta} = 16 e^{ - s }
\ea
where we used $\rho_0 = \sqrt{\epsilon} v_0$ and the result \nref{thetall}, and the definition of
$\tilde D$ in \nref{defdtilde}. Here we have split the integral in two terms, the first
receives contributions only form the small $\rho$ region and the second, which can be done
explicitly with no need to take the small $\rho_0$ limit (though we quoted here only the small
$\rho_0$ answer).

We now want to compute $L$. We will compute instead
\ba
{ L \over 4 g  } - {s \over 2 }  &= & { \epsilon \over 2 } \int_{\rho_0}^1 d \rho \rho  \sqrt{ 1 - \rho^2 } \left(
{ 1  \over
\sqrt{D} } - { 1 \over \rho^2 } \right)  -
\\
& & ~~~~ \int_{\rho_0}^1  d \rho{  \rho^3 \over \sqrt{1 -\rho^2} }
\left( { 1 \over \sqrt{D}} - { 1 \over \rho^2 } + \epsilon { (1 - \rho^2 ) \over 2  \rho^4 }
\right) - \int_{\rho_0}^1 d \rho { \rho \over \sqrt{ 1 - \rho^2 } }
\\
{ L \over 4 g } - {s \over 2 }  &= & { \epsilon \over 2 } \int_{v_0}^ \infty d v  v    \left(
{ 1  \over
\sqrt{\tilde D} } - { 1 \over v^2 } \right)  -
\\
& & ~~~~  \epsilon \int_{v_0}^\infty   d v v^3
\left( { 1 \over \sqrt{\tilde D}} - { 1 \over v^2 } +{ 1 \over 2 v^4 }
\right) - 1 + { \rho_0^2 \over 2 }
\\
{ L \over 4 g } - {s \over 2 } &=& \epsilon ( { 1 \over 4 } - { v_0^2 \over 2 } ) - 1 + \epsilon { v_0^2 \over 2 } = -1 + { \epsilon \over 4 } = -1 +  \cos \theta 4 e^{ -s}
\\
L - 2 g s &=&  - 4 g  +16  g \cos \theta  e^{ - { L \over 2 g }  - 2 }
\ea
Here we have split the integrals having in mind that we want an accuracy of order $\epsilon$.
The first has an $\epsilon$ in front and we made sure that only the small $\rho$ region
contributes.
In the second we made sure that only the small $\rho$ region contributes up to order $\epsilon$.
The last can can be evaluated exactly and we quoted here the small $\rho_0$ result.

Here we have in mind keeping $s$ fixed as we change $\theta$. Under these circumstances $\Delta$ stays
fixed, since the $AdS_2$ part of the solution would always be the same.
In addition, we know  that for $\theta =0$ the result should vanish due to the BPS condition.
Thus we find that
\be \la{alfinres}
\Delta - L  = g   ( 1 - \cos \theta ) { 16 \over e^2}  e^{ - { L \over 2 g } }
\ee

If we changed the angle in the $AdS$ part, then instead of  1 in \nref{alfinres} we would
get some function of $\phi$. However, since we know that for $\theta =\phi$ we should get
zero due to the BPS condition, we conclude that for generic angles we get
 \be \la{finres}
\Delta - L  = g   ( \cos \phi  - \cos \theta ) { 16 \over e^2}  e^{ - { L \over 2 g } }
\ee

\subsection{Strong coupling expansion of the function $F$ }
\la{funfexp}

In order to compare this to the expected answer from the Luscher type correction we need to
evaluate the function $F$ in \nref{ffun} at strong coupling.
This involves evaluating the function $\Phi$ in \nref{phasefa},  \nref{phasefacon}
at $z^{[\pm a]}$ at $q=0$. When $q=0$ we have that
\be
- 1/z^{[-a]}(0) = z^{[+a]}(0) =  i \left( \sqrt{ 1 + a^2/(16 g^2 ) } -a/(4 g ) \right) = i \left( 1 + { a \over 4 g } + \cdots \right)
\ee
which is very close to $i$, where the  strong coupling expansion is tricky, since we cannot
 use \nref{chistr}. We need
to compute
\ba \la{phaseap}
\log F &=& i \Phi(y) - i \Phi(1/y) = { 2 \over \pi } \int\limits_0^{\pi\over 2 }dt  { ( {y}^4 -1 )
\over (1 + {y}^2)^2 - 4 {y}^2 \sin^2 t } \log \left[ { \sinh 4 \pi g \sin t \over 4 \pi g \sin t } \right]
\ea
 with $y = z^{[a]}(0)$. Then the $y$ dependent factor can be well approximated by
 \be
 \left.  { ( {y}^4 -1 )
\over (1 + {y}^2)^2 - 4 {y}^2 \sin^2 t } \right|_{y = x^{[a]}(0)} \sim
{ a \over 4 g } \left[ { 1 \over \sin^2 t + { a^2 \over 16 g^2 } } \right]
\ee
We now insert this into the integral \nref{phaseap}, and split the integral into
two pieces
\ba
\log F &=& r_1 + r_2
\\
r_1  & = & { a \over 2 \pi g } \int\limits_{0}^{\pi\over 2 }dt \left[ { 1 \over \sin^2 t + { a^2 \over 16 g^2 } } \right] 4 \pi g \sin t =   2 a \log\left[  { 8 g \over a } \right] + o (1/g)
\\
r_2 &=& { a \over 2 \pi g } \int\limits_{0}^{\pi\over 2 }dt \left[ { 1 \over \sin^2 t + { a^2 \over 16 g^2 } } \right]  \log\left[ { 1 - e^{ - 8 \pi g \sin t } \over 8 \pi g \sin t } \right] =
\\
r_2 &=&  \int\limits_0^\infty dt { 4 a \over v^2 + 4 a^2 \pi^2 }
  \log\left[ { 1 - e^{-v}  \over v } \right] = 2 \left[ a \log a - a - \log \Gamma(a+1) \right]
\ea
where we have defined $t=v/(8 \pi g )$ in the integral for $r_2$ and taken the $g\to \infty$ limit. Summarizing, we get that the leading strong coupling approximation is
\be \la{ffunres}
 F(a,g) =  { 2^{ 6 a }  g^{ 2 a } \over e^{ 2 a } (a!)^2 }
 \ee

\section{Evaluating $t(q)$}
\la{ComputationOft}

Let us now evaluate $t(q)$, given in \nref{luschn}. In order to perform the trace over the matrix
indices it is convenient to write the reflection matrix in terms of the bulk $S$ matrix.
The reason is that we will be able to use the bulk crossing equation to simplify
the form of $t(q)$.
We start by writing
\be
K^{A \dot A , B\dot B }(q) =
\sigma_b(z^+,z^- ) {  1 + { 1 \over (z^+)^2}  \over 1 + { 1 \over (z^-)^2}  }   \sqrt{{
 z^+ + 1/z^+
\over   z^- + 1/z^- } } S^{A E }_{ C D } ( -p,p) T^{\dot A}_{E} T^{\dot B}_{F}
 \mathcal{C}^{C B} \mathcal{C}^{D F}\,,
 \ee
 where $S$ is the full bulk $S$ matrix for one of the $\widetilde{su}(2|2)$ factors, which
 obeys the crossing equation, with the identity in the right hand side.
  Here $p$ is the value of the momentum analytically continued
 so that we have $x^\pm$ in the mirror region. The square root arises because of a conventional
 way to define the bulk $S$ matrix. It is cancelling a square root in a the phase factor of
 the bulk $S$ matrix. The matrix $T$ converts the dotted indices into undotted indices. It arises in the precise
 implementation of the ``unfolding'' trick, where we replace
 a bulk magnon with momentum $p$
 that transforms under $\widetilde{su}(2|2)^2$ into two magnons of $\widetilde{su}(2|2)_D$,
 one with momentum $p$ and the other with momentum $-p$, see appendix \ref{Rmatrix}.
 We then see that $\bar K$ in \nref{barkdef}    is essentially the same as $K$, but
 evaluated at $ - \bar p  $. More explicitly, we can write
 \be
 \bar K_{A \dot A , B\dot B }(q) = \sigma_b\left(- { 1\over z^-} ,- { 1 \over z^+}  \right)
  {  1 + ( z^- )^2 \over 1 + (z^+ )^2 }   \sqrt{{
 z^- + 1/z^-
\over   z^+ + 1/z^+ } } S^{G H }_{ B M } ( \bar p,- \bar p) T_{\dot A}^{N} T_{\dot B}^{M}
 \mathcal{C}_{G A} \mathcal{C}_{H N}\,.
 \ee

\begin{figure}[h]
\centering
\def\svgwidth{16cm}
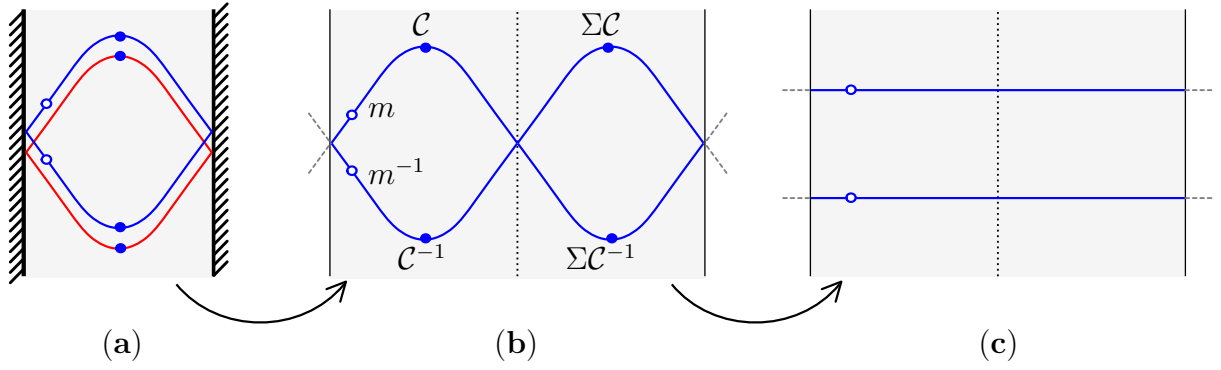
\caption{(a) Original picture. White circles represent rotation matrices, one is $m$ and the
    other is $m^{-1}$. Solid circles represent charge conjugations.
    (b) Unfolded picture. In the unfolding of the dotted indices there is a change of basis that produces the $\Sigma$'s.
    (c) Untangled picture . After using crossing we get two independent traces of the matrix $m$. }
\label{tuntangling}
\end{figure}

 We can now insert this into the expression for $t(q)$.
 We will need to use that
 \be \la{smacon}
   S^{A E }_{ C D } ( -p,p) \Sigma^{N}_{E}
 \mathcal{C}^{C B} \mathcal{C}^{D F} S^{G H }_{ B M } ( \bar p,- \bar p)  \Sigma_{F}^{M}
 \mathcal{C}_{G A} \mathcal{C}_{H N} = {\rm Tr}[\Sigma]^2\,,
\ee
here $\Sigma^N_E =  T^N_{\dot A} T_E^{\dot A }  $, where
\be
\Sigma = {\rm diag}(1,1,-1,-1)  =  (-1)^F \,.
\ee
This arises because the action of the charge conjugation
changes in the  basis given by (\ref{change2}). The charge conjugation matrix
 can be taken to be
\be
\mathcal{C}_{AB} =
\left(
\begin{array}{cc}
-i\epsilon_{ab} & 0
\\
0 & \epsilon_{\alpha\beta}
\end{array}
\right)\,.
\ee
Equation \nref{smacon} arises from the repeated use of the crossing equation.
Ignoring charge conjugation matrices the
identity we need is
\be \la{crossre}
 {\cal S}( -p,p) {\cal S}( \bar p , - \bar p) = { \cal S }(-p,p) {\cal S}(-\bar p , p)
 {\cal S}( p , - \bar p ) {\cal S }( \bar p , - \bar p ) = {\bf 1 }
 \ee
where ${\cal S}$ denotes the full $S$ matrix. When \nref{smacon} is used, we get \nref{texpr} in the main text.
The series of operations we have done are most clearly summarized by the
figure \ref{tuntangling}. First we do the unfolding trick. Then the use of the crossing relation \nref{crossre}
amounts to moving the lines, as in figure \ref{Unfolding}, and untangling them. If we introduce the rotation matrix $m$, we can do all the same steps but we have insertions of the matrix $m$ or $m^{-1}$ along some lines. This is represented in
figure \ref{tuntangling} by the solid circles. Once we untangle the lines as in figure \ref{Unfolding}, we get
get an insertion of $m$ on one line and an insertion of $m^{-1}$ on the other, leading to \nref{texprmone}
(since the trace of $m$ or $m^{-1}$ are the same).

This gave us $t(q)$ for the fundamental mirror magnon. For the bound states we can use the fusion procedure.
Given the form of $\sigma_B$ in \nref{phasefa}, which involves a ratio of a function of $x^+$ and $x^-$, then it is
clear that the fusion procedure gives a $\sigma_B$ which is the same ratio, but evaluated at $z^{[+a]}$ and $z^{[-a]}$.
The corresponding matrices are determined also by multiplying the matrices of the elementary constituents.
All the manipulations we used above can be used again for these matrices.
In particular, we can untangle the lines as in \ref{tuntangling}. The only new thing we need to understand is the set
of states of a magnon bound state and the action of the matrix $m$. The matrix $m$ is
\ba \la{mfora}
m_a \!\!&=&\!\! {\rm diag}(
\overbrace{\vphantom{e^{\frac{i}{2}p}}e^{i\theta-i (a-1)\phi},\cdots,e^{i\theta +i(a-1)\phi}}^{a},
\overbrace{e^{-i\theta- i (a-1)\phi}\vphantom{e^{\frac{i}{2}p}},\cdots,e^{-i\theta+i(a-1)\phi}}^{a},\nn\\
&&\hspace{1cm}
\overbrace{e^{-i a\phi},\cdots,e^{i a\phi}}^{a+1},
\overbrace{e^{-i (a-2)\phi},\cdots,e^{i(a-2)\phi}}^{a-1})
\ea
This can be understood as follows. The mirror magnon bound state arises from  $a$ $SL(2)$ sector fundamental magnons,
giving rise to an $SL(2)$ representation of spin $ 2j = a$, these lead to the components
with the  $a+1$ bracket. The other elements arise from acting with the supercharges in
$\widetilde{su}(2 |2)$. Then we see that the trace gives the result quoted in
\nref{tracealla}. Thus, the cancelation of the bulk $S$ matrices, plus the form for $\sigma_B$ for the
bound states, together with \nref{mfora} lead to $t_a(q)$ in \nref{texprall}.

\section{Derivation of the BTBA equations}
\la{TBAderivation}

In this appendix we derive the BTBA equations presented in the main text (\ref{1stTBA})-(\ref{asymptotic2}). We will do so in two different ways. In the first way, presented in section \ref{embedingderivation}, we will follow the original derivation of the TBA for the spectrum  \cite{GKV,GKKV} by embedding our open ABA equations into the closed ones. In the second way, presented in section \ref{directderivation}, we will take a more direct route and derive the BTBA from the thermodynamics in the mirror picture  \cite{Ghoshal:1993tm}.

\subsection{Derivation by embedding into the bulk system}\la{embedingderivation}

The spectrum of closed strings or
 single trace operators  in ${\cal N}=4$ SYM is governed
  by the so called Y-system \cite{GKV}. The Y-system is a set of functions $Y_{a,s}(u)$ characterizing the ratio of the density of bulk excitation to the density of holes in the mirror channel \cite{Zamolodchikov:1989cf}. The indices $(a,s)$ stands for rectangular representations of the bulk $SU(2,2|4)$ excitations and $u$ is the spectral parameter. The Y-function are subject to the general functional equation
\be\la{Yfunctional}
{Y_{a,s}^+Y_{a,s}^-\over Y_{a+1,s}Y_{a-1,s}}={(1+Y_{a,s+1})(1+Y_{a,s-1})\over(1+Y_{a+1,s})(1+Y_{a-1,s})}
\ee
where $f^\pm=f(u\pm i/2)$. For the bulk excitations, the Y's live in a fat hook bounded by $Y_{0,s}=\infty$, $Y_{2,|s|>2}=\infty$ and $Y_{a>2,\pm2}=0$, (see fig \ref{Ysystem}.a). The Y-system is also equivalent to the Hirota equation for the T-functions as
\be\la{Hirota}
T_{a,s}^+T_{a,s}^-=T_{a,s+1}T_{a,s-1}+T_{a+1,s}T_{a-1,s}\ ,\qquad\text{where}\qquad Y_{a,s}={T_{a,s+1}T_{a,s-1}\over T_{a+1,s}T_{a-1,s}}
\ee
and enjoy a gauge invariance under $T_{a,s}\to g^{[\pm a\pm s]}T_{a,s}$. For more details, see \cite{GKV,GKKV}. The TBA equations for the Y functions are the solution to the functional relation (\ref{Yfunctional}), subject to the relevant boundary conditions and analytic behavior. The procedure of deriving the TBA equations
 in this fashion was carried out in \cite{GKV,GKKV}.

Operators on Wilson loops  are associated to a string, or a spin chain, with
boundaries. Suppose we start with such operators and go to the mirror picture where space and time are interchanged (see figure \ref{TBAfigure}). In the mirror picture, one have exactly the {\bf same} system of mirror particles as in the closed case. In the limit of large $T$, see figure \ref{TBAfigure},
these particles live on a  large closed chain governed by the mirror asymptotic Bethe equations \cite{Arutyunov:2009ur}. The boundaries in the original physical picture are mapped to two boundary states in the mirror past and future. These boundary states are determined by  the boundary reflection matrix as we discussed in section \ref{sec2}. As opposed to the closed case where all mirror states are traced over in the partition function, in the overlap between boundary states only a
subset of mirror excitations are summed over. Moreover, the
 weights of these excitations in the summation lead to a new asymptotic behavior for their densities. We therefore expect the the Y-system and the TBA equations to
  be identical to the ones in the closed case modulo projections and new sources.

In this section we will exploit that relation to derive the boundary TBA equations. That is, we will first map the open $\widetilde{su}(2|2)_\text{D}$ ABA equations (\ref{BI})-(\ref{BIII}) into a folded version of the standard closed $\widetilde{su}(2|2)^2$ ones. That map is nothing but the embedding of the diagonal $\widetilde{su}(2|2)_\text{D}$ excitations in the full $\widetilde{su}(2|2)_\text{L}\times \widetilde{su}(2|2)_\text{R}$ by restricting to singlet excitations of the diagonal $\widetilde{su}(2|2)_\text{D}$ preserved by the boundary. Having done so, the corresponding BTBA equations will follow from the derivation of the closed TBA ones \cite{GKKV}. We will only add the angles that enter as (diagonal) twists and the boundary dressing phase that enters as a momentum dependent chemical potential
   for
   the momentum carrying excitations. As most of the details are the same as in the closed case, we will be brief.

As we discussed in section \ref{sec2}, and in appendix \ref{Rmatrix}, the diagonal $\widetilde{su}(2|2)_{{\rm D}}$ magnon excitations transforms in a tensor representation of the bulk $\widetilde{su}(2|2)_\text{L}\times \widetilde{su}(2|2)_\text{R}$. Correspondingly, the ABA equations can be embedded into the closed ones together with the addition of the reflection matrix. That is, (\ref{BI})-(\ref{BIII}) can also be written as in a redundant way as
\be\la{ABAlong}
\def\svgwidth{14cm}
\begingroup%
  \makeatletter%
  \providecommand\color[2][]{%
    \errmessage{(Inkscape) Color is used for the text in Inkscape, but the package 'color.sty' is not loaded}%
    \renewcommand\color[2][]{}%
  }%
  \providecommand\transparent[1]{%
    \errmessage{(Inkscape) Transparency is used (non-zero) for the text in Inkscape, but the package 'transparent.sty' is not loaded}%
    \renewcommand\transparent[1]{}%
  }%
  \providecommand\rotatebox[2]{#2}%
  \ifx\svgwidth\undefined%
    \setlength{\unitlength}{1440bp}%
    \ifx\svgscale\undefined%
      \relax%
    \else%
      \setlength{\unitlength}{\unitlength * \real{\svgscale}}%
    \fi%
  \else%
    \setlength{\unitlength}{\svgwidth}%
  \fi%
  \global\let\svgwidth\undefined%
  \global\let\svgscale\undefined%
  \makeatother%
  \begin{picture}(1,0.75)%
    \put(0,0){\includegraphics[width=\unitlength]{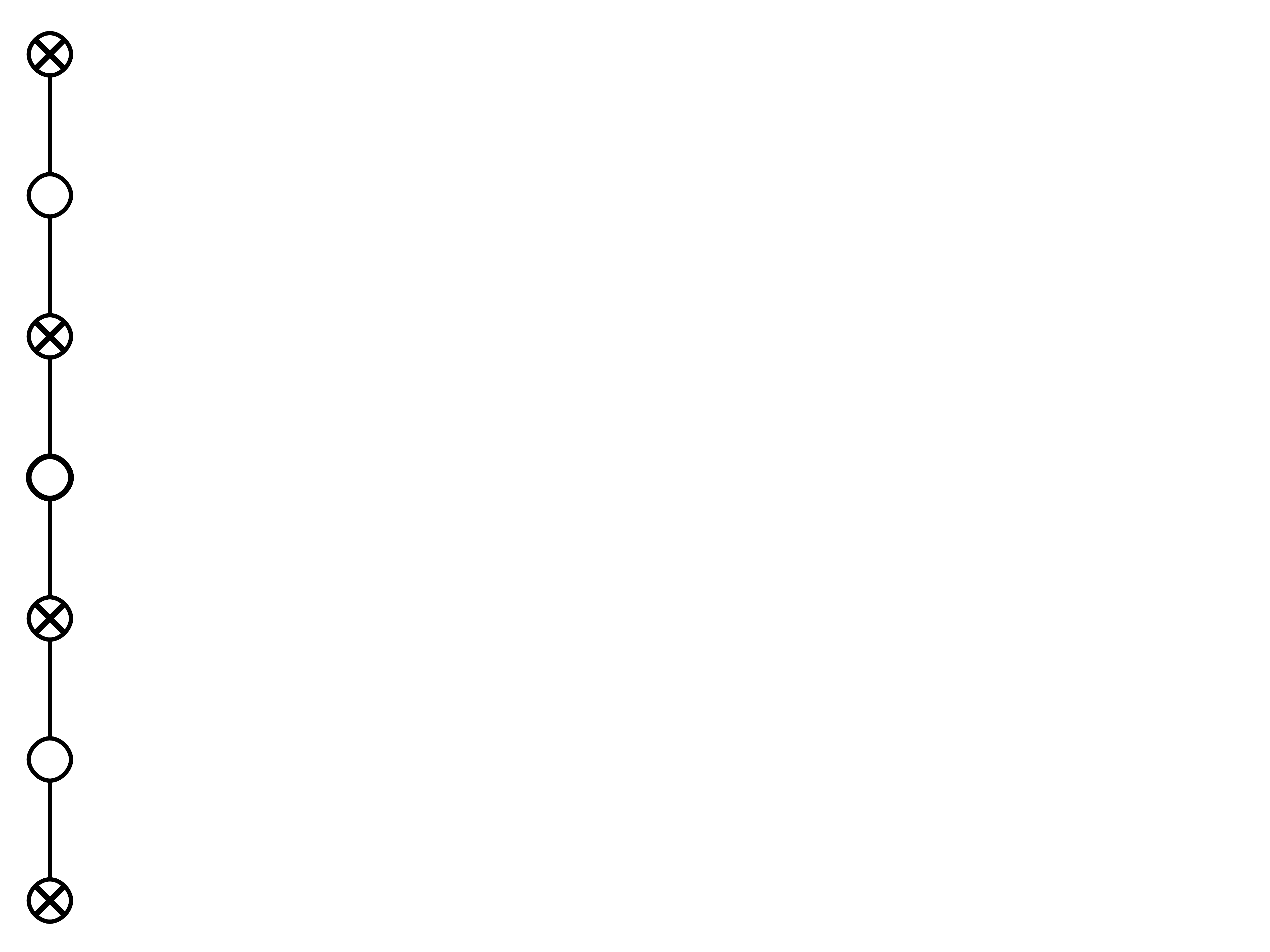}}%
    \put(0.1165813,0.5937963){\color[rgb]{0,0,0}\makebox(0,0)[lb]{\smash{$\displaystyle{-e^{2i\theta}=\left.\frac{Q_1^+Q_2^{--}Q_3^+}{Q_1^-Q_2^{++}Q_3^-}\right|_{u_{2,k}}}$}}}%
    \put(-0.00184462,0.69999998){\color[rgb]{0,0,0}\makebox(0,0)[lb]{\smash{$\scriptstyle 1$}}}%
    \put(0.11324797,0.70490741){\color[rgb]{0,0,0}\makebox(0,0)[lb]{\smash{$\displaystyle{e^{i\phi-i\theta} =\left. \frac{\mathbb B^{(-)}Q_2^+}{\mathbb B^{(+)}Q_2^-}
\right|_{u_{1,k}}}$}}}%
    \put(0.11435908,0.48268519){\color[rgb]{0,0,0}\makebox(0,0)[lb]{\smash{$\displaystyle{e^{i\phi-i\theta}=\left.\frac{\mathbb R^{(-)}Q_2^+}{\mathbb R^{(+)}Q_2^-}\right|_{u_{3,k}}}$}}}%
    \put(0.09435908,0.37157408){\color[rgb]{0,0,0}\makebox(0,0)[lb]{\smash{$\displaystyle{R_0^{-2}(u)=\left.\[\left(\frac{x^+}{x^-}\right)^LS(u)^2 \frac{R_3^-B_1^-}{R_3^+B_1^+}\]
\times\[\left(\frac{x^+}{x^-}\right)^L\bar S(u)^2\frac{\bar R_3^-\bar B_1^-}{\bar R_3^+\bar B_1^+}\]\right|_{u_{4,k}}}$}}}%
    \put(0.09769242,0.26046297){\color[rgb]{0,0,0}\makebox(0,0)[lb]{\smash{$\displaystyle{e^{-i\phi+i\theta}=\left.\frac{\mathbb R^{(-)}\bar Q_2^+}{\mathbb R^{(+)}\bar Q_2^-}\right|_{-u_{3,k}}}$}}}%
    \put(0.09769242,0.14935186){\color[rgb]{0,0,0}\makebox(0,0)[lb]{\smash{$\displaystyle{-e^{-2i\theta}=\left.\frac{\bar Q_1^+\bar Q_2^{--}\bar Q_3^+}{\bar Q_1^-\bar Q_2^{++}\bar Q_3^-}\right|_{-u_{2,k}}}$}}}%
    \put(0.09769242,0.03824075){\color[rgb]{0,0,0}\makebox(0,0)[lb]{\smash{$\displaystyle{e^{-i\phi+i\theta}=\left. \frac{\mathbb B^{(-)}\bar Q_2^+}{\mathbb B^{(+)}\bar Q_2^-}\right|_{-u_{1,k}}}$}}}%
    \put(-0.00184462,0.03333332){\color[rgb]{0,0,0}\makebox(0,0)[lb]{\smash{$\scriptstyle 7$}}}%
    \put(-0.00184462,0.14444443){\color[rgb]{0,0,0}\makebox(0,0)[lb]{\smash{$\scriptstyle 6$}}}%
    \put(-0.00184462,0.25555554){\color[rgb]{0,0,0}\makebox(0,0)[lb]{\smash{$\scriptstyle 5$}}}%
    \put(-0.00184462,0.47777776){\color[rgb]{0,0,0}\makebox(0,0)[lb]{\smash{$\scriptstyle 3$}}}%
    \put(-0.00184462,0.58888887){\color[rgb]{0,0,0}\makebox(0,0)[lb]{\smash{$\scriptstyle 2$}}}%
    \put(-0.00184462,0.36666665){\color[rgb]{0,0,0}\makebox(0,0)[lb]{\smash{$\scriptstyle 4$}}}%
  \end{picture}%
\endgroup%

\ee
where
\ba
R_{l}(u)&=&\prod_{j=1}^{K_l}\frac{x(u)-x_{l,j}}{(x_{l,j})^{1/2}}\,,
\qquad
B_{l}(u)=\prod_{j=1}^{K_l}\frac{\tfrac{1}{x(u)}-x_{l,j}}{(x_{l,j})^{1/2}}\,,
\qquad
Q_{l}(u)=\prod_{j=1}^{K_l}(u-u_{l,j})\nn\\
\bar R_{l}(u)&=&\prod_{j=1}^{K_l}\frac{x(u)+x_{l,j}}{(x_{l,j})^{1/2}}\,,
\qquad
\bar B_{l}(u) =\prod_{j=1}^{K_l}\frac{\tfrac{1}{x(u)}+x_{l,j}}{(x_{l,j})^{1/2}}\,,
\qquad
\bar Q_{l}(u) =\prod_{j=1}^{K_l}(u+u_{l,j})\nn
\\
\mathbb R^{(\pm)}(u)&=&\prod_{j=1}^{K_4}\frac{(x(u)-x^{\mp}_{j})(x(u)+x^{\pm}_{j})}{(x^{+}_{j}x^{-}_{j})^{1/2}}\,,
\qquad
\mathbb B^{(\pm)}(u)=\prod_{j=1}^{K_4}\frac{(\tfrac{1}{x(u)}-x^{\mp}_{j})(\tfrac{1}{x(u)}+x^{\pm}_{j})}{(x^{+}_{j}x^{-}_{j})^{1/2}}\nn\\
S(u)&=&\prod_{j\neq k}^{K_4}S_0(p(u),p_j)^2\,,\qquad\qquad\qquad\qquad \bar S(u)=\prod_{j\neq k}^{K_4}S_0(p(u),-p_j)^2
\ea
 where we used that $S_0(p_j,-p(u))=S_0(p(u),-p_j)$.

Note that equations for nodes 1-3 in (\ref{ABAlong}) are equivalent to the ones in nodes 5-7.\footnote{Remember that in physical kinematics, where (\ref{ABAlong}) is written, $|x^{[a]}|>1$ and therefore $x^\pm(-u)=-x^\mp(u)$, $x(-u)=-x(u)$, $p(-u)=-p(u)$ and $\epsilon(-u)=\epsilon(u)$.} However, here we will think about these as describing the left and right parts of the bulk magnon excitations correspondingly.\footnote{After rearrangement of the quantum numbers discussed in the appendix.} We see that   momentum carrying
 excitations come  in pairs with opposite momenta.
 The nested level excitations also comes in pairs of rapidities $u_5=-u_3$, $u_6=-u_2$ and $u_7=-u_1$. That is, $\bar Q_l(u)$ is nothing but $Q_{8-l}(u)$ with roots $u_{l,j}=-u_{8-l,j}$.

We will now use that embedding to derive from them the asymptotic values of the left and right T-functions. Of course, these are not independent and we only have one set of independent $\widetilde{su}(2|2)_\text{D}$ T-functions.

In the asymptotic $L\to\infty$ limit the right ($T_{a,s\ge0}$) and left ($T_{a,s\le0}$) T systems decouples\footnote{Throughout this section the word ``asymptotic'', means this large $L$ limit, which should not
be confused with the large $u$ limit.}.
The asymptotic values of $T_{1,1}$ and $T_{1,-1}$, whose analyticity would lead to equations in nodes 1-3 and 5-7 in (\ref{ABAlong}) correspondingly, are derived as in the periodic case \cite{GKV}. In what follows, we will use bold face ($\bY,\bT$) for the asymptotic values of Y's and T's. Up to a gauge transformation, the asymptotic $\bT_{1,1}^\text{R}$ and $\bT_{1,-1}^\text{L}$ are
\ba\la{T11s}
\bT_{1,1}^\text{R} \!\! &=&\!\! \frac{\mathbb R^{-(+)}}{\mathbb R^{-(-)}}\left(e^{-i\theta}\frac{Q_2^{--}Q_3^+}{Q_2Q_3^-} -e^{-i\phi}\frac{\mathbb R^{-(-)}Q_3^+}{\mathbb R^{-(+)}Q_3^-}
+e^{i\theta}\frac{Q_2^{++}Q_1^-}{Q_2 Q_1^+}-e^{i\phi}\frac{\mathbb B^{+(+)}Q_1^-}{\mathbb B^{+(-)}Q_1^+}
\right)
\label{t1-1angles}
\\
\bT_{1,-1}^\text{L} \!\! &=&\!\! \frac{\mathbb R^{-(+)}}{\mathbb R^{-(-)}}\left(e^{i\theta}\frac{\bar Q_2^{--}\bar Q_3^+}{\bar Q_2\bar Q_3^-} -e^{i\phi}\frac{\mathbb R^{-(-)}\bar Q_3^+}{\mathbb R^{-(+)}\bar Q_3^-}
+e^{-i\theta}\frac{\bar Q_2^{++}\bar Q_1^-}{\bar Q_2 \bar Q_1^+}-e^{-i\phi}\frac{\mathbb B^{+(+)}\bar Q_1^-}{\mathbb B^{+(-)}\bar Q_1^+}
\right)\label{t1m1angles}
\ea
The right 1-3 in (\ref{ABAlong}) ABA equations are obtained by demanding analyticity of $\bT_{1,1}^\text{R}$ when $u$ goes to $u_{1,k}-\tfrac{i}{2}$, $u_{2,k}$ and $u_{3,k}+\tfrac{i}{2}$. The equivalent right equations 5-7 in (\ref{ABAlong}) are obtained from the analyticity of $\bT_{1,-1}^\text{L}$ when $u$ goes to $-u_{1,k}-\tfrac{i}{2}$, $-u_{2,k}$ and $-u_{3,k}+\tfrac{i}{2}$. In the gauge where $\bT^\text{L/R}_{0,s}=\bT^\text{L/R}_{a,0}=1$, the other T functions of the right and left (decoupled)  $SU(2|2)$ wings are obtained from the generating functional \cite{Kazakov:2007fy,GKV}
\beaa\la{gfangles}
{\cal W}&=&\[1-e^{i\phi}\frac{\mathbb B^{+(+)}Q_1^-}{\mathbb B^{+(-)}Q_1^+}\frac{\mathbb R^{-(+)}}{\mathbb R^{-(-)}}D\]\[1-e^{i\theta}\frac{Q_2^{++}Q_1^-}{Q_2 Q_1^+}\frac{\mathbb R^{-(+)}}{\mathbb R^{-(-)}}D\]^{-1}\\&&\times\[1-e^{-i\theta}\frac{Q_2^{--}Q_3^+}{Q_2Q_3^-}\frac{\mathbb R^{-(+)}}{\mathbb R^{-(-)}}D\]^{-1}\[1-e^{-i\phi}\frac{Q_3^+}{Q_3^-}D\]\nn
\eeaa
as
\be\la{gentoT}
{\cal W}=\sum_{s=0}^\infty \bT_{1,s}^{[1-s]}D^s\ ,\qquad{\cal W}^{-1}=\sum_{a=0}^\infty(-1)^a\,\bT_{a,1}^{[1-a]}D^a\ ,\qquad D=e^{-i\d_u}
\ee
Similarly, for negative $s$ we reverse the sign of the angles and use $\bar Q$ instead of $Q$. Their analyticity leads to the Bethe equations for the bound states.
In the vacuum, the $\bT$'s are independent of $u$. Their constant values read
\ba\la{Ts}
\bT^\text{R}_{a,1} \!\!&=&\!\! \bT^\text{L}_{a,-1}=(-1)^{a}\, 2   (\cos\phi-\cos\theta)\frac{\sin a\, \phi}{\sin\phi}\,,
\\
\bT^\text{R}_{1,s} \!\!&=&\!\! \bT^\text{L}_{1,-s}= \quad\ \, -\ 2 (\cos\phi-\cos\theta)\frac{\sin s\, \theta}{\sin\theta}\,.
\ea
Using the Hirota equation and the definition of the Y-functions in the right and left decoupled wings (\ref{Hirota}) we read the corresponding values of the asymptotic Y's
\ba\la{asymptotic}
\bY_{1,1}\!\!&=&\!\!-\frac{\cos\theta}{\cos\phi}\ ,\qquad \bY_{1,s}=\frac{\sin[(s+1)\theta]\sin[(s-1)\theta]}{\sin^2\theta}
\\
\bY_{2,2}\!\!&=&\!\!-\frac{\cos\phi}{\cos\theta}\ ,\qquad\bY_{a,1}=\frac{\sin^2\phi}{\sin[(a+1)\phi]\sin[(a-1)\phi]} \nn
\ea
For the asymptotic $Y_{a,0}$ ($\bY_{a,0}$) we have
\be
\bY_{a,0}=\left(\frac{x^{[-a]}}{x^{[+a]}}\right)^{2L} \frac{\phi^{[-a]}}{\phi^{[+a]}} \bT_{a,-1}^L\bT_{a,1}^R
\ee
Here, $\left(\frac{x^{[-a]}}{x^{[+a]}}\right)^{2L} \frac{\phi^{[-a]}}{\phi^{[+a]}}$ is a zero mode of the discrete Laplace equation ${{\cal A}_a^+{\cal A}_a^-\over{\cal A}_{a+1}{\cal A}_{a-1}}=1$ \cite{GKV}. It comes about because $\bT^\text{R}$ and $\bT^\text{L}$ in (\ref{T11s})-(\ref{gentoT}) are written in different gauges. It is determined by demanding that $Y_{1,0}(u_{4,k})=-1$ gives the 4$^{\rm th}$ node Bethe equation. We find
\be
\frac{\phi^-}{\phi^{+}} =R_0^2(u)S^2(u)\bar S^2(u)
\frac{\mathbb R^{-(-)}\mathbb B^{+(+)}}{\mathbb R^{+(+)}\mathbb B^{-(-)}}
\frac{B_1^+\bar B_1^+B_3^-\bar B_3^-}{B_1^-\bar B_1^-B_3^+\bar B_3^+}
\ee
The boundary crossing equation (\ref{crossingequation}) is then obtained by demanding that $\bY_{1,1}$ is invariant under crossing (provided that the bulk dressing phase obeys the bulk crossing equation). In particular, in the vacuum we get
\be\la{Y0aasymptotics}
\bY_{a,0}=4{ e^{i\chi(z^{[+a]}) + i \chi(1/z^{[-a]})} \over e^{ i \chi(z^{[-a]}) + i \chi(1/z^{[+a]})} } \left(\frac{z^{[-a]}}{z^{[+a]}}\right)^{2L+2}\!\!\!\!\!\!\!\! (\cos\phi-\cos\theta)^2\,\frac{\sin^2 a\, \phi}{\sin^2\phi}\,.
\ee

Finally, note that $\bT^\text{L}_{a,-s}(u)$ can be obtained from $\bT^\text{R}_{a,s}(u)$ in two steps. First reflect the sign of $u$ by considering $\bT_{1,1}(-u)$. Second, flip the signs of all the shifts and the angles. In mirror kinematics, that second step that basically amount to a complex conjugation, is equivalent to a gauge transformation \cite{Gromov:2010kf}. We therefore find that in mirror kinematics $\bT^\text{mir}_{a,-s}(u)\simeq\bT^\text{mir}_{a,s}(-u)$ and as a result $\bY^\text{mir}_{a,-s}(u)=\bY^\text{mir}_{a,s}(u)$.\footnote{To see that $\bY^\text{mir}_{a,0}$ is a symmetric function note that in mirror kinematics, $x^{[a]}(-u)=-1/x^{[-a]}$ and that $\chi(-x)=\chi(x)$.} As the closed TBA equation respects that symmetry, we expect all solutions to the BTBA equations to respect it. That is, we have a folded version of the bulk Y-system where $Y_{a,-s}(u)$ is identified with $Y_{a,s}(-u)$ as presented in the main text (\ref{FLIP}). Such a folding is also expected a priory as the Wilson lines break the bulk $SU(2,2|4)$ symmetry (at $\theta=\phi=0$) down to $OSp(4^*|4)$. For the ground state, we expect all functions to be symmetric.

We conclude that the BTBA equations, once divided by the asymptotic ones,
 are the same as in the closed case. The only two differences are the modified asymptotic solutions (\ref{asymptotic}), (\ref{Y0aasymptotics}) and the folding $Y_{a,-s}(u)=Y_{a,s}(-u)$. These are the BTBA equations (\ref{1stTBA})-(\ref{asymptotic2}) presented in the main text.

\subsection{Direct derivation of the boundary TBA }
 \la{directderivation}

The derivation of boundary TBA equations from the boundary state \nref{BdyState} is fairly
standard in relativistic theories \cite{LeClair:1995uf,Ghoshal:1993tm}. Here we need to follow the same
steps.

\begin{figure}[h]
\centering
\def\svgwidth{16cm}
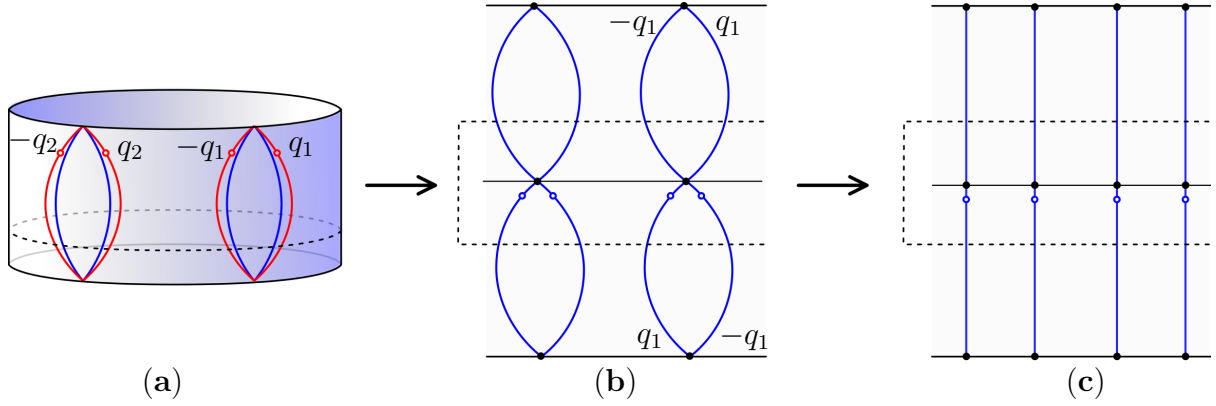
\caption{ (a)  Original computation. The boundary states create and annihilate pairs of particles.
 The blue and red lines represent the two $\widetilde{su}(2|2)$ representations of each bulk magnon.  The doted line represents the projection on to states that
obey the Bethe equations for a  chain that is closed along the horizontal direction.
The red dots are the rotation matrices $m$ in \nref{mone} and the black dots are the
 boundary dressing phase. (b) The same in the unfolded picture. We have a single $\widetilde{su}(2|2)$ group.  We continue to have insertions of the matrix $m$ and the dressing phase. The Bethe equations now involve some operation which also involves the bottom part. (c) The same in the untangled picture. Here for each line
 we take the trace over all four states of the magnon.}
\label{TBAUnfolding}
\end{figure}

After doing the flip between space and time, see figure \ref{TBAfigure},
 we have a past and
a future boundary characterized by the matrices $K$ and $\bar K$ which give the
probability amplitude for creating or annihilating a pair of particles \nref{BdyState}, \nref{BdyStatetwo}.
 This pair has opposite momenta
and it is in the singlet representation of an $SU(2|2)_D$, since this is a
symmetry preserved by the boundary state.
By independent creation events we can create a multiparticle state.
We can graphically represent the quantity we want to evaluate as in figure \ref{TBAUnfolding}(a).
Along the spatial direction we have a closed circle of length $T$, we need
to solve the Asymptotic Bethe equations on this circle and find the Bethe eigenstates.
Only the subset of momenta that corresponds
 to Bethe eigenstates can propagate. These states
propagate for Euclidean time $L$.
These asymptotic Bethe equations are the ones in the mirror theory and were
written in \cite{Arutyunov:2009zu}, following \cite{BeisertStaudacher}.
 These equations involve various roots, $u_1, \cdots u_7$,
  where
$u_4$ are the momentum carrying roots.
The only new feature is that we are considering states
which are composed of pairs of particles with opposite momenta. This imposes the condition
that the $u_4$ roots should appear in pairs. In other words a root $u_4$ should appear
together with a root $-u_4$. The boundary state is invariant under a diagonal $SU(2|2)_D$.
The condition that we only have singlets under
$SU(2|2)_D$ implies that if roots $u_a$, with $a=1,2,3$ appear, then
so should roots $u_{8-a} = - u_a$.  This will be more clearly seen below.

\begin{figure}[h]
\centering
\def\svgwidth{16cm}
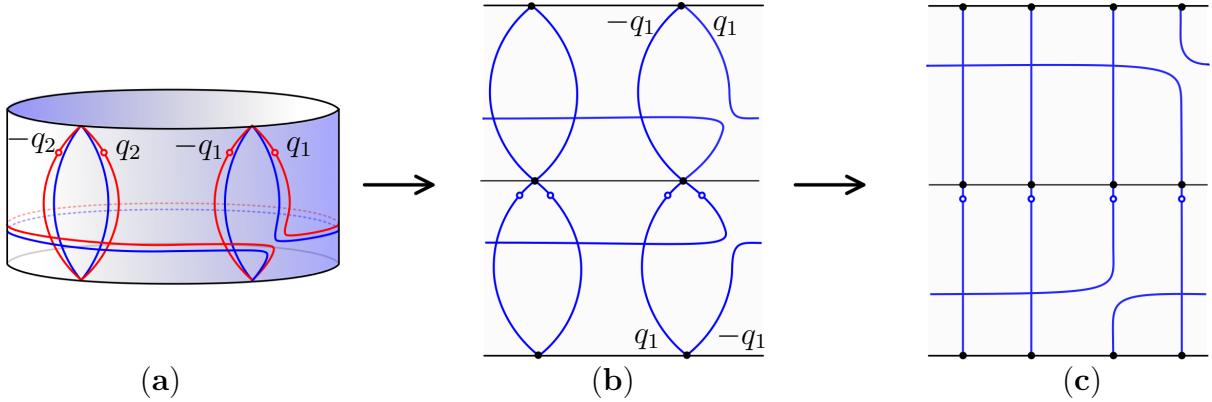
\caption{The Bethe equations in the various pictures come from demanding that each full particle
  can be taken around the chain producing a total phase of 1. (a) The Bethe equation in the original picture.
   (b) The
   Bethe equation in the unfolded picture comes from taking both a top particle with momentum $q$ and a
   bottom particle with momentum $-q$ around the chain.
   (c) The same but in the untangled picture.}
\label{TBABetheEqn}
\end{figure}

\begin{figure}[h]
\centering
\def\svgwidth{16cm}
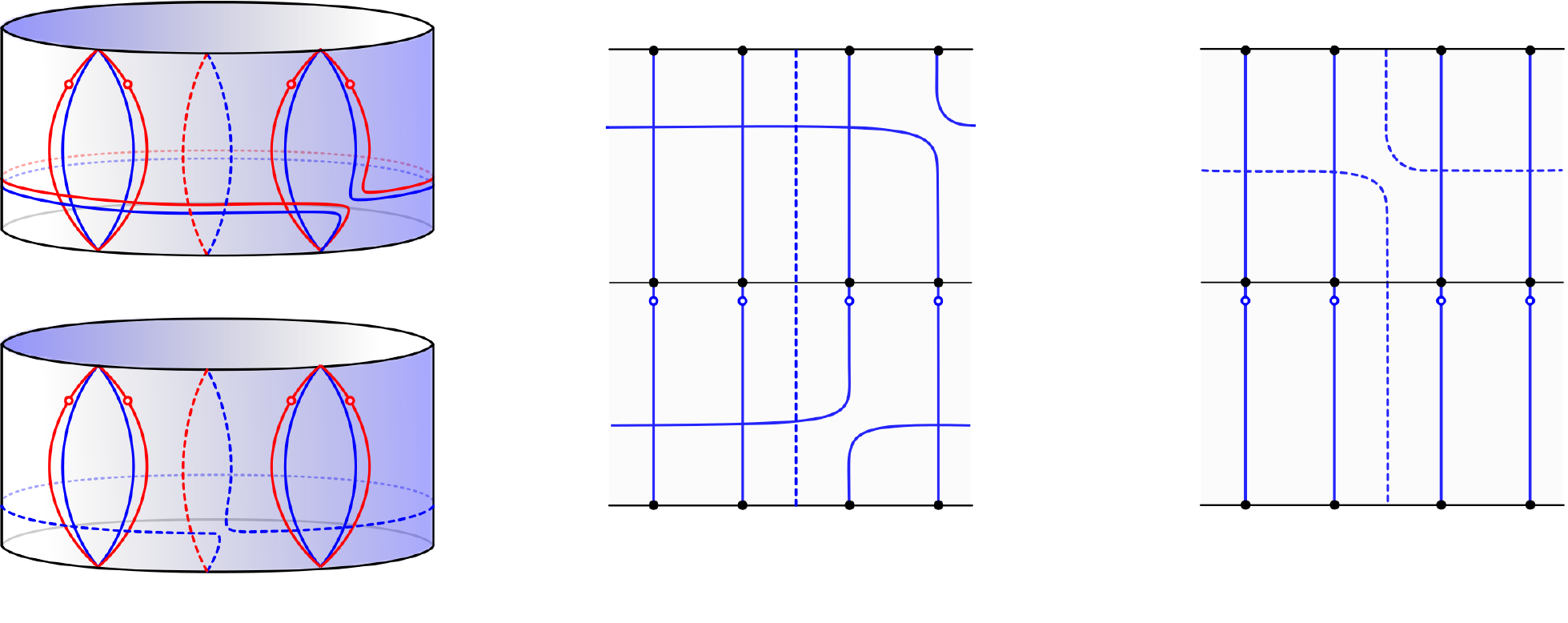
\caption{ (a) Bethe equations for the momentum carrying nodes in the original picture. The doted lines represent the various nested levels, with
red and blue for each of the $\widetilde{su}(2|2)$ factors.  Here $y$ represents the rapidity of levels $u_1$, $u_2$ or
$u_3$, and $-y$ is the rapidity of the $u_5$,$u_6$ or $u_7$ levels. (b) Bethe equation
for the nesting levels.  (c) Equations for the momentum
carrying nodes in the unfolded and untangled picture. Note that both the particle with momentum $q$ and
the one with momentum $-q$ cross the nesting doted line. (d) Bethe equation for the nested line.}
\label{Nesting}
\end{figure}

For this multiparticle state we can perform an unfolding and untangling
trick identical to the one
we did for a single particle in section \ref{sec2}, see   figure \ref{TBAUnfolding}(b),(c).
Now we have  particles for a single $\widetilde{su}(2|2)$ chain, but with insertions
associated to the rotation matrix $m$ as well as the dressing phase, $\sigma_B(q)$. We
can view these as ``chemical'' potentials in a thermodynamic computation. The
Bethe equations in this new picture look slightly more complicated because taking a single
full particle around the circle, as in figure \ref{TBABetheEqn}(a),  amounts to taking a
pair of particles around the circle, as in figure \ref{TBABetheEqn}(b),(c). We should
project onto states that obey these equations. These equations are for the full matrices.
We can do the nesting procedure and follow the unfolding for the various levels of the nesting.
The Bethe
equations for the various
 nesting levels can be graphically represented as in figure \ref{Nesting}.
The final result is that the equations are identical to the ones we would get in a situation where
we have the full mirror theory, but we impose that each root $u_a$ is accompanied by a root $-u_{8-a}$.

 The final Bethe equations in the mirror picture are

\ba \la{mBe1}
 1 &=& e^{ i p(u_4^j)  T }
 \prod_k  S_{44}(u^j_4, u_4^k) S_{44}(u_4^j,- u_4^k) \prod_{l} S_{43}(u_4^j,u_3^l)
 S_{43}(-u^j_4,u_3^l)
 \\ \la{mBe2}
 1 & = & \prod_{k} S_{34}(u_3^l,u_4^k)  S_{34}(u_3^l,-u_4^k) \prod_m S_{32}(u_3^l,u_2^m)
 \\ \la{mBe3}
 - 1 & =& \prod_{m} S_{32}(u_3^m,u_2^l)  \prod_n S_{33}(u_3^l,u_3^n)
\ea
where $S_{ab}$ are the $S$ matrices (really just phases)
between the impurities of the various kinds. We have removed the $u_1$ roots by turning them into $u_3$ roots, just to make the equations more compact. The $S_{ab}$
are the same as the one that appear in the ordinary mirror Bethe equations in \cite{Arutyunov:2009zu},
see also \cite{BeisertStaudacher}.
In this form,  these equations follow in a straightforward fashion from figure \ref{Nesting}(c)(d).
  Note, in particular, that the last factor in \nref{mBe1} can   also be
written as $S_{43}(-u^j_4,u_3^l)=S_{45}(u^j_4 , u_5^l)$ with $u_5^l = - u_3^l$. So that we can view the whole set
of equations, \nref{mBe1}-\nref{mBe3},
 as arising from the full chain, with both $\widetilde{su}(2|2)$ factors, but in a configuration with
the roots related by $u_{8-a} = - u_{a}$, as we discussed above.

We can simply add the rotation matrices to this picture. They act diagonally on the various impurities of
 the various levels. They appear  as
chemical potentials.
The potentials are as follows.
The momentum carrying modes, labeled by $u_4$,
 are in
the $SL(2)$ sector and have spacetime charges,
so we get a factor of $e^{ - 2 i \phi }$, one $e^{-i \phi}$ for
the particle of momentum $q$ and one for the one with momentum $-q$. The next root, $u_3$ or
$u_1$,
is the action of a fermionic generator that changes a fermion into a boson,
so that we need to add $e^{ i \theta + i \phi}$.
Then a bosonic generator with  $e^{ - 2 i \theta}$ is associated to the  root $u_2$. Taking into
account the number of roots of each type in a magnon bound state, we can get the chemical
potential coming from the angles for each magnon bound state \cite{Arutyunov:2009zu}.
 In addition,  the boundary dressing
phase also appears as a kind of chemical potential. For each pair of momenta $q$ and $-q$ we have
a factor of
\be
\sigma_B \bar \sigma_B \equiv \sigma_B(z^+,z^-) \sigma_B\left( -{ 1 \over z^-} , -{ 1 \over z^+} \right)
\ee

The derivation of the TBA equations for this system follows the same essential steps
as the derivation for the full TBA for the closed string. There are only minor differences due
 to the constraints on the position of the Bethe roots. For example, for the roots $u_4$ we only
 need to know the density in the region $u_4>0$, since there is always another associated root at $-u_4$.
 Thus, all integrals over $u_4$ will run over $u_4> 0$. We can define the $Y$ functions,
  as usual, in
  terms of densities of particles over densities of holes. We then run through the usual argument
  \cite{Yang:1968rm}, but
  keeping track of the range of integration, etc.
 We end up with the  boundary TBA equations 
  \ba
\label{1stTBAph}
\log{Y_{1, 1} }\!\!&=&\!\! i \theta + i \phi +  K_{m-1}*\log{1+{\baY_{1, m}}\over 1+Y_{m,1}}+\cR^{(01)}_{1\,a}* \log(1+Y_{a,0}) \\
\label{2ndTBAph}
\log{\baY_{2, 2}  }\!\!&=&\!\!\ i \theta + i \phi + K_{m-1}*\log{1+{\baY_{1, m}}\over 1+Y_{m,1}}+\cB^{(01)}_{1\,a}* \log(1+Y_{a,0})
\\
\label{3rdTBAph}
\log{\baY_{1, s} }\!\!&=&\!\!   2i (s-1) \theta - K_{s-1,t-1}*\log ( {1+\baY_{1, t} }) -K_{s-1}*\log{1+Y_{1,1}\over1+\baY_{2,2}}
\label{4thTBAph}
\\
\log{Y_{a,1} }\!\!&=&\!\!  i  2(a-1)\phi - K_{a-1,b-1}*\log ({1+Y_{b,1} })
-K_{a-1}*\log{1+Y_{1,1}\over 1+\baY_{2,2}} + \nn
\\
&&\qquad\qquad\qquad+\[\cR^{(01)}_{ab}+\cB^{(01)}_{a-2,b}\]*\log(1+Y_{b,0})
\\
\label{lastTBAph}
\log{Y_{a,0} }\!\!&=& \!\!  -  i 2 a  \phi + \log [ \sigma_B  \bar \sigma_B ] - 2 L E_m^a(u) +
\\\
&& +
  \[2{\cal S}_{a\,b}-{\cal R}_{a\,b}^{(11)}+{\cal B}_{a\,b}^{(11)}\]*\log(1+Y_{b,0})+2\[{\cal R}_{a\,b}^{(1\,0)}+{\cal B}_{a,b-2}^{(1\,0)}\]\s*\log(1+Y_{b,1} ) + \nn
\\
&&+2{\cal R}_{a\,1}^{(1\,0)}\s*\log({1+Y_{1, 1} }) -2{\cal B}_{a\,1}^{(1\,0)}\s*\log({1+\baY_{2,2} })
\ea
where the kernels are listed in appendix \ref{kernels} and are the same as in \cite{GKV,GKKV} but with $\baY_{a,s}^{(\text{here})}=1/Y_{a,s}^{(\text{there})}$.
The values of the angle dependent form of the chemical potential for each $Y_{a,s}$ function follows
from the particle content of each mirror bound state associated to each $Y_{a,s}$ function \cite{Arutyunov:2009zu}.
If we set to zero the $Y_{0,a}$ that appear in the convolutions, we get the  large $L$ solution
quoted in \nref{asymptotic1}, \nref{asymptotic2}. These equations look similar to the ones in \cite{Ahn:2011xq,deLeeuw:2012hp} for twisted boundary conditions, apart from the folding symmetry and the boundary dressing function $\sigma_B$.

Finally, the expression for the energy is the one quoted in \nref{exactE}.

\subsection{Kernels conventions}\la{kernels}

For the kernels we use the same definitions as in \cite{GKV,GKKV},
\ba
K_n(u,v) \!\!&=&\!\!  {2n/\pi\over n^2 + 4 (u-v)^2}\,, \quad K_{n,m}(u,v) =
\sum_{j=-{n-1\over 2}}^{n-1\over 2}\sum_{k=-{m-1\over 2}}^{m-1\over 2}
K_{2j+2k +2}(u,v)\,,
\\
\cR_{nm}^{(ab)}(u,v) \!\!&=&\!\! \sum_{j=-{n-1\over 2}}^{n-1\over 2}\sum_{k=-{m-1\over 2}}^{m-1\over 2}
{1\over2\pi i}{d\over dv} \log {r(u+ia/2 +i j, v-ib/2+ik)\over r(u-ia/2 +i j, v+ib/2+ik) }\,,
\\
\cB_{nm}^{(ab)}(u,v) \!\!&=&\!\! \sum_{j=-{n-1\over 2}}^{n-1\over 2}\sum_{k=-{m-1\over 2}}^{m-1\over 2}
{1\over2\pi i}{d\over dv} \log {b(u+ia/2 +i j, v-ib/2+ik)\over b(u-ia/2 +i j, v+ib/2+ik) }\,,
\\
{\cal S}_{nm}(u,v) \!\!&=&\!\! {1\over2\pi i}{d\over dv} \log \sigma\(x^{[\pm n]}(u),x^{[\pm m]}(v)\)\,,
\\
r(u,v)\!\!&=&\!\!{ x(u) - x(v) \over \sqrt{ x(v) } } ~,~~~~~~~~b(u,v)={ 1/x(u) - x(v) \over \sqrt{ x(v) } }
\ea
Fourier transformations of $K_n(u,0)$ and $K_{n,m}(u,0)$ are used at different stages,
\be
\widetilde{K}_n(w)=\sign(n)e^{-|nw|/2}\ ,\quad \widetilde{K}_{m,n}(w)=\coth{|w|\over2}\[e^{-{|w|\over2}|m-n|}-e^{-{|w|\over2}(m+n)}\]-\delta_{m,n}\,.
\ee

\section{Perturbative solution of the small $\phi$ TBA}\la{smallphisolution}

Before we start with the perturbative expansion of the small $\phi$ TBA system,
let us see in more detail how the simplified system \nref{ferdiftba}-\nref{defcnew} arises from \nref{wlanyc1}-\nref{defc}.
In order to simplify \nref{wlanyc1}, \nref{wlanyc2}   we have to convolute their
$K_{m-1}$ terms with  $\ms*\ms^{-1}$ and use the identity
\be
\ms^{-1}*K_{m-1} = K_{1,m-1}+\delta_{2,m} \,, \qquad {\rm for}\,\ \ms(u) = \frac{1}{2\cosh(\pi u)}\,.
\ee
We  then   use the TBA equation for ${\cal X}_2$ and the exact relations
\ba
&&\cR_{1\,a}^{(01)}(u,v)+\cB_{1\,a}^{(01)}(u,v)=K_{a}(u,v)\,,\\
&&\cR_{1\,a}^{(01)}(u,v)-\cB_{1\,a}^{(01)}(u,v)=\hat K_{y,a}(u,v)=K(u,v-i\tfrac{a}{2})-K(u,v+i\tfrac{a}{2})\,,
\label{kya}
\ea
where
\be
K(u,v)={1\over2\pi i}\sqrt{4g^2-u^2\over4g^2-v^2}{1\over v -u}\,.
\ee
Regarding the other TBA equations, we can simplify them by convoluting with
\be
(K_{l-1,m-1}+\delta(u)\delta_{l,m})^{-1} = \delta(u)\delta_{l,m}-\ms\; I_{l,m}\,,\qquad I_{l,m}=\delta_{l+1,m}+\delta_{l-1,m}\,.
\ee

The simplified TBA equations \nref{ferdiftba}-\nref{defcnew} are useful to solve for the Y-functions in the small $\phi$-limit perturbatively. We should begin by expanding the Y-functions in powers of $g^2$, \footnote{The different choice in the $\cY_{m}$ expansion is to have similar recurrent equations to those of $\cX_m$.},
\be
\begin{array}{rcl}
\Psi &=& \Psi^{(0)} + \Psi^{(1)} g^2 + \Psi^{(2)} g^4+\cdots\\
\Phi &=& \Phi^{(0)} + \Phi^{(1)} g^2 + \Phi^{(2)} g^4 +\cdots\\
\cY_{m} &=& \cY_{m}^{(0)}(1 + \cY_{m}^{(1)} g^2+ \cY_{m}^{(2)}g^4+\cdots)\\
\cX_m  &=&\cX_m^{(0)}+\cX_m^{(1)}g^2+\cX_m^{(2)}g^4+\cdots \\
\mC_a&=&  \mC_a^{(2)} g^2 + \mC_a^{(2)} g^4+ \mC_a^{(3)}g^6+\cdots
\end{array}
\label{allexpa}
\ee

The leading orders in the expansion \nref{allexpa} can be obtained by setting setting $\theta =0$ and $L=0$  and taking the small $\phi$ limit in the asymptotic solution \nref{asymptotic1}-\nref{asymptotic2}. Thus we have
\be
\Psi^{(0)} = \Phi^{(0)} = {1\over2}\,, \quad
\cY_{m}^{(0)}={1\over m^2-1}\,, \quad \cX_m^{(0)} = -{m^2\over3}\,,\quad
\mC_a^{(2)} = 4 (-1)^a\,.
\ee

We can obtain the higher order terms by solving the system as follows. By inspecting the equations \nref{yosimtba} and their perturbative expansions, one realizes that, to any order, the system  can be  solved in this schematic way:
$\Psi^{(k)} - \Phi^{(k)}  \rightarrow  {\cal Y}_m^{(k)} \rightarrow {\cal X}_m^{(k)}
\rightarrow \Psi^{(k)} + \Phi^{(k)} \rightarrow \mC_a^{(k)}$, provided the order $k-1$ functions are known.
In the final step of computing $\mC_a^{(k)}$  \nref{defcnew}
two kinds of contributions must be distinguished.
 On the one hand, there is $\Delta_{\rm conv}$ which is originated in from the convolutions with the Y-functions in \nref{deltaconvnew}.
On the other  hand we have contributions from the explicit functions in \nref{defc} which can be expanded to any order independently of the Y-system solution.
These give
\ba
\label{sourcescontri}
a^4 \[{z_0^{[-a]} \over z_0^{[+a]}}\]^2 F(a,g)^2  \!\!&=&\!\!
a^4 \({a-\sqrt{a^2+16g^2}\over a+\sqrt{a^2+16g^2}}\)^2 e^{i(\Phi(z_0^{[+a]})-\Phi(z_0^{[-a]})+\Phi(1/z_0^{[-a]})-\Phi(1/z_0^{[+a]}))}
\\
\!&=&\!\! 16g^4\! \[1\! +\(\pi^2-{6\over a^2}\){8g^2\over3} + \(7\pi^4-{150\pi^2\over a^2}+{630\over a^4}\){16g^4\over45}+{\cal O}(g^6)\]\nn
\ea
where we have used that $z_0^{[\pm a]}$ are given by \nref{zpm} at $q=0$. We  kept the first 3 loop orders only.

\subsection{Small $\phi$ solution at 2-loops }

Let us  now solve for the small $\phi$ Y-functions to the next to leading order in the coupling expansion.
We refer to this as the 2-loop order computation because this order gives rise to a correction ${\cal O}(g^4)$ to
the energy ${\cal E}$. Note however, that the Y-functions corresponding to this order are  ${\cal O}(g^2)$.

We start with  (\ref{ferdiftba}). Using the leading order of $\mC_a$ and the expression for $K_{y,a}$,
\be
\Phi-\Psi\approx
\sum_a{16(-1)^a g^2 \sqrt{4g^2-u^2}\over (a^2+4u^2)}\,.
\ee
Given that fermionic Y-functions are defined in the interval $(-2g,2g)$, this difference is essentially ${\cal O}(g^3)$.
Thus, it will be convenient to use the variable $\tilde u=u/(2g)$ that runs between $-1$ and $1$.  We then obtain
\be
\Phi(2g\,\tilde u)-\Psi(2g\,\tilde u)  =
32g^3\sqrt{1-\tilde{u}^2}\sum_{a=1}^\infty \frac{(-1)^a}{a^2} +{\cal O}(g^5)=
 -\frac{8\pi^2g^3}{3}\sqrt{1-\tilde{u}^2} +{\cal O}(g^5)\,.
 \label{ferdifleading}
\ee
This indicates that fermionic convolutions do not contribute to $\cX_n^{(1)}$ or $\cY_n^{(1)}$
Then, we have
\be
\cY_{m}^{(1)} - \ms * I_{m,n}\cY_{n}^{(1)}{(n^2-1)\over n^2} = 0\,.
\la{homog}
\ee
The same recurrence equations, thought with different rhs's, will repeatedly appear in the weak coupling expansion of the small angle limit TBA equations. In \ref{resolvent} we present the resolvent of the corresponding recurrence operator (demanding $\cY_{1}^{(1)} = 0$ and that $\cY_{m}^{(1)}$  remains bounded as $m\to \infty$). In this particular case, we solve (\ref{homog}) with $\cY_{m}^{(1)}=0$.

The recurrence equation that $\cX_m^{(1)}$ satisfies is non-homogeneous. It has a non-vanishing rhs
because of the term $\pi {\mathpzc s}\ \mC_m$ in \nref{xosimtba},
\be
\cX_m^{(1)} - \ms * I_{m,n}\cX_n^{(1)}{(n^2-1)\over n^2} = 4\pi (-1)^m \ms\,.
\label{recudelta0}
\ee
In Fourier space this becomes
\be
2\cosh{w\over2}\widetilde\cX_m^{(1)} - {m(m-2)\over(m-1)^2}\widetilde\cX_{m-1}^{(1)}
-{m(m+2)\over(m+1)^2}\widetilde\cX_{m+1}^{(1)} = 4\pi(-1)^{m}\, ,
\label{recudelta}
\ee
which can be solved using the resolvent of \ref{resolvent}. Here we write down only the component we need
\be
\widetilde\cX_2^{(1)}=-{8\over3}\pi\[2\cosh{w\over2}\log\(1+e^{-{|w|\over2}}\)-1-e^{-{|w|\over2}}\]\;.
\ee

Now, from  \nref{feraditba}, we have
\be
\Phi^{(1)} + \Psi^{(1)} = -{3\over2} \ms * \cX_2^{(1)} + 8\pi(-1)^n \ms * K_{n-1} - 4 \pi(-1)^a K_a\,,
\ee
where we have used
\be
\cR^{(01)}_{2b}(u,0) = K_{b-1}(u)+{8g^2\over b^2}K_{1}(u)+O(g^4)\,.
\ee
To compute the convolutions we go to Fourier space and get
\be
 \widetilde{\Phi}^{(1)} +  \widetilde{\Psi}^{(1)} = 4\pi\log(1+e^{-|w|/2})\,,
\ee
and then
\be
\Phi^{(1)} + \Psi^{(1)}  = {1\over u^2}\(1- {2\pi u\over \sinh(2\pi u)}\)\,.
\ee
To compute convolution with this fermionic Y-functions ($\Phi$ and $\Psi$)
 we shall use the aforementioned $\tilde u$, at the expense of introducing a $g$-dependence which has to be expanded as
\be
\Phi^{(1)}(2g\tilde u) + \Psi^{(1)}(2g\tilde u) =
{2\pi^2 \over 3}- {56 \pi^4 \over 45} g^2 {\tilde u}^2 + {\cal O}(g^4)\,.
\ee
For the next to leading order it is enough to keep only the constant term. Note, however, that the second term will also contribute to the next to next to leading order.

We now compute
\be
\Delta_\text{conv}={\cal R}_{a\,1}^{(1\,0)}\hat{*}\log2\Psi-{\cal B}_{a\,1}^{(1\,0)}\hat{*}\log2\Phi
+ \[{\cal R}_{a\,b}^{(1\,0)}+{\cal B}_{a,b-2}^{(1\,0)}\]*\log{1+\cY_{b}\over1+{1\over b^2-1}}\,.
\ee
and we find to the leading order
\be \la{delconvtwo}
\Delta_\text{conv} = {2\pi^2 g^2\over 3} +{\cal O}(g^4)\,.
\ee
We are now in the position to evaluate $\mC_a$ at the 2-loop order.
Inserting \nref{delconvtwo} and \nref{sourcescontri} into \nref{defcnew} we find
\be
\mC_a = 4(-1)^a g^2 + 8(-1)^a \[\pi^2-{4\over a^2}\]g^4  +{\cal O}(g^6)\,,
\ee

\subsection{Small $\phi$ solution at 3-loops}

Convolutions with fermionic Y-functions  ($\Phi$ and $\Psi$)
 start contributing to the next to next to leading order.
We have already obtained the difference of fermionic Y-functions in (\ref{ferdifleading}).
For the convolution needed in (\ref{yosimtba}) we get
\be
\ms\hat{*}\log{\Psi\over\Phi} \simeq
 {16\pi^2g^3\over 3} \ms(u-2 g\tilde{u})\hat{*}\sqrt{1-\tilde{u}^2}
\simeq
 {16\pi^3g^4\over 3} \ms(u)\,.
\ee
Thus, from (\ref{yosimtba}) using $\cY_m \simeq \frac{1}{m^2-1}(1+g^4{\cY}_m^{(2)}+\dots)$ we obtain
\be
\cY_m^{(2)} - \ms * I_{m,n} \cY_n^{(2)}{(n^2-1)\over n^2} = -\frac{16\pi^3}{3}\delta_{m,2} \ms\,.
\ee
This recursive equation in Fourier space is
\be
2\cosh{\tfrac{w}{2}} \widetilde{\cY}_m^{(2)} -\frac{m(m-2)}{(m-1)^2} \widetilde{\cY}_{m-1}^{(2)}
-\frac{m(m+2)}{(m+1)^2} \widetilde{\cY}_{m+1}^{(2)} = -\frac{16\pi^3}{3}\delta_{m,2}\,,
\label{y0equ}
\ee
for which we find the solution
\be
\widetilde{\cY}_m^{(2)} = -\frac{16 \pi ^3 }{3}{m\over m^2-1}e^{-\frac{1}{2}m|w|}\(\cosh{w\over2}+m\sinh{|w|\over2}\)\,,
\label{y0sol}
\ee
valid for $m\geq 2$, otherwise $\widetilde{\cY}_m^{(2)}$ vanishes.

Next, we consider the equation for $\cX_m$. Recall that we are expanding it as
$\cX_m = -{m^2\over3}+g^2 \cX_m^{(1)} +g^4\cX_m^{(2)} +\dots$. Here,
$\cX_m^{(1)}$ is the solution to \nref{recudelta0}, for which we have only quoted
$\cX_2^{(1)}$. This however does not contribute to the equation for $\cX_m^{(2)}$.
For the latter we get the usual recurrence equation. In the rhs there are contributions
from the solution ${\cY}_m^{(2)}$, $\mC_m$ and the fermionic convolution.
In Fourier space the recurrence equations are
\ba
2\cosh\tfrac{w}{2}\widetilde{\cX}_m^{(2)} -
\frac{m(m-2)}{(m-1)^2}\widetilde{\cX}_{m-1}^{(2)}-\frac{m(m+2)}{(m+1)^2}\widetilde{\cX}_{m+1}^{(2)}
\!\!&=&\!\! {2\over 3} \cosh{\tfrac{w}{2}} \widetilde{{\cY}}_m^{(2)} \\
&& +(-1)^m 8\[\pi^3-{4\pi\over m^2}\] - {8\over 9}\pi^3  \delta_{m,2}
\nn
\ea
where $\widetilde{{\cY}}_m^{(2)}$ is the the solution (\ref{y0sol}). This equation was also solved using the resolvent presented
in \ref{resolvent}. We write only $\widetilde{\cX}_2^{(2)}$,
\ba
\!\!\!\!\!\widetilde{\cX}_2^{(2)} \!\!\! &=&\!\!
-\frac{128\pi}{3}   \[\text{Li}_2\left(-e^{-|w|/2}\right)
   \sinh \tfrac{|w|}{2}
   +\text{Li}_3\left(-e^{-|w|/2}\right) \cosh \tfrac{w}{2}+1\]
 \\ \nn
   &&\!\!-\frac{8 \pi ^3}{27}
    \left(5 e^{-3 |w|/2}-9 e^{-|w|/2}\right)-\frac{16\pi ^3}{3}
    \[2 \cosh \tfrac{w}{2}\log \left(e^{-|w|/2}+1\right)
   -1\]\,.
\ea
which is the only component that enters in the fermionic TBA equations. Then, from the fermionic Y-functions,
we get
\ba
\Phi^{(2)} + \Psi^{(2)} \!\!&=&\!\! 2\pi\ms*\left[8\pi^2(-1)^n K_{n-1}+{32(-1)^n\over n^2}
(K_1-K_{n-1})\right]
\nn\\&& -8\pi(-1)^a\left[\pi^2 -{4\over a^2}\right]K_a
- {3\over 2} \ms*{\cX}_2^{(2)} -{1\over2} \ms*\cY_{2}^{(2)}
\ea
We compute the convolutions in Fourier space,
\be
 \widetilde\Phi^{(2)} + \widetilde\Psi^{(2)}  =
 \frac{4}{3} \pi ^3 e^{-|w|}+8\pi^3\log(1+e^{-|w|/2}) + 32 \pi \text{Li}_3(-e^{-|w|/2})\,,
\ee
and then
\be
\Phi^{(2)} + \Psi^{(2)} = {4\pi^2\over 3(1+u^2)} +{2\pi^2\over u^2}\(1- {2\pi u\over \sinh(2\pi u)}\) +
{2\over u^4}\(1- {2\pi u\over \sinh(2\pi u)}-{2\pi^2 u^2\over 3}\)\,.
\ee
Evaluating for $u=2 g \tilde u$ we obtain for the leading weak coupling limit
\be
\Phi^{(2)}(2 g \tilde u) + \Psi^{(2)}(2 g \tilde u) = {4\pi^2 \over 3} + {32\pi^4 \over 45}+
{\cal O}(g^2)
\ee

Now we have to compute up to the ${\cal O}(g^4)$ convolutions in   (\ref{deltaconvnew}).
If we use\footnote{Note the difference with $\hat K_{y,a}$ defined in (\ref{kya})}
\ba
&&\cR_{a\,1}^{(10)}(u,v)+\cB_{a\,1}^{(10)}(u,v)=K_{a}(u,v)\\
&&\cR_{a\,1}^{(10)}(u,v)-\cB_{a\,1}^{(10)}(u,v)=\hat K_{a,y}(u,v)=K(u+i\tfrac{a}{2},v)-K(u-i\tfrac{a}{2},v)
\ea
we can re-write the fermionic convolutions in the TBA equation (\ref{deltaconvnew}) as
\be
2{\cal R}_{a\,1}^{(1\,0)}\hat *\log\Psi-2{\cal B}_{a\,1}^{(1\,0)}\hat *\log\Phi
=K_a\hat *\log{\Psi\over \Phi} + K_{a,y}\hat *\log{\Psi\Phi}
\label{cachofe}
\ee
It is important to recall that we need only the $u\to 0$ limit of this.
For the order $g^4$ of the first term in (\ref{cachofe}) we get
\be
{16 \pi^3 g^4 \over 3} K_a(0) = {32 \pi^2 g^4\over 3 a}
\ee
The second term in (\ref{cachofe}) is
\ba
K_{a,y}\hat *\log{\Psi \Phi} \!\!&=&\!\!
\int\limits_{-2g}^{2g}\! dv\ K_{a,y}(0,v)\left[\log(-4) + 2g^2\left(\Psi^{(1)}(v) + \Phi^{(1)}(v) \right)
\right.\\ \nn
&&\left.
-\left(\Psi^{(1)}(v) + \Phi^{(1)}(v) \right)^2g^4
+2 \left(\Psi^{(2)}(v) + \Phi^{(2)}(v) \right) g^4 +\cdots\right]\,,
\ea
where its $g^4$ order is
\be
- {56\pi^4 g^4\over 45} \int\limits_{-1}^{1}\! d\tilde v {{2 \tilde v}^2\over \pi \sqrt{1-\tilde v^2}}
- {9\pi^4 g^4\over 4} +2 \left({4\pi^2 \over 3} + {32\pi^4 \over 45}\right) g^4
 =  -{12\pi^4 g^4\over 45} + {8\pi^2 g^4\over 3}\,.
\ee
Thus, the total $g^4$ order of \nref{cachofe} is
\be
{32\pi^2 g^4\over 3 a} + {8\pi^2 g^4\over 3}  -{12\pi^4 g^4\over 45} \,.
\ee

For the remaining convolution in (\ref{deltaconvnew}), $2\[{\cal R}_{a\,b}^{(1\,0)} +{\cal B}_{a,b-2}^{(1\,0)}\]*\log(1+\cY_{b})$
we use
\be
{\cal R}_{a\,b}^{(1\,0)}(0,v) +{\cal B}_{a,b-2}^{(1\,0)} (0,v) = K_{b-1}(v)
+\sum_{j=-{a-1\over2}}^{a-3\over 2} K_{b+2j} +{\cal O}(g^2)\,.
\ee
We go to Fourier space, where the term order $g^4$ is
\be
2\[\widetilde{K}_{b-1} + \sum_{j=-{a-1\over2}}^{a-3\over 2} \widetilde{K}_{b+2j}\] {\widetilde{\cY}_b^{(2)}\over b^2}
= -{8\pi^3\over 3} e^{-|w|} - {16\pi^3\over 3}{a-1\over a} e^{-{a|w|\over 2}}\,.
\ee
We Fourier transform back and evaluate for $u\to 0$ and we get
\be
\left.2\[{K}_{b-1} + \sum_{j=-{a-1\over2}}^{a-3\over 2} {K}_{b+2j}\]* {{{\cY}}_b^{(2)}\over b^2}\right|_{u=0}
= \frac{32 \pi ^2}{3 a^2}-\frac{32 \pi ^2}{3 a}-\frac{8 \pi ^2}{3}\,.
\ee

Thus, for $\Delta_{\rm conv}$ up the 3-loop order we have
\be
\Delta_\text{conv} = {2\pi^2 g^2\over 3} + \[ -{6\pi^4 \over 45}+\frac{16 \pi ^2}{3 a^2}\]  g^4 +{\cal O}(g^6)\,.
\ee
Which, together with \nref{sourcescontri}, gives rise to
\be
\mC_a = 4(-1)^a g^2 + 8(-1)^a \[\pi^2-{4\over a^2}\]g^4 +16(-1)^a \[{\pi^4\over 3}-{4\pi^2\over a^2}+{20\over a^4}\]g^6  +{\cal O}(g^8)\,.
\ee

\subsection{Recurrence resolvent}
\label{resolvent}
For solving perturbatively the small $\phi$ TBA equations we are faced with certain recurrence equations, all of same form
but with different inhomogeneities. They can be solved in terms of the resolvent  $\hat \chi_n^{\ m} = f(m,n)$,for $n\geq m$  and $\hat \chi_n^{\ m} = f(n,m)$  for $n<m$ with
\be
f(m,n) = - 2\sqrt{2} {(\cosh{\tfrac{w}{2}}+n\sinh{\tfrac{|w|}{2}})
(\cosh{\tfrac{w}{2}}\sinh{\tfrac{m|w|}{2}}-m\sinh{\tfrac{|w|}{2}}\cosh{\tfrac{m w}{2}}) \over  m n (\cosh w -1)^{3/2} } e^{-{n|w|\over 2}}
\ee
This resolvent $\hat \chi_n^{\ m}$ is the solution to
\be
2\cosh{w\over2} {n^2\over n^2 -1}\hat \chi_n^{\ m}-\hat \chi_{n-1}^{\ m}-\hat \chi_{n+1}^{\ m} = \delta_n^{\ m}\,.
\label{recuresolvent}
\ee
This can be transformed into our recurrence relations by setting $\chi_m = { m^2 \over m^2 -1} \hat \chi$.

 \end{document}